\documentclass[12pt]{article}
\pdfoutput=1
\usepackage[centertags]{amsmath}
\allowdisplaybreaks[1]
\usepackage{array,multirow}
\numberwithin{equation}{section}
\usepackage{amssymb}
\usepackage{float}
\usepackage{graphicx}
\usepackage{color}
\usepackage{relsize}
\usepackage{verbatim}
\usepackage{mathtools}
 
\usepackage{epsfig}

\def\tr{{\rm Tr}}

\def\Or[#1]{{\text{O}}\left({#1}\right)}
\def\dotl[#1,#2]{\left\langle #1, #2 \right\rangle}
\def\dotlb[#1,#2]{[ #1, #2 ]}
\def\dotp[#1,#2]{(#1) \cdot (#2)}
\def\aff[#1,#2]{\hat{#1}(#2)}
\def\n4sym{{\cal N}=4 SYM}
\def\>{\rangle}
\def\<{\langle}
\def\weight[#1,#2,#3]{\{(#1),#2,#3\}}
\def\ads[#1]{$\text{AdS}_{#1}$}

\newcommand{\ba}{\begin{eqnarray}}
\newcommand{\ea}{\end{eqnarray}}
\newcommand{\be}{\begin{eqnarray}}
\newcommand{\ee}{\end{eqnarray}}
\newcommand{\bq}{\begin{equation}}
\newcommand{\eq}{\end{equation}}
\newcommand{\benn}{\begin{equation*}}
\newcommand{\eenn}{\end{equation*}}
\newcommand{\bi}{\begin{itemize}}  
\newcommand{\ei}{\end{itemize}}

\newcommand{\CF}{{\cal F}}

\newcommand{\CL}{{\cal L}}

\newcommand{\CM}{{\cal M}}

\newcommand{\CO}{{\cal O}}

\newcommand{\CV}{{\cal V}}
\newcommand{\nn}{\nonumber}

\newcommand\oo\infty
\newcommand\s\sigma
\newcommand\de\delta
\newcommand\De\Delta

\newcommand\f\phi
\newcommand\g\gamma
\newcommand\x\times

\newcommand{\hol}{{\mathrm{holo}}}

\usepackage{jheppub}

\makeatletter
\def\@fpheader{\vspace{-.1cm}}
\makeatother

\title{The AdS$_3$ Propagator  and the Fate of Locality}

\author[a]{Hongbin Chen,}
\author[b]{A. Liam Fitzpatrick,} 
\author[a,c,d]{Jared Kaplan,}
\author[a]{Daliang Li}
\affiliation[a]{Department of Physics and Astronomy,  Johns Hopkins University, \\
Charles Street, Baltimore, MD 21218, USA} 
\affiliation[b]{Department of Physics, Boston University, \\
Commonwealth Avenue, Boston, MA 02215, USA}
\affiliation[c]{Center for Quantum Mathematics and Physics (QMAP) \\
University of California, Davis, California 95616, USA}
\affiliation[d]{Stanford Institute for Theoretical Physics \\
Stanford University, Palo Alto, CA 94305, USA  } 

\abstract{ 
We recently used Virasoro symmetry considerations to propose an exact formula for a bulk proto-field $\phi$ in AdS$_3$.  In this paper we study the propagator $\< \phi \phi \>$.  We show that many techniques from the study of conformal blocks can be generalized to compute it, including the semiclassical monodromy method and both forms of the Zamolodchikov recursion relations.  When the results from recursion are expanded at large central charge, they match gravitational perturbation theory for a free scalar field coupled to gravity in our chosen gauge.  

We find that although the propagator is finite and well-defined at long distances, its perturbative expansion in $G_N = \frac{3}{2c}$ exhibits UV/IR mixing effects.  If we nevertheless interpret $\< \phi \phi \>$ as a probe of bulk locality, then when $G_N m_\phi \ll 1$ locality breaks down at the new short-distance scale $\sigma_*  \sim \sqrt[4]{G_N R_{AdS}^3}$.  For $\phi$ with very large bulk mass, or at small central charge, bulk locality fails at the AdS length scale.  In all cases, locality `breakdown' manifests as singularities or branch cuts at spacelike separation arising from non-perturbative quantum gravitational effects.
}

\arxivnumber{}
 
\begin{document} 
    
\maketitle
\flushbottom
 
\section{Introduction and Summary}

General Relativity does not seem, at first glance, so very different from other effective quantum field theories.  When we study GR in a perturbative expansion about a semiclassical background, it is tempting to interpret localized gravitational and matter fluctuations as the degrees of freedom that define the space of states.  But the area-law entropy of black hole thermodynamics \cite{Bekenstein:1973ur} starkly conflicts with this picture, which was already suspect due to considerations of diffeomorphism gauge redundancy \cite{DeWitt:1967yk}.   So we must ask, to what extent can the conflicting viewpoints of local bulk effective field theory and holography be reconciled?  

Our goal is to understand the limitations of bulk locality in a concrete, quantitative way in the context of AdS$_3$/CFT$_2$.   This is a necessary step towards the resolution of the black hole information paradox   in AdS/CFT \cite{Maldacena, GKP, Witten}, because the most striking form of the paradox is a disagreement between unitarity and effective field theory in the bulk that depends on the approximate existence of  local bulk observables.

We recently proposed an exact definition \cite{Anand:2017dav} of a local bulk proto-field $\phi$ associated with a specific CFT$_2$ primary operator $\CO$.    Physically, one can think of $\phi$ as the nearest one can get to defining a free scalar field coupled to AdS$_3$ gravity in a specific coordinate system, neglecting loops of $\phi$ itself.  This bulk field operator automatically `knows' about the dynamical gravitational background, or in other words it is `gravitationally dressed'.  A simple algebraic  definition  \cite{Anand:2017dav} for $\phi$ exists because, roughly speaking, quantum gravity matrix elements in AdS$_3$ are determined by Virasoro symmetry.  

In this work we will study the simplest local bulk observable, the vacuum propagator $K = \< \phi \phi \>$.   We will compare perturbation theory in $G_N = \frac{3}{2c}$, semiclassical methods, and exact numerical results.  The computations we will present are possible because $\phi$ has a very natural definition in CFT$_2$, which means that many techniques for the efficient calculation of conformal blocks can be generalized to the study of $\phi$ correlators.  In particular, both the semiclassical `monodromy method' \cite{Monodromy, HartmanLargeC, Fitzpatrick:2014vua, Fitzpatrick:2015zha, Anous:2016kss} and the Zamolodchikov recursion relations \cite{ZamolodchikovRecursion, Zamolodchikovq, Zamolodchikov:1986gh} can be adapted and recruited to our cause.

In the remainder of this section we will separately summarize the technical machinery we have developed and the physical results we have obtained.

\subsection*{Notation}

Throughout this paper we use $h$ to refer to the conformal dimension of the primary operator $\CO$ dual to the bulk proto-field $\phi$ with mass $m^2 R_{AdS}^2 = 4h(h-1)$.  The CFT$_2$ central charge is $c = \frac{3R_{AdS}}{2G_N}$.  When discussing  the propagator $K \equiv \< \phi(X_1) \phi(X_2) \>$ (we use $K$ and $\<\phi\phi\>$ to denote the propagator interchangeably) we often use the kinematic variable $\rho = e^{-2 \sigma}$, where $\sigma(X_1, X_2)$ is the geodesic distance between the bulk points in the vacuum. In our coordinate system, the metric of  empty AdS$_3$ is
\be
ds^2 = \frac{dy^2 + dz d \bar z}{y^2}.
\ee
We compute the propagator in the AdS$_3$ vacuum throughout.  Explicitly, we have $X_1=(y_1,z_1,\bar z_1)$, $X_2=(y_2,z_2,\bar z_2)$ and $\rho \equiv  \frac{\xi^2}{\left(1 + \sqrt{1 - \xi^2}\right)^2}$ with $\xi \equiv \frac{2 y_1 y_2}{y_1^2 +y_2^2 + z_{12} \bar z_{12}}$, where $ \rho=\xi =1$ corresponds to vanishing separation in the bulk. In this coordinate system, the free field propagator, which we'll denote as $K_{\text{global}}\equiv \< \phi\phi\>_{\text{global}}$ is given by \footnote{The subscript "global" here means that $K_\text{global}$ is the 2-pt function of $\phi_\text{global}$, which is the reconstruction of a bulk field using only global conformal symmetry. }
\begin{equation}
K_{\text{global}}= \< \phi\phi\>_{\text{global}}=\frac{\rho^h}{1-\rho}.
\label{eq:PhiPhiGlobal}
\end{equation}
$K_{\text{global}}$ is the large $c$ limit of $K$, ie $K_\text{global}=\lim_{c\rightarrow\infty}K$. We also study the `holomorphic part' of $K$ due to purely holomorphic gravitons, which we denote as $K_{\text{holo}}=\<\phi\phi\>_{\text{holo}}$ and define in Section \ref{se:HoloPhiPhiDefinition}.

\subsection{Summary of Technical Developments}

In section \ref{sec:Structurephi} we briefly review $\phi$ \cite{Anand:2017dav}, and then discuss the properties of its correlators.  We introduce the technically useful notion of the `holomorphic part' $\phi_\hol$, which corresponds in perturbation theory to computing $\phi$ correlators while only incorporating holomorphic gravitons.  We show that knowledge of the $\phi_\hol$ propagator $K_\hol$ can be combined with purely global-conformal information to determine the complete $\phi$ propagator.  We also emphasize that at two-loops and beyond the full propagator is not spherically symmetric as a consequence of our gauge choice.  The full propagator depends on both $\rho$ and an angle of inclination with respect to the $z$-$\bar z$ plane.

Conformal blocks in CFT$_2$ exponentiate in the semiclassical approximation of large central charge  where ratios of conformal dimensions to the central charge, $h/c$, are held fixed.  This is dual to the semiclassical limit $G_N \to 0$ with $G_N m$ fixed in AdS$_3$.   In section \ref{sec:Semiclassical} we show that the propagator also has a semiclassical limit, and we derive a generalization of the monodromy method \cite{Monodromy, HartmanLargeC, Fitzpatrick:2014vua, Fitzpatrick:2015zha, Anous:2016kss} that computes the semiclassical $K_\hol$.  We then use this method to obtain  the semiclassical propagator to order $\frac{h^2}{c}$ in equation (\ref{eq:semiclassicalhoverc}) and at large\footnote{At large $h$ it is natural to define a variable $q$, as in equation (\ref{eq:Definitionq}); the variables $\xi, \rho, q$ play a similar role here as the variables $z, \rho, q$ used in the study of Virasoro blocks, though here we gain no advantage in convergence by using $q$ in place of $\rho$.} $h$ in equation (\ref{eq:LargehSC}).  At infinite $h$, we are able to go beyond the semiclassical limit and derive an exact expression for the block in  (\ref{eq:exactlargeh}).  Finally, while we cannot obtain an analytic expression for the semiclassical part of the correlator at general $h/c$, it is straightforward to use the monodromy method to determine it numerically.  We apply this technique to determine the critical value of the geodesic distance where the semiclassical part first develops an imaginary piece; the result is summarized in figure \ref{fig:NumMono1}.

In section \ref{sec:ExactlyfromRecursion} we derive recursion relations that compute the propagator exactly as an expansion in small $\rho$ and $q$ (these are long-distance expansions).  We find generalizations of both the $c$ and $h$ Zamolodchikov recursion relations \cite{ZamolodchikovRecursion, Zamolodchikovq, Zamolodchikov:1986gh}, though we mostly use $h$-recursion as it is more efficient algorithmically.  It is summarized by equation (\ref{eq:hRecursion}).  The large $h$ limit of equation (\ref{eq:exactlargeh}) is a crucial ingredient needed for $h$-recursion.

We discuss the perturbative expansion of the propagator in section \ref{sec:phiphiinoneoverc}. The full one-loop result is equation (\ref{eq:OneLoopResult}).  We show in section \ref{sec:OneLoop} and appendix \ref{app:PerturbativePropagator} that our result agrees with the bulk one-loop Witten diagram. We also provide an explicit unitarity-based argument in the appendix, which ultimately relates the one-loop correction to the tree-level correlator $\< \phi \CO T \>$ computed previously \cite{Anand:2017dav}.

\subsection{UV/IR Mixing in Perturbation Theory}

Before we analyze the interesting features of non-perturbative gravity, we must discuss a surprising result that is already visible at one-loop in gravitational perturbation theory!  As discussed in detail in section \ref{sec:OneLoop}, we find that in the short-distance limit $\sigma \ll R_{AdS}$, the one-loop corrected bulk propagator takes the form
\be
\< \phi \phi \>  &\approx &
   \frac{1}{\sigma} \left(1 +  \frac{3 G_N R_{AdS}^3 }{2 \sigma^4}-\frac{G_N R_{AdS} (10+m^2 R_{AdS}^2) }{4 \sigma^2} + \cdots \right) .
\ee
Notice that the one-loop correction  is very singular at short-distances, so that it competes with the free field propagator at $\sigma_* \sim \sqrt[4]{G_N R_{AdS}^3}$.   In contrast, we might have expected a one-loop correction that scaled like $\frac{G_N}{\sigma}$, so that it only became important for separations of order the Planck length.  Instead we have discovered an intermediate  scale that mixes the UV Planck scale with the IR scale of $R_{AdS}$.  This  UV/IR mixing is not what one would expect for a local observable\footnote{Of course our $\phi$ must be accompanied by `gravitational Wilson lines', in the same way that the physical electron field must be attached to a Wilson line.  So $\phi$ correlators are not truly local.} in a local theory, and it would lead to power-law IR divergences if we were to take the flat space limit of AdS. 

One could interpret this result as an indication that we should modify the definition of $\phi$ or $K$ to eliminate this UV/IR mixing.  For a variety of reasons discussed in section \ref{sec:OneLoop}, it would seem that the required modifications would have to be rather consequential. In particular, since our results agree with bulk Witten diagrams at large $c$, the same modifications also apply to these Witten diagram calculations in AdS$_3$. Nevertheless, we believe this is an interesting avenue for future exploration.  For the rest of this summary (and most of the paper) we will just study the naive vacuum $\phi$ propagator and {\it assume} that its correlators provide a meaningful probe of bulk locality, but one should keep in mind the caveat that the results could be different if we were to identify an  observable free from UV/IR mixing.

\subsection{Physics of the Exact Propagator and the Breakdown of Locality}
\label{sec:SummaryPhysics}

By construction, $\phi$ is a real scalar field and its propagator should be a real-valued function.  Both the propagator $K$ and the holomorphic part $K_\hol$ should not develop imaginary parts, because there are no states for $\phi$ quanta to decay into.\footnote{One can formalize these expectations for a local $\phi$ using the Kallen-Lehmann representation.  Readers may wonder if $\phi$ can decay into gravitons, but this is forbidden for the proto-field.  Specifically, all correlators $\< \phi T \cdots T \bar T \cdots \bar T\>$ vanish because $\phi$ is a linear combination of descendants of the Virasoro primary $\CO$. In order for these correlators to be turned on, one would have to also include dressing of $\phi$ by operators that are not Virasoro descendants of $\CO$.}  So if we find that the exact $K$ or $K_\hol$ develop imaginary parts at spacelike separation, then we may interpret this as a violation of bulk unitarity, even though the CFT itself remains perfectly healthy; potentially, the proto-field $\phi$ may be indicating the presence of an instability that arises when two $\phi$s are brought close together in the full bulk gravitational theory, where a complete bulk field would include not only $\CO$ and its descendants but other states as well.  In order to develop an imaginary part at a distance $\sigma_*$,  $K$ must exhibit a singularity at spacelike separation, which also represents a direct violation of bulk locality.

In the global or $c=\infty$ limit we have $K_{\text{global}}=\lim_{c\rightarrow \infty}K = \frac{\rho^h}{1-\rho}$, which is real and finite for all spacelike geodesic separations $\sigma = -\frac{1}{2} \log \rho$.  Furthermore, \emph{to all orders in perturbation theory}, the propagator remains real and finite.  However, we find that in various limits the exact propagator develops new singularities (branch cuts) indicating violations of bulk locality.  Specifically:
\begin{itemize}
\item When studying light bulk fields with $h \ll c$, we can resum the the full $1/c$ expansion in the short-distance limit (section \ref{sec:AllOrdersoneoverc}).  The result is ambiguous, but generically includes an imaginary piece associated with the length scale $\sigma_* \propto c^{-1/4}$.  We obtain substantial numeric evidence (section \ref{sec:phiphinumerical}) that the light-field propagator develops a singularity at a finite separation that scales as $\sigma_* \propto c^{-1/4}$.  In figure \ref{fig:HoloVsMixedFit} we display evidence that the full and holomorphic propagators show the same scaling of $\sigma_*$ and $c$.
\item Semiclassical results at $c \to \infty$ with $\frac{h}{c} \ll 1$ fixed indicate an apparent breakdown of locality at $\sigma_* \propto \left( \frac{h}{c} \right)^{\frac{1}{3}}$ (section \ref{subsec:Monodromyhc}).  This is corroborated by semiclassical numerics (section \ref{sec:NumMono}) and by exact numerics (figure \ref{fig:SemiclassicalNumericalPlot}), which also demonstrate that our semiclassical results are reliable at large $c$ and spacelike separation.
\item In the heavy bulk field limit $h \gg c$, we find the exact propagator analytically (section \ref{sec:LargehLimit}) and demonstrate that it develops a branch cut at $\sigma_* = R_{\rm AdS} \log(2+\sqrt{3}) \approx 1.32 R_{AdS}$.  Thus for heavy bulk fields, locality breaks down at the AdS scale.  We find numerically that in the limit of large $c$ and fixed $h/c$, our results smoothly interpolate (figure \ref{fig:SemiclassicalNumericalPlot}) between large $h$ and the fixed $h \ll c$ scaling  $c^{-1/4}$.  Moreover, we show that  the behavior at large $h$ and at very small $c$ appear to be identical (figure \ref{fig:NumericFitsFinalDataZeroh}), with locality breaking down at the same numerical multiple of the AdS scale in both cases.
\end{itemize}
Aside from the surprising UV/IR mixing effect and associated emergent scale $\sigma_*$ discussed above, this is roughly what one might have expected.  Bulk locality makes approximate sense in gravitational perturbation theory, but breaks down due to non-perturbative gravitational effects in an explicitly quantifiable way.  Light fields in theories with a large separation of scales between $G_N$ and $R_{AdS}$ can be local to a high degree of precision, but outside this regime bulk locality breaks down at the AdS length.

\section{Structure of  $\phi$ Correlators}
\label{sec:Structurephi}

In recent work \cite{Anand:2017dav} we provided an exact definition for the bulk scalar proto-field $\phi(y, z, \bar z)$ as a linear combination of a primary CFT$_2$ scalar $\CO$ and its Virasoro descendants.\footnote{There have been many approaches to reconstruction, for an incomplete sample see \cite{Banks:1998dd, Bena:1999jv, Hamilton:2005ju, Hamilton:2006az, Kabat:2011rz, Kabat:2012av, Kabat:2013wga, Heemskerk:2012mn, Papadodimas:2012aq, Kabat:2015swa, Nakayama:2015mva, Guica:2015zpf, Guica:2016pid, Kabat:2016zzr, Czech:2016xec, Nakayama:2016xvw, Faulkner:2017vdd, Almheiri:2017fbd,Verlinde:2015qfa,Lewkowycz:2016ukf}.
}
We refer to $\phi$ as (merely) a proto-field because its existence follows entirely from symmetry considerations in AdS$_3$/CFT$_2$.  One might expect that full scalar fields\footnote{Full bulk scalar fields may not exist, and to the extent that they do exist, their definition may be ambiguous.  These are interesting issues but we will not be addressing them here, as we will only be studying  proto-fields and their correlators.  } can be represented as infinite sums of proto-fields \cite{Kabat:2011rz, Kabat:2012av, Kabat:2013wga, Kabat:2016zzr}.  We can also think of the proto-field as a free scalar field in AdS$_3$ coupled to pure quantum gravity, where loops of $\phi$ itself have been neglected.

The proto-field is interesting because it encodes an infinite sum of quantum gravitational effects, which involve Virasoro (CFT stress tensor) matrix elements.  For example, we will see that the  propagator $\< \phi(X_1) \phi(X_2) \>$ includes graviton loops to all-orders.   The proto-field is labeled by a bulk point $(y,z, \bar z)$ associated with a specific coordinate system (or gauge choice) where AdS$_3$ vacuum metrics take the form \cite{Banados:1998gg, Roberts:2012aq}
\be
\label{eq:MetricwithT}
ds^2 = \frac{dy^2 + dz d \bar z}{y^2} - \frac{6 T(z)}{c} dz^2   - \frac{6 \bar T(\bar z)}{c} d \bar z^2 + y^2 \frac{36 T(z) \bar T( \bar z)}{c^2} dz \bar dz
\ee
for holomorphic functions $T(z), \bar T(\bar z)$.  We emphasize that the proto-field operator depends in an essential way on this gauge choice; were we to choose a different gauge, we would obtain a different bulk operator.  The dependence on the gauge will appear explicitly later on, where we will see that in our gauge, the full propagator $\< \phi(X) \phi(Y) \>$ is not spherically symmetric.  

We will briefly review the definition of $\phi$ in section \ref{sec:Review}; for detailed explanations and derivations we refer the reader to \cite{Anand:2017dav}.  The operator $\phi$ and its correlators do not factorize into a product of holomorphic and anti-holomorphic parts.  However, it is possible to define a `holomorphic' part  $\phi_\hol$, by which we mean that we only include the effects of holomorphic gravitons on $\phi$.  We explain these facts and define $\phi_\hol$ in section \ref{sec:HolomorphicDefinitions}.  This notion is useful because full $\phi$ correlators can be  determined from $\phi_\hol$ correlators using additional data that only depends on global conformal information.  Throughout this paper we will primarily be studying $\phi_\hol$ correlators.

\subsection{Brief Review of the AdS$_3$ Proto-field $\phi$}
\label{sec:Review}

We define the operator $\phi(y,z, \bar z)$ using a Boundary Operator Expansion (BOE)
\be \label{eq:BOEofphi}
\phi\left(y, z, \bar z\right)  = \sum_{N=0}^{\infty} y^{2h+2N} \phi_N(z, \bar z) .
\ee
Each operator $\phi_N(z, \bar z)$ can be defined by first translating $z \to 0$ and then applying the operator/state correspondence to study the state $| \phi \>_N = \phi_N(0,0) | 0 \>$.  These states are then defined by the bulk primary conditions
\be
L_{m} \left|\phi\right\rangle_{N}=0,\quad\overline{L}_{m}\left|\phi\right\rangle _{N}=0,\qquad\text{for }m\ge2.
\ee
along with a normalization condition
\be
L_{1}^{N}\overline{L}_{1}^{N}\left|\phi\right\rangle _{N}=\left(-1\right)^{N}N!\left(2h\right)_{N}\left|\CO \right\rangle .
\ee
The bulk primary condition can be given the simple, physical interpretation that the line $(y, 0, 0)$ is fixed by $L_{m \geq 2}$ in our gauge.  The normalization condition simply guarantees that we recover the global conformal bulk reconstruction when $c = \infty$.  These conditions have a unique solution \cite{Anand:2017dav}, which can be conveniently written 
\be \label{eq:SolutionInMathemcalL}
\phi(y,0,0)  = \sum_{N=0}^\infty y^{2h + 2N} \lambda_N \CL_N \bar \CL_N \CO(0) ,
\ee
where $\lambda_N \equiv \frac{(-1)^N}{(2h)_N N!}$ and the $\CL_N$ are a certain linear combination of holomorphic Virasoro generators\footnote{For example, the explicit solution at level 2 is
 \be
 \label{eq:MathematicalLMinus2}
\mathcal{L}_{-2}=\frac{(2 h+1) (c+8 h)}{\left(2h+1\right)c+2 h (8 h-5)}\left(L_{-1}^2-\frac{12h}{c+8h}L_{-2}\right)
 \ee
with $\bar \CL_{-2}$ only differing by $L_{-n} \to \bar L_{-n}$.  } at level $N$, and similarly for the anti-holomorphic $\bar \CL_N$. When $c \to \infty$ with other parameters held fixed, our prescription reduces to the global conformal bulk reconstruction  of $\phi$ that can be obtained from the `HKLL kernel' \cite{Banks:1998dd, Hamilton:2005ju, Nakayama:2015mva}, and we have the simplification $\CL_N \to L_{-1}^N$. 

This prescription for $\phi$ can be motivated in a number of ways; for details see \cite{Anand:2017dav}.   When Virasoro transformations are realized as bulk diffeomorphisms preserving the gauge choice of equation (\ref{eq:MetricwithT}), our definition emerges by demanding that $\phi(y, z, \bar z)$ transforms as a bulk scalar field.  Alternatively, one can arrive at our prescription by studying correlators of $\phi$ with $\CO(x)$ and any number of stress tensors $T(z_i)$ and $\bar T(\bar z_i)$.  After gauge fixing, Virasoro symmetry appears to determine these correlators exactly \cite{Fitzpatrick:2016mtp, Anand:2017dav}, and their specification is equivalent to our definition of $\phi$.  In more conventional terms, our definition of $\phi$ should agree with bulk gravitational perturbation theory to all orders in $G_N = \frac{3}{2c}$, and this has been verified explicitly to order $1/c^3$ for some observables.  In section \ref{sec:OneLoop} we will verify the agreement between one-loop gravitational perturbation theory and our prescription for the propagator $\< \phi \phi \>$.

\subsubsection*{Solution for $\phi$ Using Quasi-Primaries}

For various purposes it is useful to solve for $\phi_N$ explicitly in terms of quasi-primary states, which are annihilated by $L_1$ but not $L_m$ with $m\ge 2$.  Importantly, we will take the quasi-primaries to be orthogonal, and we fix their overall normalization by demanding that a level $M$ quasi-primary includes the term $L_{-1}^M$ with overall coefficient $1$.   In this basis, we showed\footnote{For clarity, by $\mathcal{L}_{-N}^{\text{quasi}}$ we mean $\mathcal{L}_{-N}^{\text{quasi}}$ acting on $\mathcal{O}$ creates a level $N$ quasi-primary, while $\mathcal{L}_{N}$ defined in equation (\ref{eq:SolutionInMathemcalL}) is the sum of all level $N$ contributions to $\phi$ and it's given by
\begin{equation}
\mathcal{L}_{N}=L_{-1}^{N}+\frac{\left|L_{-1}^{N}\mathcal{O}\right|^{2}}{\left|L_{-1}^{N-2}\mathcal{L}_{-2}^{\text{quasi}}\mathcal{O}\right|^{2}}L_{-1}^{N-2}\mathcal{L}_{-2}^{\text{quasi}}+\frac{\left|L_{-1}^{N}\mathcal{O}\right|^{2}}{\left|L_{-1}^{N-3}\mathcal{L}_{-3}^{\text{quasi}}\mathcal{O}\right|^{2}}L_{-1}^{N-3}\mathcal{L}_{-3}^{\text{quasi}}+\cdots
\end{equation}
and $\left|L_{-1}^{N}\mathcal{O}\right|^{2}=\left|\overline{L}_{-1}^{N}\mathcal{O}\right|^{2}=\left(2h\right)_{N}N!=\frac{1}{\left|\lambda_{N}\right|}$. 
} that  \cite{Anand:2017dav}
\begin{align}
\label{eq:SolutionforPhiQP}
\phi_N & =\left(-1\right)^{N}\left|L_{-1}^{N} \mathcal{O}\right|^2  \left(\frac{L_{-1}^{N}}{\left|L_{-1}^{N}\mathcal{O}\right|^2 }+\frac{\mathcal{L}_{-N}^{\text{quasi}}}{\left|\mathcal{L}_{-N}^{\text{quasi}}\mathcal{O}\right|^2}+\frac{L_{-1} \mathcal{L}_{-\left(N-1\right)}^{\text{quasi}}}{\left|L_{-1}\mathcal{L}_{-\left(N-1\right)}^{\text{quasi}}\mathcal{O}\right|^2}+\cdots\right) 
\nn \\
 & \quad  \times \left(\frac{ \bar L_{-1}^{N}}{\left| \bar L_{-1}^{N}\mathcal{O}\right|^2 }+\frac{\bar \CL_{-N}^{\text{quasi}}}{\left|\bar \CL_{-N}^{\text{quasi}}\mathcal{O}\right|^2}+\frac{\bar L_{-1} \bar \CL_{-\left(N-1\right)}^{\text{quasi}}}{\left| \bar L_{-1} \bar \CL_{-\left(N-1\right)}^{\text{quasi}}\mathcal{O}\right|^2}+\cdots\right) \CO ,
\end{align}
where the notation is slightly schematic, as each term represents a sum over all quasi-primaries  at the indicated level.  

Once we establish the overall coefficient of the quasi-primary contributions at level $(N, \bar N)$, the contributions of all global conformal descendants of these quasi-primaries are fixed.  Thus much of the non-trivial information required to define correlators of $\phi(X)$ is encoded in sums over inverse normalization factors
\be
\label{eq:CNDefinition}
C_{N} \equiv \sum_{i=1}^{p\left(N\right)-p\left(N-1\right)}\frac{1}{\left|\mathcal{L}_{-N}^{\text{quasi},i}\mathcal{O}\right|^{2}},
\ee
where the sum includes all quasi-primaries at level $N$ ($p(N)$ denotes the number of integer partitions, and the super-script $i$ denotes the $i$-th quasi-primary at level $N$).   We can take advantage of this fact by finding efficient methods for isolating and determining the $C_N$ \cite{Cho:2017oxl}, and then recombining them to compute $\phi$ correlators.

\subsection{`Holomorphic' Parts Determine Full Correlators}
\label{sec:HolomorphicDefinitions}

In CFT$_2$, many observables can be  decomposed into holomorphic and anti-holomorphic parts.  For example, the conformal partial waves or conformal blocks involve sums over all states related by conformal symmetry.  Since the symmetry algebra is a product of holomorphic and anti-holomorphic Virasoro algebras, conformal blocks can thus be written as products  $\CV \times \bar \CV$.  This feature leads to many convenient simplifications.  Due to the $y$-dependence of $\phi(y, z, \bar z)$, this property \emph{does not} hold for $\phi$ correlators, but we can still take advantage of something almost as useful, which can be summarized by equations (\ref{eq:phiphifromCnn}), (\ref{eq:CnnFactorizes}), and (\ref{eq:PhiPhiHoloDefinition}). 
 
\subsubsection{The Full Correlator $\<\phi\phi\>$}
\label{sec:FromHoloToFull}

Computing $\left\langle \phi\phi\right\rangle$ using the quasi-primary
decomposition in equation (\ref{eq:SolutionforPhiQP}) is useful because distinct quasi-primaries
(and their global descendants) have vanishing two-point correlators.
So we can write $\left\langle \phi\phi\right\rangle $ as a sum over contributions
from different quasi-primaries, that is, 
\begin{equation}
\left\langle \phi(y_1,z_1,\bar z_1)\phi(y_2, z_2, \bar z_2)\right\rangle =\sum_{n,\overline{n}}\sum_{i,j}\left\langle \phi_{i,j}^{n,\overline{n}}(y_1,z_1,\bar z_1)\phi_{i,j}^{n,\overline{n}}(y_2,z_2,\bar z_2)\right\rangle, \label{eq:PhiPhiSum} 
\end{equation}
where the sum $\left(n,\overline{n}\right)$ is over different
levels for the quasi-primaries, and the sum $\left(i,j\right)$
is over all of the different quasi-primaries at level $\left(n,\overline{n}\right)$.
By $\phi_{i,j}^{n,\overline{n}}$ we denote the contribution to $\phi$ from the quasi-primary
$\mathcal{L}_{-n}^{\text{quasi},i}\overline{\mathcal{L}}_{-\overline{n}}^{\text{quasi},j}\mathcal{O}$
and all its global descendants
\begin{small}
\begin{equation}
\phi_{i,j}^{n,\overline{n}}(y,z,\bar z)\equiv y^{2h+2n}\sum_{m=0}^{\infty}\left(-1\right)^{n+m}y^{2m}\left|L_{-1}^{n+m}\mathcal{O}\right|^{2}\frac{L_{-n}^{m}\mathcal{L}_{-n}^{\text{quasi},i}}{\left|L_{-1}^{m}\mathcal{L}_{-n}^{\text{quasi},i}\mathcal{O}\right|^{2}}\frac{\overline{L}_{-1}^{m+n-\overline{n}}\mathcal{\overline{L}}_{-\overline{n}}^{\text{quasi},j}}{\left|\overline{L}_{-1}^{m+n-\overline{n}}\overline{\mathcal{L}}_{-\overline{n}}^{\text{quasi},j}\mathcal{O}\right|^{2}}\mathcal{O}(z,\bar z),
\end{equation}
\end{small}
where without loss of generality, we assume $n\ge\overline{n}$. The above equation can be read off from equation (\ref{eq:SolutionInMathemcalL}) and equation (\ref{eq:SolutionforPhiQP}).  As we'll show in Appendix \ref{app:ComputingKdF}, $\left\langle \phi_{i,j}^{n,\overline{n}}\phi_{i,j}^{n,\overline{n}}\right\rangle $
is given by 
\begin{equation}
\left\langle \phi_{i,j}^{n,\overline{n}}\phi_{i,j}^{n,\overline{n}}\right\rangle =\frac{1}{\left|\mathcal{L}_{-n}^{\text{quasi},i}\overline{\mathcal{L}}_{-\overline{n}}^{\text{quasi},j}\mathcal{O}\right|^{2}}\mathcal{F}_{n,\overline{n}}\left(h\right),
\end{equation}
where $\mathcal{F}_{n,\overline{n}}\left(h\right)$ only depends on
the level of the quasi-primary $\left(n,\overline{n}\right)$ and it's symmetric under exchange of $n$ and $\bar n$. So we can 
write $\left\langle \phi\phi\right\rangle $ as 
\begin{equation}
\label{eq:phiphifromCnn}
\left\langle \phi\phi\right\rangle =\sum_{n,\overline{n}}\left(\sum_{i,j}\frac{1}{\left|\mathcal{L}_{-n}^{\text{quasi},i}\overline{\mathcal{L}}_{-\overline{n}}^{\text{quasi},j}\mathcal{O}\right|^{2}}\right)\mathcal{F}_{n,\overline{n}}\left(h\right)\equiv\sum_{n,\overline{n}=0}^{\infty}C_{n,\overline{n}}\mathcal{F}_{n,\overline{n}},
\end{equation}
where we define $C_{n,\overline{n}}$ to be the sum over the inverse
of all quasi-primaries at level $\left(n,\overline{n}\right)$, and
it factorizes as 
\begin{equation}
\label{eq:CnnFactorizes}
C_{n,\overline{n}}=C_{n}C_{\overline{n}}=\left(\sum_{i=1}^{p\left(n\right)-p\left(n-1\right)}\frac{1}{\left|\mathcal{L}_{-n}^{\text{quasi},i}\mathcal{O}\right|^{2}}\right)\left(\sum_{j=1}^{p\left(\overline{n}\right)-p\left(\overline{n}-1\right)}\frac{1}{\left|\mathcal{\overline{L}}_{-\overline{n}}^{\text{quasi},j}\mathcal{O}\right|^{2}}\right).
\end{equation}
So we only need to compute $C_{n}$ to determine $C_{n,\overline{n}}$.
In section \ref{sec:ExactlyfromRecursion}, we'll show that $C_{n}$ can be obtained by modifying
Zamolochikov's recursion relations for Virasoro blocks.  In the semiclassical limit, the $C_n$ can also be determined by the monodromy method of section \ref{sec:Semiclassical}.

To get $\left\langle \phi\phi\right\rangle $, we also need to compute
$\mathcal{F}_{n,\overline{n}}$. We will show in Appendix \ref{app:ComputingKdF} that $\mathcal{F}_{n,\overline{n}}$
is given by the Kampe de Feriet (KdF)\footnote{The general form of a KdF series is given by 
\begin{equation}
\resizebox{\textwidth}{!}{$
F_{r:s}^{p:q}\left(\begin{array}{cc}
a_{1},\cdots,a_{p}: & b_{1},b_{1}^{'};\cdots;b_{q},b_{q}^{'};\\
c_{1},\cdots,c_{r}: & d_{1},d_{1}^{'};\cdots;d_{s},d_{s}^{'};
\end{array},x,y\right)=\sum_{m=0}^{\infty}\sum_{n=0}^{\infty}\frac{\left(a_{1}\right)_{m+n}\cdots\left(a_{p}\right)_{m+n}}{\left(c_{1}\right)_{m+n}\cdots\left(c_{r}\right)_{m+n}}\frac{\left(b_{1}\right)_{m}\left(b_{1}^{'}\right)_{n}\cdots\left(b_{q}\right)_{m}\left(b_{q}^{'}\right)_{n}}{\left(d_{1}\right)_{m}\left(d_{1}^{'}\right)_{n}\cdots\left(d_{s}\right)_{m}\left(d_{s}^{'}\right)_{n}}\frac{x^{m}y^{n}}{m!n!}
$}
\end{equation}
so these can be viewed as a generalization of hypergeometric functions.  We only need some of the simplest examples of these functions.}  series $F_{0,3}^{2,2}$:
\begin{align}
\mathcal{F}_{n,\overline{n}}\equiv & \left(\frac{y_{1}y_{2}}{z_{12}\overline{z}_{12}}\right)^{2h_{n}}\left(2h_{\overline{n}}\right)_{2l}\left[\frac{n!\left(2h\right)_{n}}{l!\left(2h_{\overline{n}}\right)_{l}}\right]^{2}\\
 & \times F_{0,3}^{2,2}\left(\begin{array}{cc}
2h_{n},2h_{n}: & 2h_n-n,2h_n-n; n+1,n+1;\\
-: & 2h_{n},2h_{n};2h_{n}-l,2h_{n}-l;l+1,l+1;
\end{array},-\frac{y_{1}^{2}}{z_{12}\overline{z}_{12}},-\frac{y_{2}^{2}}{z_{12}\overline{z}_{12}}\right),\nonumber 
\end{align}
with 
\begin{equation}
h_{n}  \equiv h+n,\qquad h_{\overline{n}}\equiv h+\overline{n},\qquad l\equiv n-\overline{n}.
\end{equation}
As far as we know, there is no closed form expression\footnote{In Appendix \ref{app:ComputingKdF} we present an integral expression for $F_{0,3}^{2,2}$ in terms of hypergeometric functions.} for the general KdF series $F_{0,3}^{2,2}$. But in the case that $\overline{n}=0$ (or $n=0$), the above KdF series is given by an Appell $F_{4}$ function, which in our case greatly simplifies to 
\begin{equation}
\mathcal{F}_{n,0}=\left(2h\right)_{2n}\frac{\rho^{h+n}}{1-\rho}.
\end{equation}
where $\rho=e^{-2\sigma}$ with $\sigma=\log\frac{1+\sqrt{1-\xi^{2}}}{\xi}$ and $\xi=\frac{2y_{1}y_{2}}{y_{1}^{2}+y_{2}^{2}+z_{12}\overline{z}_{12}}$.  
Here, $\sigma$ is the geodesic separation between the two bulk operators in pure AdS$_3$. Note that the global (or free field) bulk-bulk propagator $\left\langle \phi\phi\right\rangle _{\text{global}}$
is given by $\mathcal{F}_{0,0}$:
\[
\left\langle \phi\phi\right\rangle _{\text{global}}=\mathcal{F}_{0,0}=\frac{\rho^{h}}{1-\rho} .
\]
This means that the general $\mathcal{F}_{n,0}$ is just proportional to the global bulk-bulk propagator with a shifted bulk mass $h \to h+n$. We'll see that these $\mathcal{F}_{n,0}$ can be summed to give the the holomorphic correlator $\<\phi\phi\>_{\text{holo}}$, which we define next.

\subsubsection{The Holomorphic Correlator $\<\phi\phi\>_{\text{holo}}$}
\label{se:HoloPhiPhiDefinition}
The definition of $\phi$ involves a sum over products of Virasoro
generators $\mathcal{L}_{N}$ and $\overline{\mathcal{L}}_{N}$, which
are related by $L_{n}\leftrightarrow\overline{L}_{n}$. We can define
the non-trivial holomorphic part of $\phi$ as
\begin{equation} \label{eq:DefinitionOfPhiHolo}
\phi_{\text{holo}}\left(y,z,\overline{z}\right)=\sum_{N=0}^{\infty}y^{2h+2N}\lambda_{N}\mathcal{L}_{N}\left(\overline{L}_{-1}\right)^{N}\mathcal{O}\left(z,\overline{z}\right).
\end{equation}
by replacing $\overline{\mathcal{L}}_{N}$ with its $c\rightarrow\infty$
limit $\overline{L}_{-1}^{N}$. This simplified operator $\phi_{\text{holo}}$ is useful because, roughly speaking, it encodes all of the non-trivial quantum gravity information in $\phi$.  As we will show, its two-point function 
\begin{equation}
\left\langle \phi\phi\right\rangle _{\text{holo}}\equiv\left\langle \phi_{\text{holo}}\left(y_{1},z_{1},\overline{z}_{1}\right)\phi_{\text{holo}}\left(y_{2},z_{2},\overline{z}_{2}\right)\right\rangle  \label{eq:HoloPhiPhiDefinition}
\end{equation}
involves all the $C_{n}$ coefficients. 
In large $c$ perturbation theory, the holomorphic part $\phi_{\text{holo}}$ can be understood as the result of including only holomorphic gravitons $h_{zz}$ while neglecting $h_{\bar z \bar z}$. Thus $\phi_{\text{holo}}$ will have valid correlators of the form $\left\langle \phi_{\text{holo}}\mathcal{O}T\cdots T\right\rangle $, but it will not have valid correlators with the anti-holomorphic stress tensor $\overline{T}$. This means that the holomorphic propagator  $\left\langle \phi\phi\right\rangle _{\text{holo}}$ includes holomorphic graviton loops, but not mixed or anti-holomorphic loops.

As one can easily see, $\left\langle \phi\phi\right\rangle _{\text{holo}}$
defined in equation (\ref{eq:HoloPhiPhiDefinition}) can be written as 
\begin{equation}
\left\langle \phi\phi\right\rangle _{\text{holo}}=\sum_{n=0}^{\infty}C_{n}\mathcal{F}_{n,0},
\label{eq:PhiPhiHoloDefinition}
\end{equation}
since we defined $\phi_{\text{holo}}$ in equation (\ref{eq:DefinitionOfPhiHolo}) such that it contains no information about anti-holomorphic Virasoro generators (thus $\bar n=0$ and $C_0=1$). So just as with $\left\langle \phi\phi\right\rangle _{\text{global}}$, the holomorphic propagator $\left\langle \phi\phi\right\rangle _{\text{holo}}$ will only depend
on $\rho$, which means that in our Fefferman-Graham gauge, $\left\langle \phi\phi\right\rangle _{\text{holo}}$
is spherically symmetric. This is not true for $\left\langle \phi\phi\right\rangle$, which will depend on another variable besides $\rho$, specifically an angle with respect to the $z$-$\bar z$ plane, captured for example by the ratio $y_1/y_2$.

Since the contribution to $\left\langle \phi\phi\right\rangle $
from $\mathcal{F}_{0,\overline{n}}$ is the same as $\mathcal{F}_{n,0}$,
we can write $\left\langle \phi\phi\right\rangle $ as 
\begin{equation}
\left\langle \phi\phi\right\rangle =2\left\langle \phi\phi\right\rangle _{\text{holo}}-\left\langle \phi\phi\right\rangle _{\text{global}}+\left\langle \phi\phi\right\rangle _{\text{mixed}},
\label{eq:phiphiParts}
\end{equation}
where $\left\langle \phi\phi\right\rangle _{\text{mixed}}$ is the
contribution from $\mathcal{F}_{n,\overline{n}}$ with $n,\overline{n}>0$
and the substraction of $\left\langle \phi\phi\right\rangle _{\text{global}}=\mathcal{F}_{0,0}$
is necessary because we count it twice in the first term.

In this paper, we will focus mostly on  $\left\langle \phi\phi\right\rangle _{\text{holo}}$. In section \ref{sec:Semiclassical}, we will use monodromy method to obtain the semiclassical limit of $\left\langle \phi\phi\right\rangle _{\text{holo}}$, and in section \ref{sec:ExactlyfromRecursion}, we will provide two recursion relations for computing $\left\langle \phi\phi\right\rangle _{\text{holo}}$ exactly.  We provide some discussion of the mixed and holomorphic terms in section \ref{subsec:FullVSHolo}, and we provide one important and physically relevant comparison, restricted to the $z$-$\bar z$ plane, in figure \ref{fig:HoloVsMixedFit}.

\section{Semiclassical Limit}
\label{sec:Semiclassical}

When studying quantum gravity, it is always important to make contact with the semiclassical limit of general relativity, where $G_N \to 0$ with products like $G_N M$ fixed.  This limit of GR appears directly at the kinematical level in CFT$_2$ \cite{Monodromy, HartmanLargeC, Fitzpatrick:2014vua, Fitzpatrick:2015zha, Anous:2016kss}. Conformal blocks in CFT$_2$ (which are determined by Virasoro symmetry) have a semi-classical limit of the form $e^{c f}$ as we take $c \to \infty$ with the ratios of scaling dimensions to the central charge, $h/c$, held fixed.  This has a beautiful connection with AdS$_3$ gravity via $G_N = \frac{3}{2 c}$ in AdS units, with scaling dimensions playing the role of AdS$_3$ masses.  

Correlators of $\phi$ also behave nicely in this semiclassical limit.  The bulk propagator can be approximated by
\be
\< \phi(y_1, z_1, \bar z_1) \phi(y_2, z_2, \bar z_2) \> \approx e^{c \, g \left( \frac{h}{c}, \xi, r \right)}
\label{eq:phiphiSCform}
\ee
for some function $g$ at large $c$.  In this section we will show how to compute the semiclassical $g_{holo}$ using a generalization of the `monodromy method' \cite{Monodromy, HarlowLiouville, HartmanLargeC, Fitzpatrick:2014vua} that has been used to compute conformal blocks.  Then we will apply our method to calculate $g_{holo}$ perturbatively in small $\frac{h}{c}$, and more importantly, to obtain $g_{holo}$ in the limit $h \to \infty$ in section \ref{sec:LargehLimit}.  In fact, we will be able to determine the large $h$ limit of $\< \phi \phi \>_\hol$ exactly, and this will be an important seed for very efficient recursion relations discussed in section \ref{sec:ExactlyfromRecursion}.  As one might expect for the trans-Planckian $h\gg c$ regime, the large $h$ limit of the propagator exhibits a breakdown of bulk locality.  In section \ref{subsec:Monodromyhc} we also obtain some explicit analytic results to all-orders in $h/c$ in the short-distance limit.

\subsection{Generalizing the Monodromy Method to $\< \phi \phi \>$}

The monodromy method for Virasoro conformal blocks was developed by Zamolodchikov in \cite{ZamolodchikovRecursion,Zamolodchikovq}; for some recent reviews see \cite{HarlowLiouville, Fitzpatrick:2014vua}.  The basic idea is that the $\CO(c)$ piece ``$g$'' in the exponent of \ref{eq:phiphiSCform} is unaffected by adding extra `light' operators with $\CO(1)$ conformal weights inside the correlator.  Therefore, one can add a degenerate operator
\be
\hat{\psi}_{2,1}(z)
\ee
that has a null Virasoro descendant at level 2.  Correlators of this degenerate operator must obey a second order differential equation.  In the case of $\phi$, let us define the ``wavefunction'' $\psi$ to be the three-point function
\be
\psi \equiv \< \hat{\psi}_{2,1}(z) \phi(X_1)\phi(X_2) \>.
\ee
Because of $\hat{\psi}_{2,1}$'s null descendant, $\psi$ obeys the following differential equation
\be
\partial_z^2 \psi(z,X_1, X_2)+ \frac{6}{c} T(z,X_1,X_2) \psi(z,X_1,X_2) =0, 
\label{eq:nullEOM}
\ee
where $T(z,X_1,X_2)$ is the stress tensor evaluated in the presence of the two $\phi$s:
\be
 T(z,X_1,X_2) = \frac{\< T(z) \phi(X_1) \phi(X_2) \>}{\< \phi(X_1) \phi(X_2) \>}.
 \ee
In (\ref{eq:nullEOM}),  $T(z,X_1,X_2)$ acts like a Schrodinger potential for $\psi$.  It is fixed in terms of the $\< \phi \phi\>$ correlator by recursion relations that follow from the $T \phi$ OPE.  Unlike boundary primary operators, $\phi$ has a third-order pole in its OPE with $T$, due to the fact that it transforms non-trivially under special conformal transformations $L_1$ at the origin.  When both holomorphic and anti-holomorphic stress tensors contribute, the action of $L_1$ is somewhat complicated:
\be
L_1 \phi(y,0,0) = - y^2 \frac{\bar{\partial} + y^2 \frac{6}{c} \bar{T}(0) \partial}{1-y^4 \frac{36}{c^2} T(0)\bar{T}(0)} \phi(y,0,0). 
\ee
We will just develop the monodromy method for the ``holomorphic'' $\< \phi \phi \>$ correlator, where $\bar{T}$ contributions are absent (it would be interesting to study the full case, which is more complicated, in the future).  Then, $L_1$ acts much more simply, and the singular terms in the OPE of $T$ and $\phi$ are the following:
\be
T(z) \phi(y,w,\bar{w}) \sim -\frac{y^2 \partial_{\bar{w}} \phi(y,w,\bar{w})}{(z-w)^3} + \frac{1}{2} \frac{y \partial_y \phi(y,w,\bar{w})}{(z-w)^2} + \frac{\partial_w \phi(y,w,\bar{w})}{z-w}.
\ee
Another significant simplification of the holomorphic correlator is that it depends only on $\rho$ or equivalently $\xi = \frac{2 y_1 y_2}{y_1^2 +y_2^2 + z_{12} \bar z_{12}}$; in other words, the holomorphic correlators are still invariant under the AdS isometries, despite the gauge fixing.  

We can evaluate $T(z,X_1, X_2)$ by summing over its poles at $z_1$ and $z_2$, and the residues are given by derivatives of the exponent $g$\ \footnote{Since we will be focusing on $g_{\text{holo}}$ from this point on, we will denote it using $g$ to reduce clutter.} in (\ref{eq:phiphiSCform}):
\be
\frac{T}{c} &=& -\frac{y_1^2 \partial_{\bar{z}_1} g }{(z-z_1)^3} + \frac{1}{2} \frac{y_1 \partial_{y_1} g }{(z-z_1)^2} + \frac{\partial_{z_1} g }{z-z_1}
  -\frac{y_2^2 \partial_{\bar{z}_2} g }{(z-z_2)^3} + \frac{1}{2} \frac{y_2 \partial_{y_2} g}{(z-z_2)^2} + \frac{\partial_{z_2} g}{z-z_2}.
 \ee

Finally, without loss of generality we can take $z_1=0, y_1=1$, and $z_2, y_2 \rightarrow \infty$ with $z_2/y_2 = 1$ fixed.\footnote{Any non-zero value of $z_2/y_2$ is allowed without loss of generality.  Taking different positions for the two $\phi$s in the correlator leads to different forms of the potential $T$, and consequently different solutions for the wavefunction $\psi$.  However, as long as the geodesic distance between the positions is the same, the monodromy of the solutions does not depend on the specific values of the coordinates.   Another, slightly more complicated limit we could take is $z_1 = \bar{z}_1 =0, z_2 =1$ and $y_1 = y_2=1$, in which case the potential takes the form
\be \label{eq:TzForAppendix}
 T(z) = \frac{\xi  (\xi +(2 \xi +1) (z-1) z) g'(\xi )}{2 (z-1)^3 z^3} .
\ee
And in Appendix \ref{app:OPEblockCalculation}, we also show that we can use the bulk-bulk OPE to obtain the leading term of the above equation.
} In this limit, using the fact that $g$ depends on the coordinates $X_i$ only through the invariant combination $\xi$, $T(z,X_1, X_2)$ simplifies to
\be
\frac{T}{c} &=& \xi g'(\xi) \left( \frac{z - \xi (z^2+1)}{2z^3} \right).
\label{eq:Tphiphi}
\ee
The solutions for $\psi$ are given by the differential equation (\ref{eq:nullEOM}) with this potential. 

The final input into the monodromy method is that the solutions for $\psi$ have fixed monodromy when $z$ is taken along closed paths that encircle other operators in the correlator. This follows from the fact that when $\hat{\psi}_{2,1}$ fuses with an operator, only two possible operator dimensions are allowed in its OPE.   In our case, when $\hat{\psi}_{2,1}$ fuses with one of the $\phi$s, it can only produce operators $\CO_\beta$ that have weight $h_\beta$ satisfying\footnote{One may ask how is it possible for the three-point function $\< \hat{\psi}_{2,1} \phi \phi\>$ to be non-zero at all if $\hat{\psi}_{2,1}$ can only fuse with $\phi$ to produce operators of weight $h_\beta$.  To make sense of this puzzle, one should remember that we are just computing the semiclassical piece of $\< \phi \phi\>$, which is insensitive to additional light operators in the correlator.  So, one can think of the correlator as really being $\< \hat{\psi}_{2,1} \phi \phi \CO'\>$, where $\CO'$ is another light operator whose OPE with $\phi$ contains $\CO_\beta$.}
\be
h_\beta - h_\phi - h_\psi = \frac{1}{2} \left( 1 \pm \sqrt{ 1- \frac{24 h_\phi}{c}} \right) .
\ee
The LHS above is the power of the leading singularity of $\CO_\beta$ in the $\psi \times \phi$ OPE, so when $z$ circles one of the $\phi$s, the monodromy matrix of the two solutions to (\ref{eq:nullEOM}) must have eigenvalues
\be
M_{\pm} = - e^{\pm i \pi \Lambda_h} , \qquad \Lambda_h \equiv \sqrt{ 1- \frac{24 h_\phi}{c}}.
\label{eq:monoevals}
\ee
In summary, the monodromy method for $\< \phi \phi\>_{\rm holo}$ is that one solves (\ref{eq:nullEOM}) for $\psi(z,\xi)$ with $T$ given by (\ref{eq:Tphiphi}), and then fixes $g(\xi)$ by demanding that the monodromy matrix for the two solutions have eigenvalues given by (\ref{eq:monoevals}) as $z$ encircles the origin. 

\subsection{Perturbation Theory in $\frac{h}{c}$ and an All-Orders Resummation}
\label{subsec:Monodromyhc}

Let us see how to apply the monodromy method in the limit of small $h/c$.  To first order, $g \sim \CO(\frac{h}{c})$, and since $c g \sim h$ is independent of $c$ at this order we should just rederive the $h$-dependence of the free scalar propagator in AdS$_3$.  

At zero-th order in $h/c$, $\frac{T}{c}$ vanishes, so the solutions for $\psi$ are just
\be
\psi^{(1)} = 1, \qquad \psi^{(2)} = z.
\ee
Both of these have trivial monodromy, consistent with $-e^{i \pi \Lambda_h} = 1 + \CO(h/c)$ at leading order.  At next order, we demand monodromies of $-e^{\pm i \pi \Lambda_h} = 1 \pm \frac{12 i \pi h }{c}$.  We will use the method of separation of variables.  The Wronskian of the zero-th order solutions is trivial
\be
W = \psi^{(1)} \psi^{(2) \prime} - \psi^{(1)\prime} \psi^{(2)}=1 .
\ee
The monodromy matrix $M_{ij}$ of the first-order solutions as $z$ goes around the origin are given by the following residue formula:
\be
M_{ij} &=& \delta_{ij} -2 \pi i {\rm res}_{z=0} \left[ \frac{\frac{6}{c} T(z)}{W(z)} \tilde{\psi}^{(i)} \psi^{(j)} \right], \quad \tilde{\psi} \equiv \left( \begin{array}{c} - \psi^{(2)} \\ \psi^{(1)} \end{array} \right).
\ee
The eigenvalues of $M$ are
\be
{\rm evals}(M) =1 \pm  6 i \pi   \xi \sqrt{1- \xi^2} g'(\xi) .
\ee
Equating ${\rm evals}(M) =1 \pm 12 i \pi \frac{h}{c}$, we obtain
\be
e^{c g(\xi)} &=& \left( \frac{\xi}{1 + \sqrt{1-\xi^2}} \right)^{2h} = \rho^h .   
\ee
which is indeed the right answer.

We can continue to higher orders in $h/c$ as well. It becomes somewhat nicer to write expressions in terms of the variable $\rho$ rather than $\xi$.  At higher orders, rather than writing $M_{ij}$ in terms of residues of a matrix, one must solve order-by-order for the solutions $\psi^{(1)}$ and $\psi^{(2)}$; the non-trivial monodromies arise from logarithms in $\psi^{(i)}$, and these are fairly easy to deal with by hand.  At $\CO(h^2/c)$, we find
\be
\label{eq:semiclassicalhoverc}
c g =h \log \rho +  \frac{12 h^2}{c} \left( \frac{\rho}{(1-\rho)^2} + \log(1-\rho) \right) + \CO \left(\frac{h^3}{c^2} \right).
\ee 
in the semiclassical limit.  This agrees with bulk gravitational perturbation theory (see section \ref{sec:OneLoop} and appendix \ref{app:PerturbativePropagator}) and the methods of section \ref{sec:ExactlyfromRecursion}.

 After working to sufficiently high  order using the recursion relation of sectio†n \ref{sec:ExactlyfromRecursion} , a pattern emerges and one can conjecture the following ansatz for the all-orders result:
\be
\label{eq:semiclassicalexact}
& & \log\frac{\left\langle \phi\phi\right\rangle _{\text{holo}}}{\left\langle \phi\phi\right\rangle _{\text{global}}}\label{eq:SemiPhiPhiHolo}\\
&= & \sum_{n=1}^{\infty}\frac{h^{n+1}}{c^{n}}\left(\frac{2\times12^{n}(2n-1)!!}{(n+1)!}\log\left(1-\rho\right)+\frac{24^n (3n-3)!!}{(n+1)!(n-1)!!}\frac{g_{n}\left(\rho\right)}{\left(1-\rho\right)^{3n-1}}\right), \nonumber 
\ee
with the first three $g_n({\rho})$ taking the form
\begin{align}
g_{1}\left(\rho\right) & =\rho , \nonumber \\
g_{2}\left(\rho\right) & =\frac{1}{12}\rho\left(7\rho^{4}-41\rho^{3}+73\rho^{2}-33\rho+6\right), \\
g_{3}\left(\rho\right) & =\frac{1}{192}\rho\left(-42\rho^{7}+366\rho^{6}-1205\rho^{5}+1758\rho^{4}-1112\rho^{3}+606\rho^{2}-209\rho+30\right),\nonumber \end{align}
and where we have normalized $g_{n}\left(\rho\right)$ so that $g_{n}\left(1\right)=1$.  The leading term involving $g_1$ matches equation (\ref{eq:semiclassicalhoverc}).

The second term in equation (\ref{eq:SemiPhiPhiHolo}) can be summed if we work to leading order in the short-distance limit $\rho\rightarrow1$, i.e. by setting $g_{n}\left(\rho\right)=1$.\footnote{The logarithmic terms in equation (\ref{eq:SemiPhiPhiHolo}) can also be summed to give 
\be
\sum_{n=1}^{\infty}\frac{2 h^{n+1}}{c^{n}}\frac{12^{n}(2n-1)!!}{(n+1)!}\log\left(1-\rho\right)=\frac{c}{6}\left(1-\frac{12h}{c}-\sqrt{1-\frac{24h}{c}}\right)\log\left(1-\rho\right) .
\label{eq:semiclassicallogpart}
\ee
 Note that for $h > \frac{c}{24}$, i.e. above the BTZ black hole threshold, equation (\ref{eq:semiclassicallogpart}) develops an imaginary piece.  Unfortunately, we cannot conclude anything from this fact alone, since that result is subdominant to the other terms in equation (\ref{eq:semiclassicalexact}).}
We find
\begin{small}
\begin{align}\label{eq:semiclassicalshortdistanceexact}
 & \sum_{n=1}^{\infty}\frac{24^n (3n-3)!!}{(n+1)!(n-1)!!}\frac{h^{n+1}}{c^{n}\left(1-\rho\right)^{3n-1}}\\
= & \frac{c(1-\rho)^{4}}{576}\left(\,_{2}F_{1}\left(-\frac{2}{3},-\frac{1}{3};\frac{1}{2};\frac{15552h^{2}}{c^{2}(1-\rho)^{6}}\right)-1\right)+h\left(1-\rho\right)\left(1-\,_{2}F_{1}\left(-\frac{1}{6},\frac{1}{6};\frac{3}{2};\frac{15552h^{2}}{c^{2}(1-\rho)^{6}}\right)\right)\nonumber 
\end{align}
\end{small}for the function that appears in the exponent of the semiclassical propagator in the short-distance limit.

Both hypergeometric functions in equation (\ref{eq:semiclassicalshortdistanceexact}) can be expanded as $\rho \to 1$ with fixed $h/c$, and both develop complex parts in this limit.  More generally, the hypergeometric functions both have branch cuts running from the point where $(1- \rho)^6 = \frac{15552 h^2}{c^2}$ to $\rho=1$.  This indicates an apparent breakdown of bulk  locality in the semiclassical part of the propagator.  Note that in terms of the geodesic length $\sigma$, this breakdown occurs at the critical value
\be
\sigma_* \approx \left( 9 \sqrt{3} \   \frac{h}{c} \right)^{1/3} R_{AdS}
\label{eq:SigmacAnalyticSCSmallHC}
\ee
at large $c$ and small $h/c$.  This formula only applies in the regime where $\sigma_* \ll R_{AdS}$, because we were only able to compute the semiclassical result analytically to leading order as $\rho \to 1$.

\subsection{Exact Large $h$ Limit  and the Breakdown of Locality}
\label{sec:LargehLimit}

We can also solve the monodromy method in an expansion about large $h$.  This limit is interesting for two very different reasons.  The first  is that large $h$ corresponds with a large bulk mass for $\phi$. 
So in this case we expect $\phi$ to have a very large effect on the local geometry, potentially leading to the breakdown of bulk locality at macroscopic distances.\footnote{One might have expected to see indications of black hole physics, since we are studying the limit $h \gg c$ where $\phi$ would have to be interpreted as a sort of `black hole field'.  We do not see any direct indications of the Hawking temperature or Schwarzschild radius in the $\phi$ propagator, though these parameters must appear in higher-point semiclassical correlators.} 
Our second motivation is more technical:  as we will demonstrate in section \ref{sec:ExactlyfromRecursion}, the infinite $h$ limit of the correlator is the necessary ``seed'' for a very efficient recursion relation that can be used to numerically compute the correlator exactly and at any $h$. 

At large $h$,  the potential $T$ should become large and therefore one can solve the Schrodinger equation for $\psi$ using a WKB approximation. This approach (used for the blocks in \cite{Zamolodchikovq}), is easiest to implement if we change variables according to
\be
\psi = (y'(z))^{-1/2} \Psi(y) ,
\ee
bringing the Schrodinger equation into the form
\be
\Psi''(y) + U(y) \Psi(y) = \xi g'(\xi) \Psi(y).
\label{eq:newSchrEq}
\ee 
  The coordinate $y(z)$ that achieves this is    
  \be
  y(z) &=&\sqrt{3}  \int^z dt \frac{\sqrt{-t+\xi (1+t^2)}}{t^{3/2}} \nn\\
   &=& -\frac{2 \sqrt{6} \left(\frac{(1-z) \sqrt{s (z+1)^2-4 z}}{2 \sqrt{z} (z+1)}+E(\varphi |s)\right)}{\sqrt{2-s}},
  \ee  
  where $E$ is an elliptic integral, $\sin \varphi \equiv \frac{2 z^{1/2}}{(1+z) s^{1/2}}$, and  we have introduced the new coordinate $s$:
  \be 
  s \equiv \frac{4 \xi}{1+2 \xi}.
  \ee
The new potential $U(y)$ is
\be
U(y) = \frac{3 y''(z)^2-2 y^{(3)}(z) y'(z)}{4 y'(z)^4} .
\ee
Now, the advantage is that $y$ is a periodic variable - under a monodromy cycle in $z$,  $y$ shifts because of a corresponding shift in the elliptic integral: 
\be
E(\varphi + n \pi,s) = E(\varphi, s)+ 2 n E(s).
\label{eq:EllipticEshift}
\ee
Consequently, the $y$ variable lives in a box of size $R$ given by 
 \be
R = 4 \sqrt{6}  \frac{E(s)}{\sqrt{2-s}}.
\ee
The new form (\ref{eq:newSchrEq}) of the Schrodinger equation is for a particle in a box, having energy $\xi g'(\xi)$; the monodromy condition is that the particle should have quasimomentum $\Lambda_h$.  Therefore, in the limit $\Lambda_h \rightarrow \infty$, we have 
\be
c \xi g'(\xi) = -c \frac{(\pi \Lambda_h)^2}{R^2} = \pi^2\frac{(2-s) (h -\frac{c}{24})}{4E^2(s)}.
\ee
Equivalently,
\be
c g'(s) =\frac{ \pi^2 (h -\frac{c}{24}) }{2 sE^2(s)}+ \CO(c) . 
\ee
The subleading in $1/h$ correction is an $\CO(c)$ correction, as indicated above.  We can obtain this correction as follows. Because $\xi g'(\xi)$ is the eigenvalue in the Schrodinger equation  (\ref{eq:newSchrEq}), in the WKB approximation its subleading correction enters as
\be
\log \Psi(y) \approx \int dy \sqrt{ \xi (g'(\xi) + \delta g'(\xi)) - U(y)) } \approx \sqrt{ \xi g'(\xi)} y + \frac{1}{2 \sqrt{\xi g'(\xi)}} \int dy (\xi \delta g'(\xi) - U(y)).\nn\\
\ee
The monodromy of our leading order solution above is already the correct value, $\Lambda_h$.  Demanding that the correction to the monodromy vanish, one therefore obtains\footnote{We performed the $dy$ integral in (\ref{eq:Udy}) by changing variables to $dz$ and doing the indefinite integral to get a result involving elliptic integrals.  The definite integral $\int_0^R dy$ then corresponds to the shift in the $\int dz$ integral under a $z$ monodromy cycle, which is easy to read off using (\ref{eq:EllipticEshift}) together with $F(\varphi + n \pi,s) = F(\varphi, s)+ 2 n K(s)$.  }
\be
\xi \delta g'(\xi) = \frac{1}{R} \int_0^R U(y) dy = \frac{2-s}{144} \left( \frac{7 K(s)}{E(s)} - \frac{2-s}{2(1-s)} \right).
\label{eq:Udy}
\ee
Putting this correction together with the leading piece, we obtain the full semiclassical part of the exponent $c g$ at $h=\infty$:
\be
c g = \left( h - \frac{c}{24} \right)  \int \frac{ds}{2s} \frac{\pi^2}{E^2(s)}  + \frac{c}{144} \left[ \log \left( \frac{ 16(1-s)}{s^2}  \right) + 14 \int \frac{ds}{s} \frac{K(s)}{E(s)}\right]  . 
\ee   
The integrals over $s$ can all be done in closed form.\footnote{For reference, $\int \frac{ \pi ds}{s E^2(s)} = 4\frac{E(1-s)-K(1-s)}{E(s)}$  and $\int \frac{ds K(s)}{s E(s)}= \log \left( \frac{s}{E^2(s)} \right)$.}    We fix the integration constants by matching to the  known, small $\xi$ behavior.  The result is that the semiclassical part of the correlator at large $h$ is given by
\be
e^{c g} \stackrel{h \gg c}{=}  q^{h - \frac{c}{24}} \left( \frac{s}{8} \right)^{\frac{c}{12}} (1-s)^{\frac{c}{144}} \left( \frac{2E(s)}{\pi} \right)^{\frac{-7c}{36}} ,  
\label{eq:LargehSC}
 \ee
 where we have defined the new variable $q$:
 \be
 \label{eq:Definitionq}
 q \equiv 4 e^{ 2 \pi \frac{E(1-s)-K(1-s)}{E(s)} - 4}.
 \ee
In addition to the semiclassical part above, the full holomorphic correlator $\<\phi \phi\>_{\rm holo}$ at infinite $h$ has a residual piece that is independent of $h$ and $c$.  We have not been able to derive this residual piece from first principles, but we believe we were able to obtain the correct formula as follows. 
The corresponding residual piece in the conformal blocks simply results in a shift $c\rightarrow c-1$ in the formula as compared to the semiclassical result.  We tried an ansatz of the form of (\ref{eq:LargehSC}) where  in each of the four places $c$ appears, we allow a separate shift in $c$.  Comparing to an exact calculation of the leading small $\xi$ expansion, we fixed these four new parameters and checked that the Ansatz reproduced the correct result to high order in $\xi$.  The final result is
\be
\label{eq:exactlargeh}
\lim_{h \rightarrow \infty} \< \phi \phi\>_{\rm holo} = q^{h - \frac{c-1}{24}} \left( \frac{s}{8} \right)^{\frac{c-1}{12}} (1-s)^{\frac{c-13}{144}} \left( \frac{2E(s)}{\pi} \right)^{\frac{19-7c}{36}}.
\ee
We emphasize that this is not merely a semiclassical result, but the exact answer at large $h$.

The large $h$ limit of the holomorphic correlator has an important feature.  As  is evident in equation (\ref{eq:Definitionq}), the $q$ variable becomes complex when $s > 1$, which corresponds to $\rho_c = 7 - 4 \sqrt{3} \approx 0.072$, or a physical geodesic separation 
\be
\frac{\sigma_c}{R_{AdS}} \approx  \log(2+\sqrt{3}) = 1.32 
\label{eq:SigmacAnalyticSCLargeHC}
\ee
 in the bulk (for emphasis, we have written the AdS radius explicitly).  This represents a breakdown of  bulk locality at the AdS scale.  Note that this is not merely a relic of the semiclassical approximation, since it applies to the exact holomorphic propagator in the large $h$ limit.

Physically, this failure of bulk locality is not very surprising.  We certainly would not have expected to have healthy bulk correlators for a field $\phi$ with extremely large (trans-Planckian!) bulk mass.  But it is nevertheless reassuring that we can identify the breakdown of bulk locality in a precise, quantitative way.

\begin{figure}[th!]
\begin{center}
\includegraphics[width=0.48\textwidth]{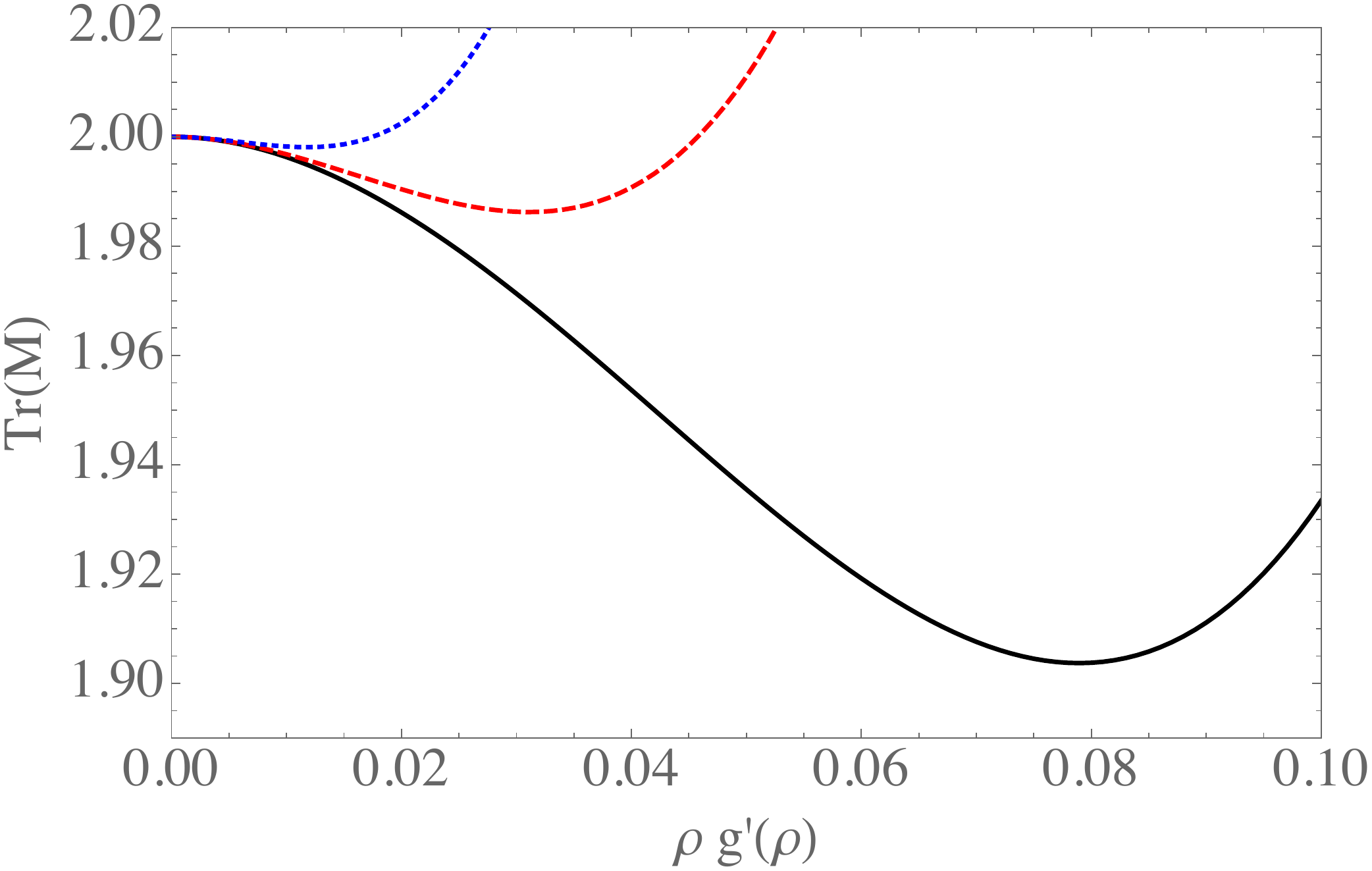}
\includegraphics[width=0.48\textwidth]{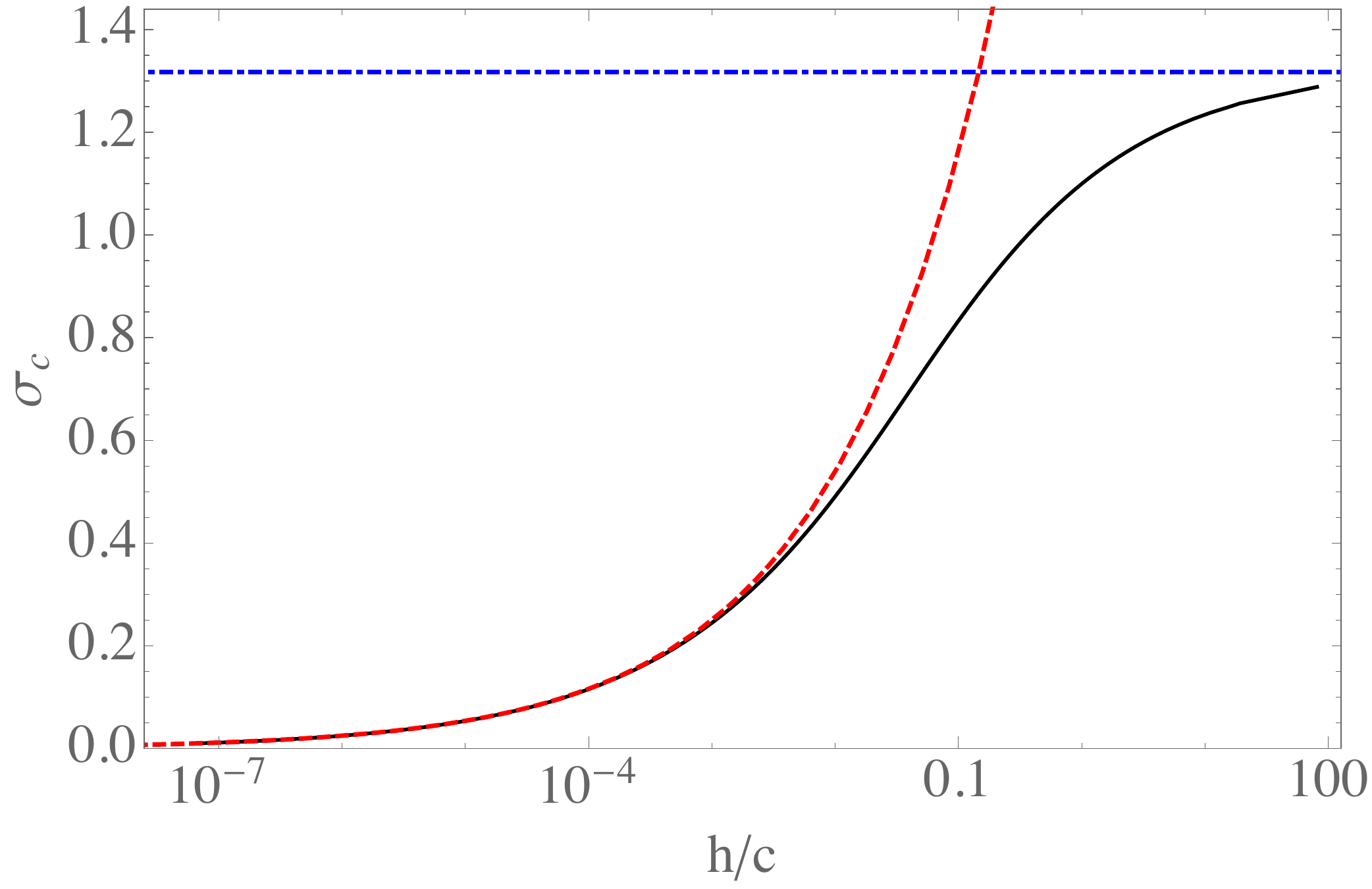}
\caption{{\it Left:} The trace of the monodromy matrix $M$ computed numerically as a function of $\rho$ and $\rho g'(\rho)$, for three values of $\rho$: $\rho=0.4$ {\it (black, solid)}, $\rho=0.5$, {\it (red, dashed)}, and $\rho=0.6$, {\it (blue, dotted)}. {\it Right:} The critical value of $\sigma_c$  {\it (black, solid)}  where the semiclassical part of the $\< \phi \phi\>$ correlator first develops an imaginary piece, as a function of $h/c$. For comparison, we show {\it (red, dashed}) the analytic small $h/c$ behavior, $\sigma_c \approx (9 \sqrt{3} h/c)^{1/3}$ from (\ref{eq:SigmacAnalyticSCSmallHC}), and {\it (blue, dot-dashed}) the large $h$ behavior $\sigma_c \approx \log (2+\sqrt{3})$, from (\ref{eq:SigmacAnalyticSCLargeHC}).     }
\label{fig:NumMono1}
\end{center}
\end{figure}

\subsection{Numeric Monodromy Results}
\label{sec:NumMono}

Away from the limits of large and small $h/c$, we have not been able to solve the monodromy method for the semiclassical piece $g(\rho)$ of the $\< \phi \phi\>_{\rm holo}$ correlator in closed form.  However, it is straightforward to compute $g$ numerically.  Converting to $\rho$ coordinates, 
\be
\frac{T}{c} = C_\rho \left( \frac{(\rho +1) z-2 \sqrt{\rho } \left(z^2+1\right)}{(1-\rho ) z^3} \right),
\ee
where numerically we fix the parameters $\rho$ and $C_\rho \equiv \rho g'(\rho)$, and numerically integrate the wavefunction $\psi$ along a cycle around the origin and match in order to compute the monodromy matrix $M$.  Once $C_\rho$ is known for all $\rho$, it can be integrated to obtain $g(\rho)$.

 In the left plot of Fig. \ref{fig:NumMono1}, we  show $\tr(M)$ computed in this way as a function of $C_\rho$ for a few values of $\rho$, and for real $C_\rho$.  One can invert the relation to find $C_\rho$ as a function of $h/c$ and $\rho$ by looking at the point where 
 \be
 \tr(M_h) \equiv -2 \cos (\pi \Lambda_h) \qquad \left (\Lambda_h \equiv \sqrt{1 - \frac{24 h}{c}} \right) 
 \ee
 intersects the curve for any specific $\rho$.  For small enough $\rho$, the curve will cross $\tr(M_h)$ at multiple values of $C_\rho$, but by continuity with the small $\rho$ limit, one should take the smallest value for $C_\rho$ in these plots. 
 
 The most physically important feature of these numeric results is that each curve has a minimum at some value of $C_\rho$:
 \be
 \Big[ \tr(M) \Big]_{\rm min}(\rho) \equiv {\rm min}_{C_\rho} \tr(M).
 \ee
    Therefore, if $\tr(M_h)$ is below this minimum value $ \left[ \tr(M) \right]_{\rm min}(\rho)$, then $C_\rho$ must become complex.  Note also that the minimum value is an increasing function of $\rho$, so for fixed $\tr(M_h)$, there is a critical value of $\rho$ where $\left[ \tr(M) \right]_{\rm min}(\rho) = \tr(M_h)$; for larger $\rho$, $C_\rho$ develops an imaginary piece. The right plot of Fig. \ref{fig:NumMono1} shows  the resulting critical value for $\sigma_c = -\frac{1}{2} \log \rho_c $ in Fig. \ref{fig:NumMono1}. We obtain a satisfying agreement with the results of section \ref{subsec:Monodromyhc}  in the limit of small $h/c$.  We compare these results with the exact methods of section \ref{sec:ExactlyfromRecursion} in figure \ref{fig:SemiclassicalNumericalPlot}.

\section{Computing the Propagator Exactly}
\label{sec:ExactlyfromRecursion}

In this section we will develop a generalization of the Zamolodchikov recursion relations that make it possible to compute the bulk propagator exactly.  The relations produce the exact coefficients for a series expansion of $K_\hol$ in the variable $\rho$ (and $q$).  This means that we obtain a long-distance expansion for the propagator, since $\rho = 0$ corresponds to geodesic separation $\sigma \to \infty$.

\subsection{Generalizing the Zamolodchikov Recursion Relations to $\< \phi \phi \>$}

The recursion relations that we will develop for computing $\left\langle \phi\phi\right\rangle_{\text{holo}}$ are very similar to Zamolodchikov's recursion relations for computing Virasoro blocks \cite{ZamolodchikovRecursion, Zamolodchikovq}. They are based on the large $c$ limit and large $h$ limit of the $\left\langle \phi\phi\right\rangle_{\text{holo}}$, as well as the pole structure of $\left\langle \phi\phi\right\rangle_{\text{holo}}$ as a function of $c$ or $h$, respectively. In the following, we'll denote them as the $c$-recursion and $h$-recursion.

Let us first write the central charge $c$ in terms of a variable $b$
as $c=13+6\left(b^{2}+b^{-2}\right)$ and define an function $A_{m,n}^{c}$ given by
\begin{equation}
A_{m,n}^{c}=\frac{1}{2}\prod_{k=1-m}^{m}\prod_{l=1-n}^{n}\frac{1}{kb+\frac{l}{b}},\qquad\left(k,l\right)\ne\left(0,0\right),\left(m,n\right)
\end{equation}
which will be an ingredient of both $c$-recursion and $h$-recursion.
It was determined in \cite{Zamolodchikov:2003yb} that $A_{m,n}^{c}$ is equal to\footnote{The coefficient of $(L_{-1})^{mn}$ in $\mathcal{L}_{-mn}^{\text{quasi}}\mathcal{O}^{h}$ in \cite{Zamolodchikov:2003yb} is also normalized to 1, which is the same as the convention of this paper.}
\begin{equation}
A_{m,n}^{c}=\lim_{h\rightarrow h_{m,n}\left(c\right)}\left(\frac{\left\langle \mathcal{L}_{-mn}^{\text{quasi}}\mathcal{O}^{h}|\mathcal{L}_{-mn}^{\text{quasi}}\mathcal{O}^{h}\right\rangle }{h-h_{m,n}\left(c\right)}\right)^{-1},\label{eq:Amnc}
\end{equation}
where $h_{m,n}$ is the degenerate-state dimensions (which will be
given below) and we put a superscript $h$ on $\mathcal{O}$ to emphasize
that if we send $h\rightarrow h_{m,n}\left(c\right)$ then $\mathcal{L}_{-mn}^{\text{quasi}}\mathcal{O}^{h}$
becomes a level $mn$ null-state. 
\subsubsection{$c$-recursion relation}
To obtain the $c$-recursion, we need to know the large $c$ limit
of $\left\langle \phi\phi\right\rangle _{\text{holo}}$, which is
simply 
\[
\lim_{c\rightarrow\infty}\left\langle \phi\phi\right\rangle _{\text{holo}}=\left\langle \phi\phi\right\rangle _{\text{global}}=\frac{\rho^{h}}{1-\rho}.
\]
As in the Virasoro block case, as a function of $c$ the correlator $\left\langle \phi\phi\right\rangle _{\text{holo}}$
has simple poles at $c=c_{m,n}\left(h\right)$.
The residue of the pole at $c=c_{m,n}\left(h\right)$ must be proportional
to the two-point function of $\phi_\text{holo}$ with dimension $h+mn$ and central
charge $c_{m,n}\left(h\right)$. As shown in Section \ref{sec:HolomorphicDefinitions}, $\left\langle \phi\phi\right\rangle _{\text{holo}}$
can be written as 
\begin{equation}\label{eq:PhiPhiHoloQuasiPrimaryContribution}
\left\langle \phi\phi\right\rangle _{\text{holo}}=\sum_{N=0}^{\infty}C_{N}\left(2h\right)_{2N}\frac{\rho^{h+N}}{1-\rho}=\frac{\rho^{h}}{1-\rho}\sum_{N=0}^{\infty}\left(\sum_{i=0}^{p\left(N\right)-p\left(N-1\right)}\frac{1}{\left|\mathcal{L}_{-N}^{\text{quasi},i}\mathcal{O}\right|^{2}}\right)\left(2h\right)_{2N}\rho^{N},
\end{equation}
where we have written $\left\langle \phi\phi\right\rangle _{\text{holo}}$
explicitly as a sum over contributions from different quasi-primaries
and their global descendants. So if we write $\left\langle \phi\phi\right\rangle _{\text{holo}}$
as 
\begin{equation}
\left\langle \phi\phi\right\rangle _{\text{holo}}=\frac{\rho^{h}}{1-\rho}F\left(h,c\right) ,
\end{equation}
then in $F\left(h,c\right)$, the residues at $c=c_{m,n}\left(h\right)$
will include a factor $\rho^{mn}F\left(h+mn,c_{m,n}\left(h\right)\right)$.
At the poles, the residues should also includes a factor that will
give $C_{N}$. But this is precisely given by $-\frac{\partial c_{m,n}\left(h\right)}{\partial h}A_{m,n}^{c_{m,n}}$,
where $\frac{\partial c_{m,n}\left(h\right)}{\partial h}$ is the
Jacobian factor, because we are considering the poles at $c=c_{m,n}\left(h\right)$
while equation (\ref{eq:Amnc}) is at the poles of $h=h_{m,n}\left(c\right)$. So combining all these facts, we find that $F\left(h,c\right)$ is given by the following $c$-recursion
relation:
\begin{equation}
F\left(h,c\right)=1+\sum_{m\ge2,n\ge1}-\frac{\partial c_{m,n}\left(h\right)}{\partial h}\frac{A_{m,n}^{c_{m,n}}\left(2h\right)_{2mn}}{c-c_{m,n}\left(h\right)}\rho^{mn}F\left(h+mn,c_{m,n}\left(h\right)\right),
\label{eq:cRecursion}
\end{equation}
where poles $c_{m,n}\left(h\right)$ are given by 
\begin{equation}
c_{m,n}\left(h\right)=13+6\left[\left(b_{m,n}\left(h\right)\right)^{2}+\left(b_{m,n}\left(h\right)\right)^{-2}\right],
\end{equation}
with 
\begin{small}
\begin{equation}
\left(b_{m,n}\left(h\right)\right)^{2}=\frac{2h+mn-1+\sqrt{\left(m-n\right)^{2}+4\left(mn-1\right)h+4h^{2}}}{1-m^{2}},m=2,3,\cdots,n=1,2\cdots.
\end{equation}
\end{small}The super-script in $A_{m,n}^{c_{m,n}}$ means that the $b$ in $A_{m,n}^{c}$ (equation (\ref{eq:Amnc})) should be substituted
by $b_{m,n}\left(h\right)$. The factor $\left(2h\right)_{N}$ in
equation (\ref{eq:PhiPhiHoloQuasiPrimaryContribution}) is accounted for by the $\left(2h\right)_{2mn}$
in the residues of this $c$-recursion (\ref{eq:cRecursion}). 

Compared to the $c$-recursion relation for Virasoro block, we see
that besides adding the factor $\left(2h\right)_{2mn}$, we simply get
rid of the factor in the residues that encodes the information about
the three point function between the intermediate state and the external
operators. The existence of this $c$-recursion relation can be traced
back to the similarity of our definition of $\phi$ and a projection
operator to project the four-point function into Virasoro blocks.
We've checked this recursion relation by directly computing the $\left\langle \phi\phi\right\rangle _{\text{holo}}$
up to level\footnote{For example, at level 2 the $c$-recursion gives 
\be
C_2=-\frac{\partial c_{1,2}\left(h\right)}{\partial h}\frac{A_{1,2}^{c_{1,2}}}{c-c_{1,2}\left(h\right)}=\frac{9}{2 (2 h+1) (2 c h+c+2 h (8 h-5))}
\ee 
which is exactly equal to $\frac{1}{|\mathcal{L}^{\text{quasi}}_{-2}\CO|^2}$ with $\mathcal{L}^{\text{quasi}}_{-2}=L_{-1}^{2}-\frac{2\left(2h+1\right)}{3}L_{-2}$.
} $N=5$. We also used this recursion relation to obtain
the semiclassical limit of $\left\langle \phi\phi\right\rangle _{\text{holo}}$,
and the results agree with those obtained from monodromy method of Section \ref{subsec:Monodromyhc}.

\subsubsection{$h$-recursion relation}
The $h$-recursion relation is obtained by considering $\left\langle \phi\phi\right\rangle _{\text{holo}}$
as a function of $h$ with simple poles at $h=h_{m,n}\left(c\right)$.
In Section \ref{sec:LargehLimit}, we already obtained the large $h$ limit of $\left\langle \phi\phi\right\rangle _{\text{holo}}$
by the monodromy method and a bit of guesswork; we found 
\begin{equation}
\lim_{h\rightarrow\infty}\left\langle \phi\phi\right\rangle _{\text{holo}}=q^{h-\frac{c-1}{24}}\left(\frac{s}{8}\right)^{\frac{c-1}{12}}\left(1-s\right)^{\frac{c-13}{144}}\left(\frac{2E\left(s\right)}{\pi}\right)^{\frac{19-7c}{36}} .
\end{equation}
So if we write $\left\langle \phi\phi\right\rangle _{\text{holo}}$
as
\[
\left\langle \phi\phi\right\rangle _{\text{holo}}=q^{h-\frac{c-1}{24}}\left(\frac{s}{8}\right)^{\frac{c-1}{12}}\left(1-s\right)^{\frac{c-13}{144}}\left(\frac{2E\left(s\right)}{\pi}\right)^{\frac{19-7c}{36}}H\left(h,c\right),
\]
then $H\left(h,c\right)$ is given by the following recursion relation:
\begin{equation}
\label{eq:hRecursion}
H\left(h,c\right)=1+\sum_{m,n\ge1}^{\infty}\frac{q^{mn}\left(2h_{m,n}\right)_{2mn}A_{m,n}^{c}}{h-h_{m,n}\left(c\right)}H\left(h_{m,n}+mn,c\right),
\end{equation}
where the poles are given by $h_{m,n}=\frac{1}{4}\left(b+\frac{1}{b}\right)^{2}-\frac{1}{4}\left(mb+\frac{n}{b}\right)^{2}$,
with $c=13+6\left(b^{2}+b^{-2}\right)$. The residues of the $h$-recursion
are just those of the $c$-recursion but now evaluated at the poles
$h_{m,n}\left(c\right)$. 

The $c$-recursion and $h$-recursion can be solved numerically, in the sense that we can obtain higher order coefficients from lower order coefficients, analogously to the blocks \cite{Chen:2017yze,Perlmutter:2015iya}. We discuss the algorithm for implementing these recursions in Appendix \ref{app:AlgorithmForRecursion} and we have also attached our Mathematica code.

In each iteration of the $c$-recursion,
we need to change both $h\rightarrow h+mn$ and $c\rightarrow c_{m,n}\left(h\right)$,
whereas in the $h$-recursion, we only need to change $h\rightarrow h_{m,n}+mn$.
Thus, the implementation of the $h$-recursion is faster than the $c$-recursion
by roughly a factor of $N$. Although obtaining $C_{N}$ from the
$c$-recursion is straightforward, one can also use the $h$-recursion
to obtain $\left\langle \phi\phi\right\rangle _{\text{holo}}$ and
then expand the result in terms of $\rho$ to obtain $C_{N}$, which is faster for higher order coefficients.

\subsection{Comparison of Full and Holomorphic Propagators}
\label{subsec:FullVSHolo}

In this section, we exhibit a numerical result comparing the full and holomorphic propagators. First we recall a convenient definition from equation (\ref{eq:phiphiParts})
\begin{equation}
\left\langle \phi\phi\right\rangle =2\left\langle \phi\phi\right\rangle _{\text{holo}}-\left\langle \phi\phi\right\rangle _{\text{global}}+\left\langle \phi\phi\right\rangle _{\text{mixed}},
\label{eq:phiphiParts-2}
\end{equation}
where $\langle\phi\phi\rangle _{\text{global}}$ is given in (\ref{eq:PhiPhiGlobal}). Most of the analytic tools we have developed in this paper apply directly to $\langle\phi\phi\rangle _{\text{holo}}$.  But we can use the recursion relations to numerically compute both $\langle\phi\phi\rangle _{\text{holo}}$ and $\langle\phi\phi\rangle _{\text{mixed}}$ to high order, as explained in section \ref{sec:FromHoloToFull}.

\begin{figure}
\centering
\includegraphics[width=0.95\textwidth]{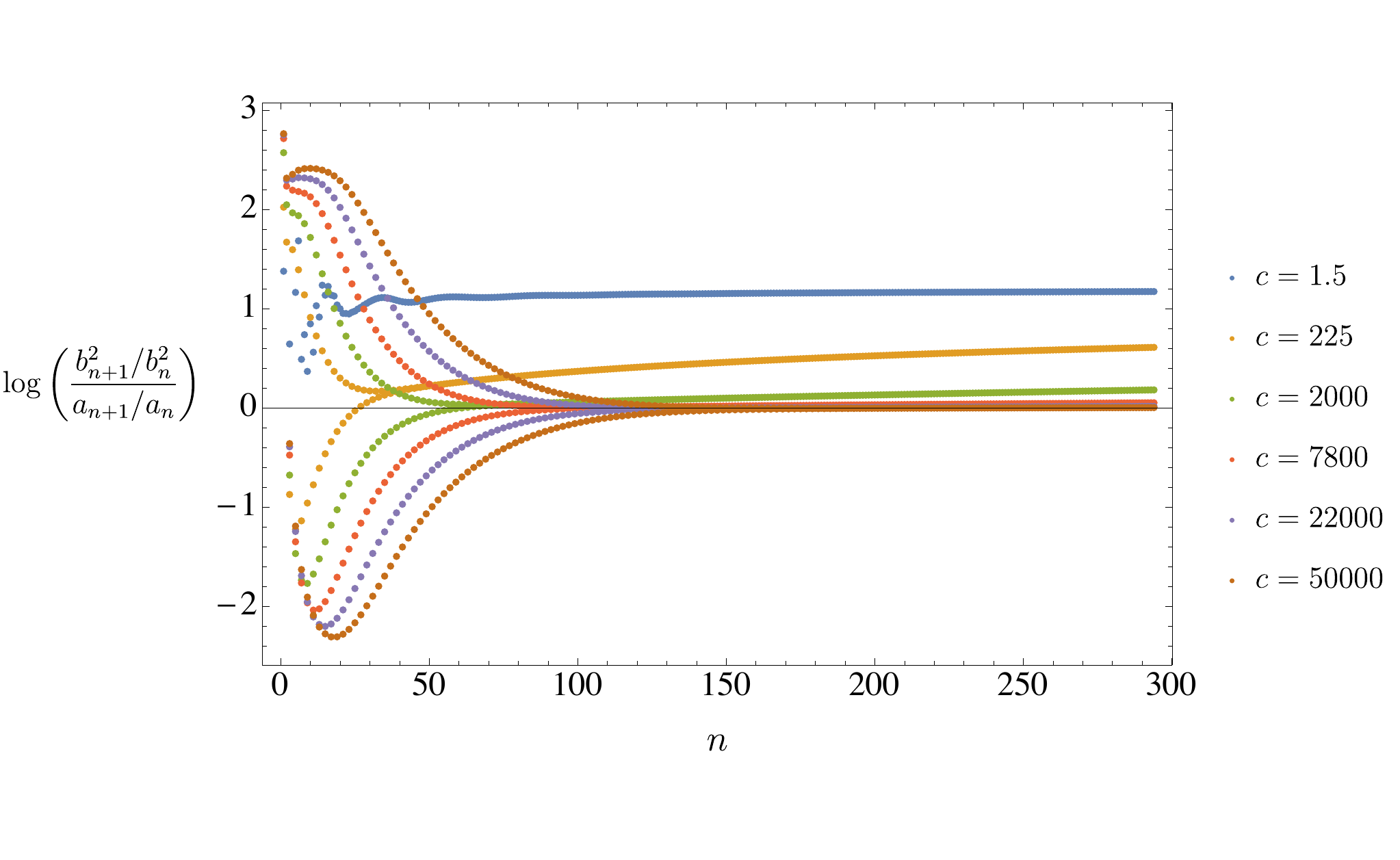}	
\caption{ This plot  compares the ratios of successive coefficients in the holomorphic and mixed terms contributing the the full propagator. We see that at large $c$, the coefficients of $\rho^n$ grow at the same rate, meaning that the holomorphic propagator provides a good estimate for the behavior of the full propagator.}
\label{fig:holoVsMixedCoefficientsRatio}
\end{figure}

The  coordinate system we specified in equation (\ref{eq:MetricwithT}) is not invariant under  the isometries of vacuum AdS$_3$. Therefore $\langle \phi(y_1,z_1, \bar z_1)\phi(y_2,z_2, \bar z_2) \rangle$ can depend on both the geodesic separation between two points and an angular variable with respect to the $z$-$\bar z$ plane, such as the ratio $y_1/y_2$. However, the holomorphic propagator $\langle\phi\phi\rangle _{\text{holo}}$ is invariant under the isometries of vacuum AdS$_3$.  Specifically, we found\footnote{Compared to equation (\ref{eq:PhiPhiHoloQuasiPrimaryContribution}), we see that $\sum_{n=0}^{\infty} a_n \rho^{n}=\frac{1}{1-\rho}\sum_{n=0}^\infty C_n(2h)_{2n}\rho^n$, but the effect of the factor $\frac{1}{1-\rho}$ is negligible in the following discussion.}
\be \label{eq:PhiPhiHoloanExpand}
\langle\phi(y_1,z_1, \bar z_1)\phi(y_2,z_2, \bar z_2)\rangle _{\text{holo}}=\rho^h\sum_{n=0}^{\infty} a_n \rho^{n}.
\ee 
where $\rho = e^{-2\sigma}$ with $\sigma$ the geodesic separation. This nice property does not hold for $\langle\phi\phi\rangle _{\text{mixed}}$. We will  leave  detailed discussion of the dependence of $\< \phi \phi \>$ on $y_1/y_2$ to the future.\footnote{In Appendix \ref{app:ComputingKdF} we discuss the properties of the KdF series in general configurations, giving some further information on the angular dependence of the propagator.} In this section, we focus on computing $\langle\phi\phi\rangle _{\text{mixed}}$ when the two points are in the same $z$-$\bar z$ plane, ie when $y_1 = y_2$, so that 
\be\label{eq:PhiPhiMixedbnExpand}
\langle\phi(y,z_1,\bar z_1)\phi(y,z_2, \bar z_2)\rangle _{\text{mixed}}=\rho^h\sum_{n=0}^{\infty} b_n \rho^{\frac{n}{2}}.
\ee
The coefficients $b_n$ can be computed exactly using the method outlined in section \ref{sec:HolomorphicDefinitions}. Notice that the coefficients $a_n$ in equation (\ref{eq:PhiPhiHoloanExpand}) (which are related to $C_n$ in equation (\ref{eq:PhiPhiHoloQuasiPrimaryContribution})) are always positive, but the coefficients $b_n$ in equation (\ref{eq:PhiPhiMixedbnExpand}) can be negative. We have displayed the ratios of the growth rates of the coefficients $a_n$ and $b_n$ in figure \ref{fig:holoVsMixedCoefficientsRatio}. These coefficients grow exponentially at large $n$, indicating that $\langle\phi\phi\rangle$ has a finite radius of convergence in $\rho$. In other words, there is a singularity in $\langle\phi\phi\rangle$ when the two point are separated by a finite distance, signally a break-down of locality. We will discuss this phenomenon in great detail in section \ref{sec:NumericsandLocality}. 

Comparing these coefficients, we see numerically that for sufficiently large $c$ and $n$, the two types of coefficients seem to satisfy a rough empirical relation $b_{n}^2 \sim a_n$.
Since $a_n$ and $b_n$ are approximated by exponentials at large $n$, this relation would indicate that, roughly, $b_{2n} \sim a_n$. This is the condition for the holomorphic and full correlators to have similar radii of convergence. Therefore we believe that although many of our analytical results are explicitly obtained by studying $\langle\phi\phi\rangle _{\text{holo}}$, our conclusions about the physics should also hold approximately for $\langle\phi\phi\rangle$.  In figure \ref{fig:HoloVsMixedFit} we compare the convergence rates of the holomorphic and full propagators (in the $z$-$\bar z$ plane) explicitly.

\section{Perturbation Theory in $\frac{1}{c}$}
\label{sec:phiphiinoneoverc} 

We are using CFT$_2$ to learn about AdS$_3$ quantum gravity, so it  is very natural to study the expansion of observables in $G_N = \frac{3}{2c}$. In this section we will present the first $1/c$ correction to the propagator, and then a conjectured all-orders formula for light bulk proto-fields in the short-distance limit.

However, we find a surprising and potentially disturbing result, which appears already at one-loop: there are `UV/IR mixing' effects, by which we mean that singular, short-distance terms in $\< \phi \phi \>$ are enhanced by powers of the AdS scale $R_{AdS}$.  Specifically, at one-loop and at short distances $\sigma \ll R_{AdS}$, we find
\be
\< \phi \phi \>  &\approx &
   \frac{1}{\sigma} \left( \frac{3 G_N R_{AdS}^3 }{4 \sigma^4}-\frac{G_N R_{AdS} (10+m^2 R_{AdS}^2) }{8 \sigma^2} + \cdots \right) .
  \label{eq:perturbativephiphishortidstance}
\ee
Although this is a finite result in AdS$_3$, it does not have a good flat space limit as $R_{AdS} \to \infty$.  We believe there are two plausible responses to this state of affairs:
\begin{enumerate}
\item One can interpret this UV/IR mixing effect as a signal that $\< \phi \phi \>$ is too non-local in perturbation theory, and thus requires modification in order to obtain an IR safe quantity.   Likely this would involve summing over external graviton states in place of the vacuum.  We will not pursue this avenue of investigation here, but we believe it is interesting and important to consider, and we plan to return to it in the future.  
\item One can `bite the bullet' and simply study $\< \phi(X_1) \phi(X_2) \>$, the exact vacuum propagator.  In AdS$_3$ this observable  is finite and well-defined in perturbation theory, since AdS$_3$ acts as an IR regulator, and our results for it accord with naive gravitational perturbation theory in our gauge.   
We will take this approach for the remainder of this work, with the caveat that conclusions about $\< \phi \phi \>$ could change if we instead found a modified observable with an IR safe flat limit.
\end{enumerate}
In section \ref{sec:OneLoop} we will discuss the results of a one-loop gravity calculation, with the technical details relegated to appendix \ref{app:PerturbativePropagator}.  Then in section \ref{sec:AllOrdersoneoverc} we will present analytic results for the holomorphic propagator with fixed $h \ll c$, to all orders in $\frac{1}{c}$, but in the leading short-distance limit.  Our one-loop results exactly match those of the recursion relation of section \ref{sec:ExactlyfromRecursion}, and our all-orders results were obtained by extrapolating from the recursion relations.  If one takes the vacuum propagator $\< \phi \phi \>$ seriously as an observable, then our all-orders results suggest that bulk locality breaks down due to non-perturbative effects at a length scale $\sigma_{*} \sim c^{-1/4}$.  We will obtain corroborating evidence for this conclusion numerically in section \ref{sec:NumericsandLocality}.

\begin{figure}
\centering
\includegraphics[width=0.45\textwidth]{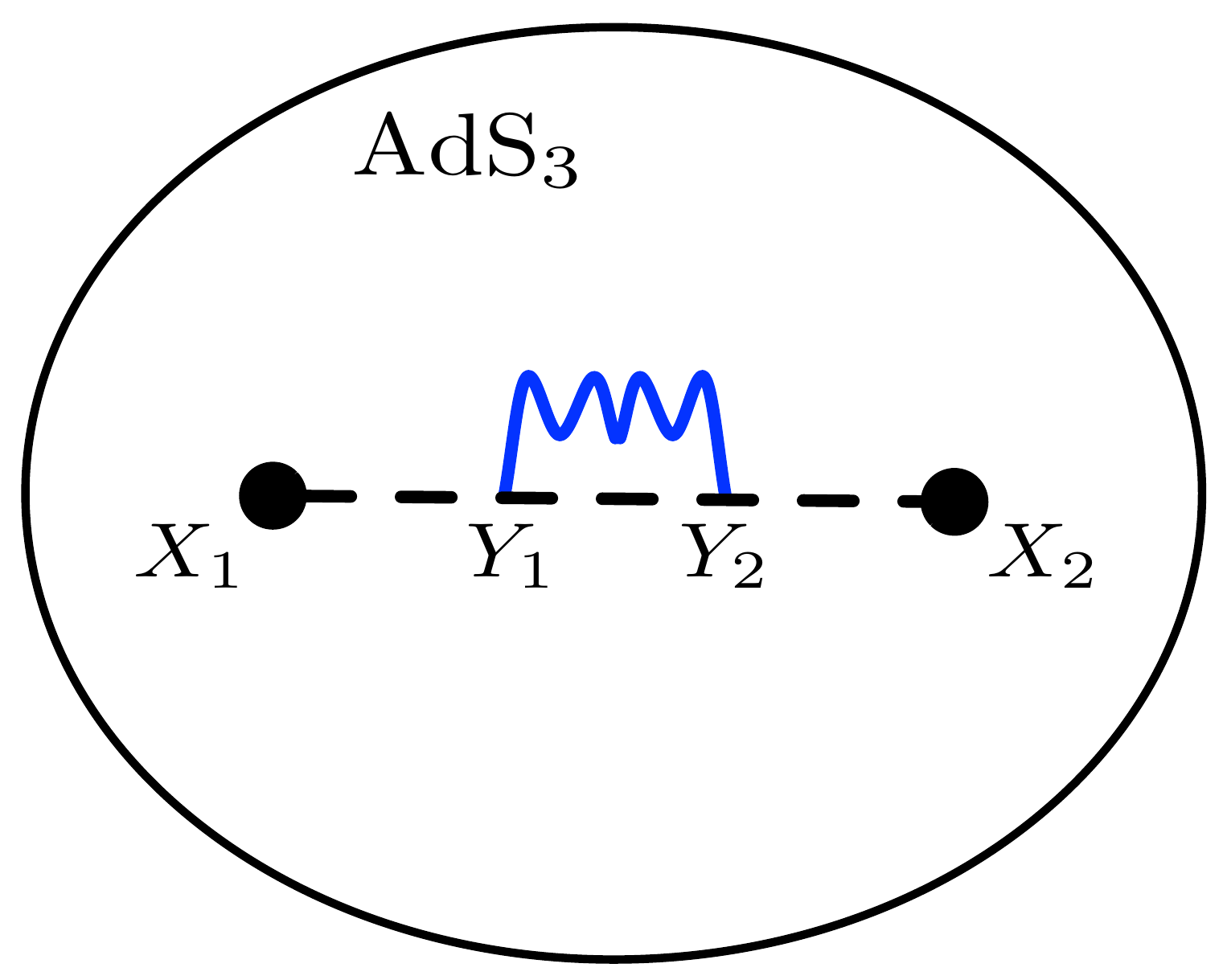}	
\caption{This figure displays the scalar-graviton one-loop diagram that contributes to $\< \phi(X_1) \phi(X_2) \>$ at order $1/c$.  There is also a contact interaction, but the associated diagram vanishes. The computation is performed in appendix \ref{app:GravitonLoop}.}
\label{fig:ScalarGravitonLoopDiagram}
\end{figure}

\subsection{One-Loop Bulk Gravity and UV/IR Mixing}
\label{sec:OneLoop}

We find that bulk perturbation theory matches the recursion relations developed in section \ref{sec:ExactlyfromRecursion}, and to leading non-trivial order in $1/c$, both give\footnote{Note  that the $1/c$ correction to $\<\phi\phi\>$ is just twice the $1/c$ correction to $\<\phi\phi\>_\hol$, because the full propagator gets corrections from both holomorphic and anti-holomorphic gravitons, but no mixed terms, at this order.}
\begin{align}
\label{eq:OneLoopResult}
\< \phi \phi \> =&\frac{\rho^n}{1-\rho}\left[1+ \frac{12 }{c } \left(\frac{\rho  \left(2 h^2 (\rho -1)^2+h (\rho  (3 \rho
   -11)+2) (\rho -1)+\rho ^2 ((\rho -5) \rho +10)\right)}{(1-\rho )^4} \right.\right.
\nn \\   & \left.\left. +
   2 h \rho ^2 \Phi (\rho ,1,2 h+1)+h \rho ^{1-2 h} B_{\rho }(2 h+1,-1)+2 (h-1) h
   \log (1-\rho )\right)\right] ,
\end{align}
where $B_\rho$ is the incomplete beta function and $\Phi$ is the Hurwitz Lurch function.
In appendix \ref{app:PerturbativePropagator} we explicitly perform the bulk loop calculation, and we also show how a part of this result can be obtained directly from unitarity.  

The formula above is complicated, but it simplifies in the short distance limit of $\sigma \ll 1$ with $\rho = e^{-2 \sigma}$.  The most singular terms are
\be
\< \phi \phi \>  &\approx &
   \frac{1}{\sigma} \left( \frac{3 G_N R_{AdS}^3 }{2 \sigma^4}-\frac{G_N R_{AdS} (10+m^2 R_{AdS}^2) }{4 \sigma^2} + \cdots \right).
  \label{eq:perturbativephiphishortidstance}
\ee
where we have used $G_N = \frac{3}{2c}$ and $m^2 = 2h(2h-2)$, and we have also included factors of the AdS scale. 
This result suggests a new length scale
\be
\sigma_{*} \sim \sqrt[4]{G_N R_{AdS}^3} .
\ee
Although this follows straightforwardly from perturbative gravitational field theory, the emergence of this new scale is quite surprising.  It is indicative of UV/IR mixing and the presence of IR divergences in the flat space limit $R_{AdS} \to \infty$.  We do not expect a result like equation (\ref{eq:perturbativephiphishortidstance}) from a well-defined local observable
in a local theory.  At a computational level, the scale $\sigma_*$ arises  from the $\sigma^{-5}$ short distance singularity in equation (\ref{eq:perturbativephiphishortidstance}), which can itself be traced to the fact that the bulk `graviton' propagator  \cite{Kabat:2012hp}  is proportional to $\frac{1}{(z_1 - z_2)^4}$ and independent of anti-holomorphic coordinate $\bar z$ and the radial direction $y$.   This AdS$_3$ graviton propagator has been used successfully in other calculations; for example the results of \cite{Fitzpatrick:2016mtp} can be re-interpreted \cite{Anand:2017dav} as geodesic Witten diagrams  \cite{Hijano:2015zsa} that use this graviton propagator to compute conformal blocks.  But we would expect a quite different graviton propagator in higher dimensions \cite{LiuTseytlin}.

Note that the less singular terms in equation (\ref{eq:perturbativephiphishortidstance}) also display UV/IR mixing.  There is both a semiclassical effect $\sim \frac{G_N m^2 R_{AdS}^3}{\sigma^2}$ and a quantum effect $\sim \frac{G_N R_{AdS}}{\sigma^2}$ which are enhanced by $R_{AdS}$.  The former has also been obtained from the monodromy method of section \ref{sec:Semiclassical}.  This suggests that it may be quite non-trivial to define a fully IR safe modification of $\< \phi \phi \>$.  We should also emphasize that because 
equation \ref{eq:perturbativephiphishortidstance} has been obtained directly from unitarity in appendix \ref{app:PerturbativePropagator}, modifying it may require a different choice for the $\< \phi \CO T \>$ correlator, which itself follows \cite{Anand:2017dav} from a simple tree-level  calculation.  Modifying $\< \phi \CO T \>$ might also jeopardize the ability of $\phi$ to `know its location' \cite{Anand:2017dav} in general semiclassical geometries.

Finally, it is natural to ask whether the one-loop corrected  propagator can be used as an ingredient in a complete and gauge-invariant calculation of a CFT correlator.  In this way one might approach the short-distance behavior of the propagator indirectly.  For example, we could attach $\< \phi(X) \phi(Y)\>$ to a pair of bulk-boundary scalar propagators at $X$ and another pair at $Y$, and then integrate over $X$ and $Y$ to obtain a complete Witten diagram for a CFT 4-pt correlator, though it is not clear exactly what CFT quantity such a diagram should correspond to when the fully dressed $\phi$ propagator is used.   When computing Virasoro conformal blocks using Wilsons lines  \cite{Fitzpatrick:2016mtp}, bulk diagrams like figure \ref{fig:ScalarGravitonLoopDiagram} were not included.  The connection between the recursion relations of section \ref{sec:ExactlyfromRecursion} and the Zamolodchikov relations for conformal blocks might also provide further clues.  It would be interesting to study these issues further.

\subsection{All-Orders in $\frac{1}{c}$ in the Short Distance Limit}
\label{sec:AllOrdersoneoverc}

Now let us study $1/c$ perturbation theory to all orders.  We are interested in light fields with $h \ll c$.  In fact, the correlator $\< \phi \phi \>$ remains very non-trivial even when $h \to 0$, so for definiteness and simplicity we will focus\footnote{This does not imply that the identity operator/vacuum has a bulk dual, as infinitesimal $h$ differs from $h=0$ identically.  Even in the $c=\infty$ limit the propagator is the non-trivial $\frac{1}{1 - \rho}$ as $h \to 0$.} on this case, which we have found (numerically) to be representative of the light field regime.  Using the recursion relations of section \ref{sec:ExactlyfromRecursion}, we find that the holomorphic part $K_\hol(\rho) = \< \phi \phi \>_\hol$ of the bulk propagator takes the form\footnote{These results are really conjectural, as they were discovered by computing the $\rho$ expansion to high orders using the recursion relations of section \ref{sec:ExactlyfromRecursion}  and then identifying a pattern in the result.}
\be
K_\hol = \frac{1}{1 - \rho} \left(1 + \sum_{n=1}^\infty \rho^3 f_n(\rho) \frac{(4n-1)!!}{n!}  \left( \frac{12}{c (1-\rho)^4} \right)^n \right),
\label{eq:KholSeries}
\ee
where $f_n(\rho)$ are polynomials of order $4n-2$ in $\rho$.  For definiteness, the first three are
\be
f_1(\rho) &=&  \frac{1}{6} \left(\rho ^2-5 \rho +10\right), \\
f_2(\rho) &=& \frac{1}{1260} \left(13 \rho ^6-117 \rho ^5+468 \rho ^4-1112 \rho ^3+1833 \rho ^2+195 \rho -20\right), \nn \\
f_3(\rho) &=& \frac{1}{99786} \left(41 \rho ^{10}-533 \rho ^9+3198 \rho ^8-11718 \rho ^7+29226 \rho ^6-56454 \rho ^5+105078 \rho ^4 \right. \nn\\
 && \left. +34722 \rho ^3-3687 \rho ^2-89 \rho +8\right) , \nn
\ee
and we have computed  $f_1(\rho)$ perturbatively in appendix \ref{app:PerturbativePropagator}. 
We have chosen the normalizations so that $f_n(\rho \to 1) = 1$ in order to ensure that the $f_n$ become trivial in the short-distance limit. This means that to leading order in that limit, $K_\hol$ takes the very simple form
\be
\label{eq:Kholoshortdistanceseries}
K_{\text{holo}}(\rho \to  1) \approx \frac{1}{1 - \rho} \left(1 + \sum_{n=1}^\infty  \frac{(4n-1)!!}{n!}  \left( \frac{12}{c (1-\rho)^4} \right)^n \right) .
\ee
So we see that the quantity $c (1-\rho)^4 \propto c \sigma^4$ indicative of the new bulk length scale $\sigma_* \sim c^{-1/4}$ appears in every term.  

The series expansion in $1/c$ has zero radius of convergence because the coefficients grow factorially.  But this series is not very exotic, and in fact it  can be obtained from the $1/c$ expansion of the well-studied quartic integral
\be
Z = \frac{1}{\sqrt{2\pi}}\int_{-\infty}^\infty dz \, e^{-\frac{1}{2}z^2 + \frac{12}{c(1-\rho)^4} z^4} .
\ee
This integral can be re-summed either via a Borel transform or by noting that it obeys a second order differential equation.\footnote{If we define $x = \frac{72}{c (1-\rho)^4}$ then the integral obeys the differential equation
\be\label{eq:DifferentialEquationNonPertur}
16 x^2 Z'' + (32x-6)Z' + 3 Z = 0
\ee
which can be solved in terms of incomplete Bessel functions.}  The correspondence between the quartic integral and the combinatorics of the series is easy to explain by considering the computation of $\< \phi \phi \>$ in gravitational perturbation theory.  The leading terms at  short-distances come from summing all diagrams generated by a bulk cubic coupling of schematic form $\frac{1}{\sqrt{c}} (\partial_{\bar z} \phi)^2 h_{zz}$ and ignoring graviton self-interactions.  We can count  the diagrams in this theory by integrating out the graviton, which leads to a pure quartic interaction for $\phi$ and explains the combinatorics of our result.

One might expect that one could take this leading order diagrammatic argument further, working to all orders in the effective action after integrating out the graviton.  Since we are dealing with $\< \phi \phi\>_{\rm holo}$, one should include only the holomorphic modes of the graviton.  We can obtain additional evidence that such an effective action for $\phi $ is possible by computing finite-distance corrections to the correlator.  More precisely, we look at small $\sigma$ corrections to the limit with $c \sigma^4$ fixed at large $c$ (equivalently, these are $1/c$ corrections to the large $c$, fixed $c \sigma^4$ limit).  In terms of the representation (\ref{eq:KholSeries}), these corrections are the subleading series coefficients in the $f_n(\rho)$s in an expansion around $\rho = 1$.  We find empirically that these subleading terms are correctly reproduced up to the fourth derivative $f_n^{(4)}(1)$ by the following integral expression:
\be
(1-\rho) K_{\rm holo}& \sim& e^{- (\sigma - \frac{\sigma^2}{2} + \frac{\sigma^4}{12} )}\left[ -1+ \sqrt{\frac{c \sigma^4 }{\pi}}
\int_{-\infty}^\infty e^{-c \sigma^4 (z^2 + a_4(\sigma) z^4 + a_6(\sigma) z^6 + a_8(\sigma) z^8 )} dz \right], \nn\\
\ee
where we have determined the first few $a_n$ coefficients  to be
\be
a_4(\sigma) &=& -3+ 6 \sigma^2  - \frac{151}{15} \sigma^4 , \nn\\
a_6(\sigma) &=& - 27 \sigma^2 + \frac{617}{5} \sigma^4,  \nn\\
a_8(\sigma) &=& =\frac{3519}{10} \sigma^4 ,
\ee
up to higher order corrections in $\sigma$.  What is notable about this expression is that, by fitting only a few numbers in the $a_n(\sigma)$ coefficients, we correctly reproduce the first several terms  in the $f_n(\rho)$ expansion around $\rho=1$ for all $n$.  It would be very interesting if these $a_n$ coefficients could be determined directly by integrating out the graviton modes in AdS$_3$.\footnote{The effective actions in \cite{Mertens:2017mtv} may be a useful tool for such a derivation.}

Coming back to the leading order expression (\ref{eq:Kholoshortdistanceseries}), we can attempt to transform the asymptotic series into an exact function.  
Either by solving the differential equation (\ref{eq:DifferentialEquationNonPertur}) or by Borel resumming,  we obtain a one-parameter family of possible results, which are linear combinations of modified Bessel\footnote{They have series expansions $I_\nu(x) = \sum_{k=0}^\infty  \frac{1}{\Gamma(k+\nu+1)k!} \left( \frac{x}{2} \right)^{2k+\nu}$ and are real for $\nu=\pm \frac{1}{4}$ and $x>0$.} functions,
\be
\label{eq:ExactShortDistancephiphi}
\lim  \Big[ (1-\rho) K_\hol (\rho) \Big] 
 &=&e^{-X} \sqrt{2\pi X} \left(\left(1 + \kappa \right) I_{\frac{1}{4}}(X)-\kappa  I_{-\frac{1}{4}}(X)\right),
\ee
where  $X \equiv \frac{c(1-\rho)^4}{384}$ and the limit is $c \to \infty$ and $\rho \to 1$ with $X$ fixed.  One can verify that this function reproduces equation (\ref{eq:Kholoshortdistanceseries}) when expanded in large $c$.  

The parameter $\kappa$ in equation (\ref{eq:ExactShortDistancephiphi}) is arbitrary, as $K_\hol$ has the correct perturbative expansion around $c=\infty$ for any value of this parameter.  Thus $\kappa$ represents a non-perturbative ambiguity in the definition of the correlator; it arises because there is a branch cut on the positive real axis in the Borel plane. 

If $\kappa$ is real, then equation (\ref{eq:ExactShortDistancephiphi}) will be real for positive $X$, i.e. for positive $c$ and real $\rho$.  
On the other hand, if  the correct  choice is not $\kappa \in \mathbb{R}$, then $K_\hol$ will be complex.  If the propagator has a Kallen-Lehmann representation, then it would seem that its spectral function must develop an imaginary part in this case.  A complex value for a scalar propagator usually signals the presence of an instability where $\phi$  quanta decay into other states. However, it is less clear what the precise interpretation is in our case since our $\phi$ is a linear combination of descendants of the scalar primary $\CO$.  In CFT$_2$ such operators cannot mix with the vacuum sector (i.e. with `gravitons'), as correlators like $\< \phi T\cdots T\>$ vanish, and the only interactions we have included are those of $\phi$ with gravity. Thus any Im$[\kappa] \neq 0$ suggests a non-perturbative violation of unitarity at short distances.

We cannot determine the value of $\kappa$ with the methods of this section.  However, in section \ref{sec:NumericsandLocality} we will take a numeric approach, and argue that the $\< \phi \phi\>$ correlator develops a singularity and likely an imaginary piece at short distances.

\section{Numerics and Locality}
\label{sec:NumericsandLocality}

Arguments based on black hole thermodynamics and the gauge redundancies of general relativity suggest that local observables do not exist in quantum gravity.  However, we have introduced an exact bulk proto-field operator $\phi(X)$ and provided various techniques for computing its correlation functions.  While $\phi(X)$ is in some sense a non-local operator,\footnote{In part $\phi$ is non-local simply because it includes gravitational dressing; this is analogous to the way that the electron operator is non-local because it must be attached to a Wilson line.  But we should also recall the caveat (discussed in section \ref{sec:phiphiinoneoverc}) that even in bulk gravitational perturbation theory $\< \phi \phi \>$ exhibits a surprising UV/IR mixing, so perhaps $\sigma_*$ can be modified or eliminated by identifying a different observable with better IR behavior.} one may nevertheless wonder if its correlation functions exhibit pathologies associated with the failure of bulk locality in quantum gravity. 

The propagator depends on the central charge $c$,  on the kinematic configuration, and on the conformal dimension $h$ of the CFT$_2$ scalar primary $\CO$ dual to $\phi$.    As we have explained in sections \ref{sec:HolomorphicDefinitions}, the full $K$ depends on two independent kinematic variables, although most of the non-trivial information in the full $K$ can be obtained from the holomorphic part $K_\hol$.  This part only depends on the variable $\rho = e^{-2 \sigma}$, where $\sigma(X,Y)$ is the geodesic separation between $X$ and $Y$ in the vacuum.  Throughout this section we will mostly focus on $K_\hol$, as it is easier to obtain high-orders numerical results for this object, though in figure \ref{fig:HoloVsMixedFit} we provide evidence that our conclusions concerning $K_\hol$ should also apply to the full $K$.

Recall that in section \ref{sec:LargehLimit} we already observed a sharp conflict between bulk locality  when in the limit of large $h$.  The conflict arose because $K_\hol$ developed an imaginary part at the geodesic separation $\sigma_* = R_{AdS}\log(2 +  \sqrt{3}) \approx 1.32 R_{AdS}$.  Of course we would not have expected correlators of fields with trans-Planckian masses to be local, so this result was not too surprising.  In this section we mostly focus on light bulk fields, though the form of the pathologies we uncover will be very similar.

We have studied short-distance locality in three ways.  First, in section \ref{sec:phiphiinoneoverc}  we discussed the $1/c$ expansion of $K_\hol$, and observed that the $1/c^n$ corrections can be determined exactly in the short distance limit $\rho \to 1$.  However, the $1/c$ expansion was asymptotic (it has zero radius of convergence), and Borel resumming the series led to a non-perturbative ambiguity.  Generically, this means that $\< \phi \phi \>$ develops an instability or unitarity-violating imaginary piece, though there does exist a reality-preserving resolution of the ambiguity. 

Second, in section \ref{sec:Semiclassical}, we developed methods that allow us to compute the semiclassical part of the correlator numerically and, in some cases, analytically.  In particular, we numerically computed the critical value $\sigma_c$ where the semiclassical part develops an imaginary piece.  At infinite $h$, where we know the exact (not just semiclassical) correlator, this critical value for the semiclassical part matches that of the exact result.  At smaller $h$, the exact correlator could in principle develop additional singularities or imaginary pieces at even larger values of $\sigma$, but it seems very unlikely that quantum effects  could cancel the imaginary part of the semiclassical propagator.\footnote{We cannot prove this does not happen.  However, the semiclassical piece scales differently ($\CO(c)$ in the exponent) from the residual piece ($\CO(1)$), so a potential cancellation cannot be as simple as the residual piece contributing an exactly opposite phase. }  

Third,  in the next section, \ref{sec:phiphinumerical}, we will obtain additional numeric evidence that is complementary to the first and second methods, by  evaluating $K_\hol$  to high orders in the $\rho$ expansion.  This numeric high-order behavior provides abundant evidence that the $\rho$-series has a finite radius of convergence, breaking down when $\sigma \propto -\log \rho \propto c^{-1/4}$ when $h \sim \CO(c^0)$, and at $\sigma \propto (h/c)^{1/3}$ when $h/c$ is fixed but small in the large $c$ limit.   Assuming our numerical extrapolations are correct, this implies that $\< \phi \phi \>$ becomes singular at a finite separation,  which is a harbinger of the failure of bulk locality.  It may be possible to analytically continue the propagator  to shorter distances, but one would expect it to develop an imaginary part.  In section \ref{sec:InterpretationBreakdown} we discuss the interpretation of these results.

\begin{figure}
\centering
\includegraphics[width=0.65\textwidth]{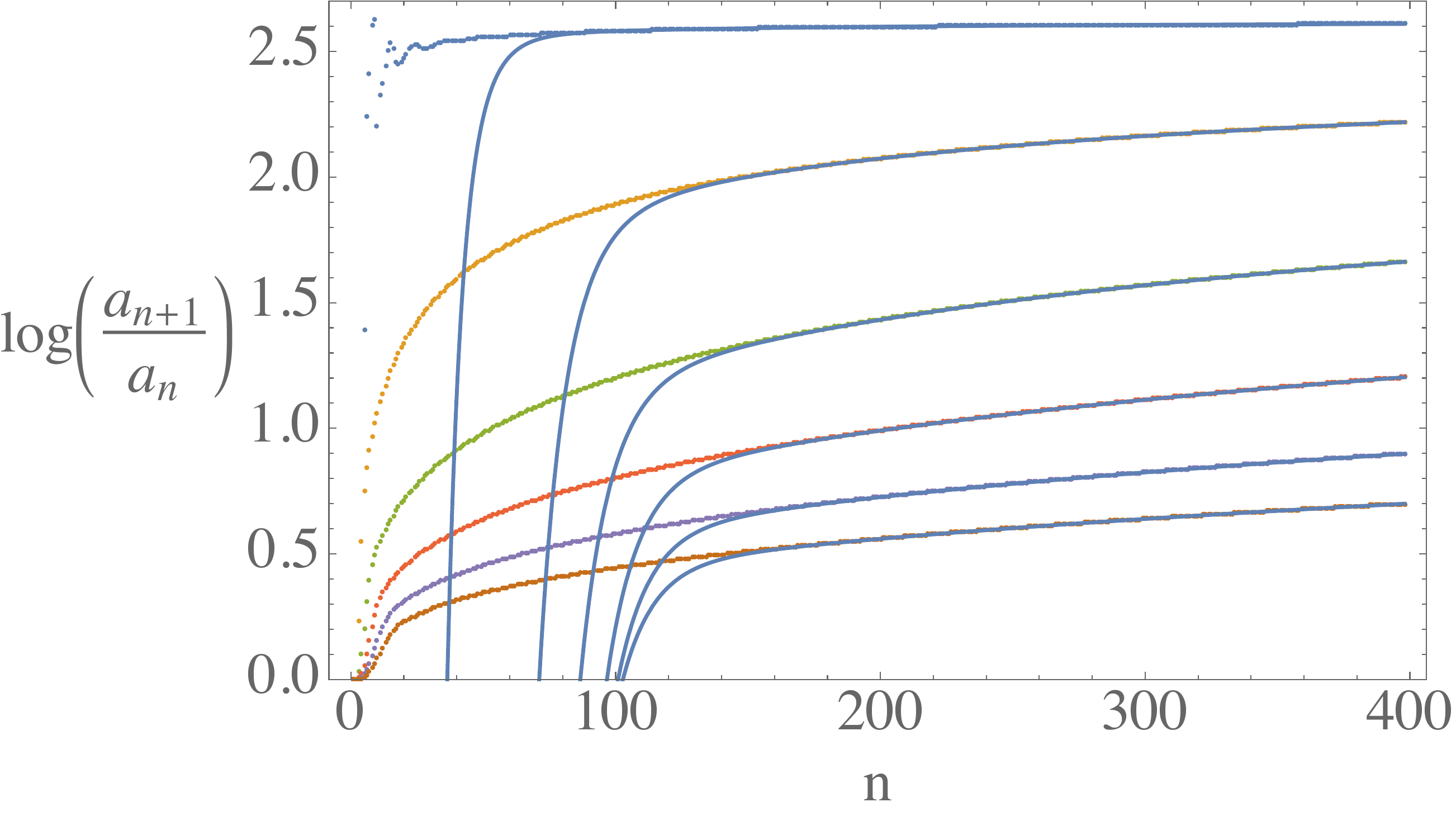}	
\caption{
This figure displays fits to logarithms of ratios of successive coefficients in the $\rho$ expansion of equation (\ref{eq:DefinitionSeriesCoeffs}) up to the $400$th order.  In all cases we have set $h = 0$ identically, and the value of $c$ increases from the top to the bottom of the plot, ranging from $1.5$ to $10^5$.   Each line corresponds to one of the points on Fig. \ref{fig:NumericFitsFinalDataZeroh}, but for legibility we have only included every fifth point.} 
\label{fig:NumericFitsZeroh}
\end{figure}

\begin{figure}
\centering
\includegraphics[width=0.7\textwidth]{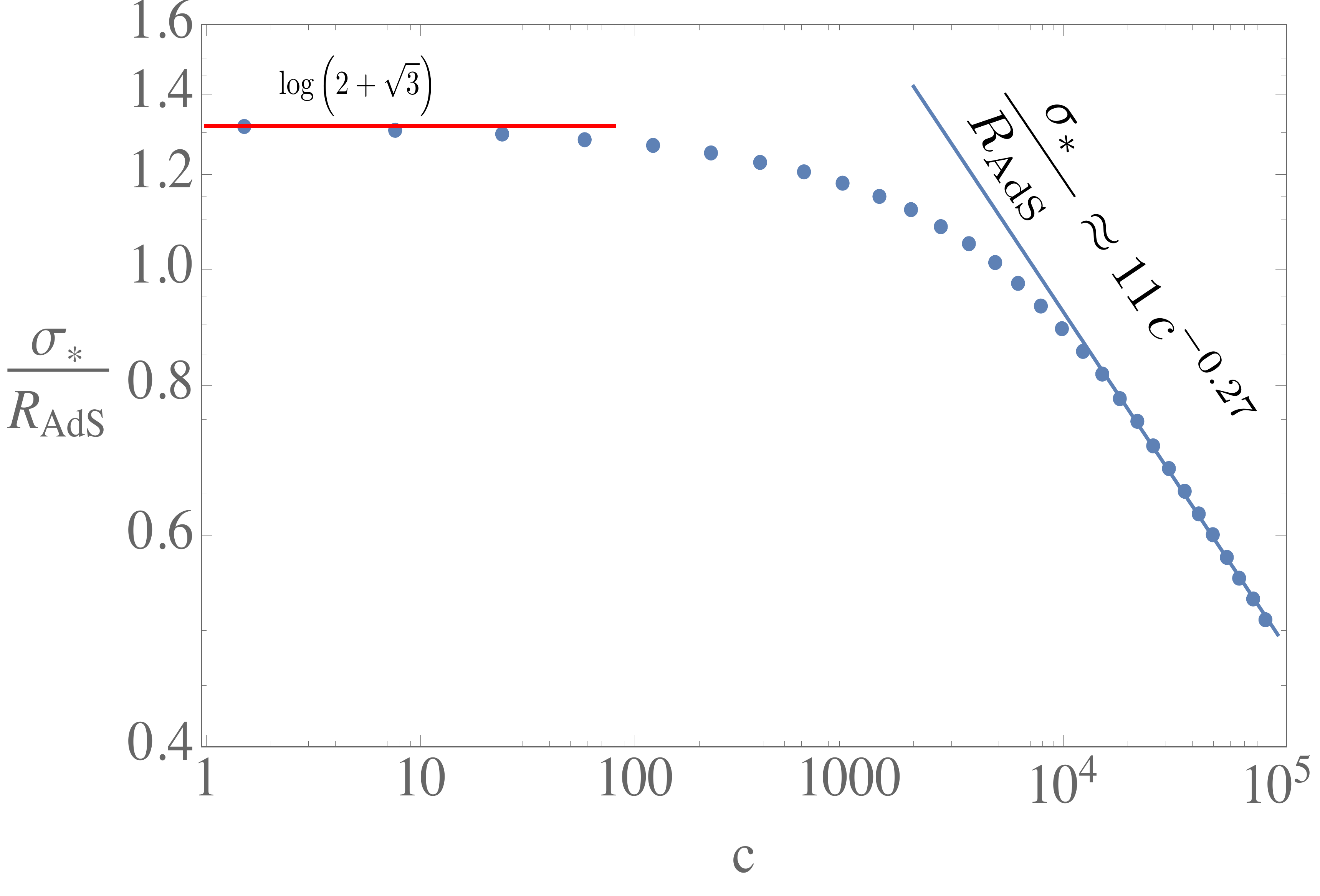}	
\caption{ In this plot we used the fits of Fig. \ref{fig:NumericFitsZeroh}  to extract an approximate asymptotic ratio $\frac{a_{n+1}}{a_n}$, which was then used to identify $\sigma_*$, the scale at which bulk locality appears to break down, for each value of $c$.   For very small values of $c$ we find $\sigma_*$ of order the AdS scale, so that at $c \to 1$ we smoothly match the large $h$ results of section \ref{sec:LargehLimit}, as indicated by the red line.  At large $c$ we enter the flat space regime of small $\sigma_*$, where we extract the fit $\sigma_* \propto c^{-0.27}$.  Varying the details of the fitting shifts the exponent, but we consistently find that it lies between $0.25$ and $0.28$.}
\label{fig:NumericFitsFinalDataZeroh}
\end{figure}

\subsection{Numerical Results for the Exact $\rho$ Expansion}
\label{sec:phiphinumerical}

\begin{figure}
\centering
\includegraphics[width=0.9\textwidth]{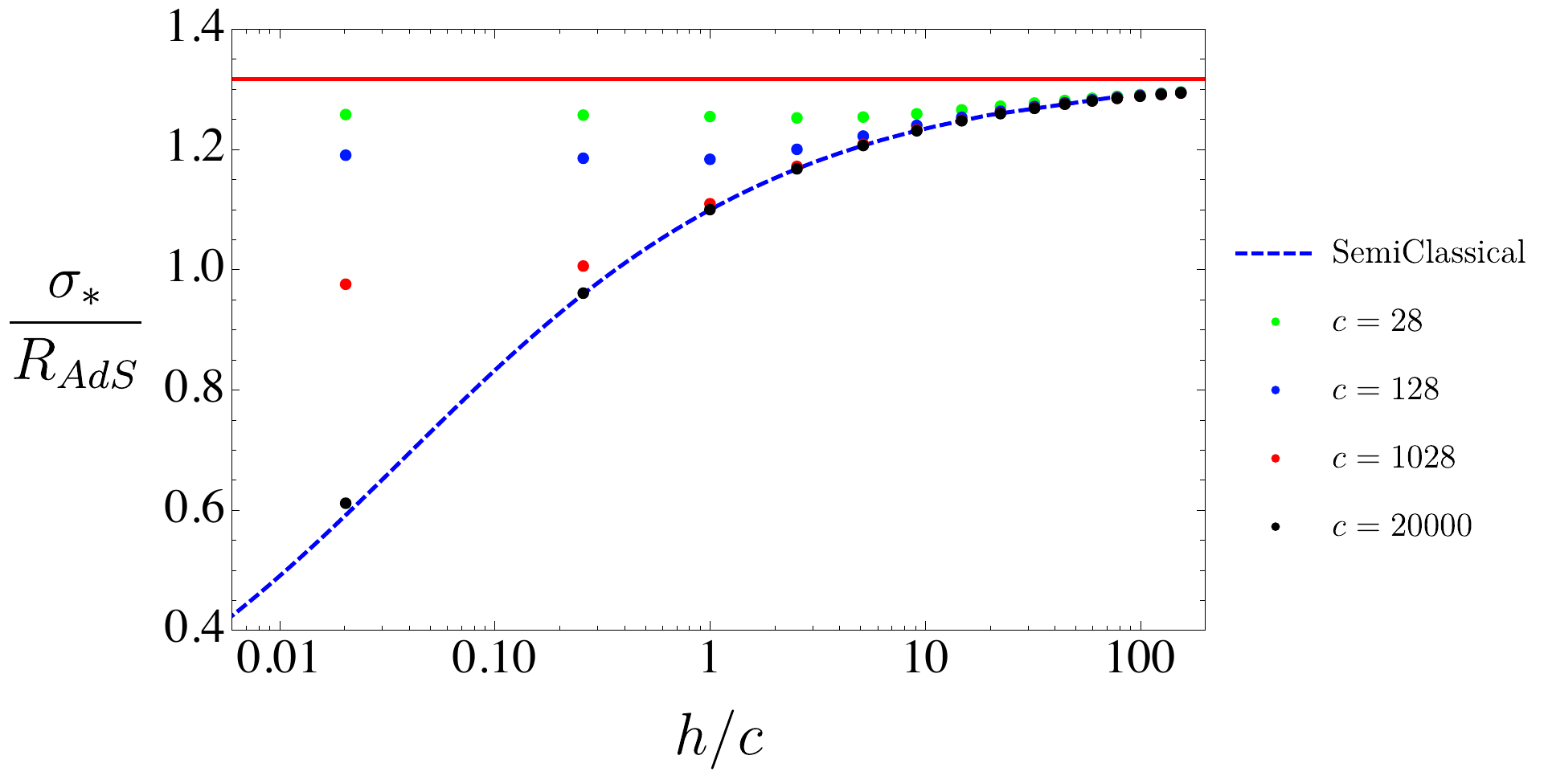}	
\caption{This figure displays the scale at which the propagator breaks down as we approach the semiclassical limit; for each value of $c$, we've taken a range of values for the ratio $\frac{h}{c}$.  The data was extracted in the same way as in Fig. \ref{fig:NumericFitsFinalDataZeroh}. We see that at large $h$ we approach the convergence bound $\sigma_* = R_{AdS} \log( 2 + \sqrt{3})$ from the exact result of section \ref{sec:LargehLimit}. We have also shown ({\it blue, dashed}) the result from the numeric semiclassical computation in section \ref{sec:NumMono}, and find that it agrees with the radius of convergence analysis for the large $c$ (= 20,000) points shown above. }
\label{fig:SemiclassicalNumericalPlot}
\end{figure}

In this section we will study the AdS$_3$ proto-field propagator numerically to high orders in the $\rho$ expansion.  Since $\rho = e^{-2 \sigma}$ and $\sigma$ is the geodesic separation between the points, this is an expansion around the long-distance limit.  So on physical grounds, we should expect the propagator to be well-behaved as $\rho \to 0$.  If bulk locality did not break down, then we would expect the radius of convergence of the $\rho$-series to be $1$, as is the case for the free field propagator $\frac{\rho^h}{1-\rho}$.  Instead we will present evidence that:
 \begin{itemize}
 \item The radius of convergence in $\rho$  is strictly less than $1$ at finite $c$, which means that $K_\hol$ develops a singularity\footnote{Pad\'e approximants to the $\rho$ series expansion display a `condensation' of poles that suggest that at distances shorter than $\sigma_c$, the correlator will develop a branch cut.  %(This footnote was written on Halloween, explaining the invocation of black magic.)
 % Sorry this was just too silly for my taste :P
 } at some finite critical distance  $\sigma_*(c) > 0$.
 \item  The failure of convergence occurs at a physical separation in AdS$_3$  that scales as  $\sigma_*(c) \propto c^{-p}$ at large $c$.  We find $0.25 < p < 0.28$, which approximates the expected $p \approx \frac{1}{4}$ from section \ref{sec:phiphiinoneoverc}  but appears slightly larger, as shown in figure \ref{fig:NumericFitsFinalDataZeroh}.  This behavior holds throughout the $h \ll c$ regime.
 \item When $h \sim c \gg 1$, convergence fails at a physical separation of order the AdS$_3$ length.  The behavior as $h \gg c$ connects smoothly with the results of section \ref{sec:LargehLimit}, as shown in figure \ref{fig:SemiclassicalNumericalPlot}.  We also find that for any $h$, when $c \approx 1$ the propagator breaks down at roughly the same distance scale as in the large $h$; this is indicated with the red line in Fig. \ref{fig:SemiclassicalNumericalPlot}.
 \end{itemize}
Since these results only follow from a numerical analysis, they are not theorems.  Readers are encouraged to conduct their own investigations with the attached code implementing the recursion relations of section \ref{sec:ExactlyfromRecursion}.

The radius of convergence in $\rho$ can be analyzed by studying the growth of the coefficients $a_n$ in the expansion  
\be
\label{eq:DefinitionSeriesCoeffs}
K_\hol = \rho^h \sum_{n=0}^\infty a_n \rho^n ,
\ee
where the $a_n$ depend implicitly on $h$ and $c$.  If the radius of convergence in $\rho$ is less than $1$, then the $a_n$ must grow exponentially, which means that as $n \to \infty$ we must have $\frac{a_{n+1}}{a_n} \to r$ for some $r > 1$.  However, there will likely be a subleading power-law behavior as well, so that $a_n \approx  n^v r^n$ for some $v$.  We display a fit to this behavior  for $30$ values of $c$, ranging from $1.5$ to $10^5$ in figure \ref{fig:NumericFitsZeroh}.

The convergence radius in $\rho$ and thus the value of $r$ will correspond with a physical geodesic distance scale in the bulk $\sigma = -\frac{R_{AdS}}{2} \log r$.  Since $r$ depends implicitly on $c$, if the physical separation is proportional to $c^{-1/4}$, then we should find $\log [r(c)] \propto c^{-1/4}$ at large $c$.  We can test this hypothesis by identifying $r(c)$ for a large range of values of $c$, and then fitting a line to $\log [\log r(c)]$ vs $\log c$, as the slope of this line measures the exponent $-\frac{1}{4}$.  We have provided such a fit in figure \ref{fig:NumericFitsFinalDataZeroh}.  Varying the details of the fit changes the exponent $p$ of $c^{-p}$, but in all cases we find that $p$ ranges between about $0.25$ and $0.28$.  Thus the exponent appears systematically slightly larger than would be expected from the analysis of section \ref{sec:phiphiinoneoverc}.  This may be due to the fact that for large values of $c$, we simply do not have enough coefficients $a_n$ to get to the asymptotic regime of very large $n$ necessary to correctly identify the exponent.  But this discrepancy may be worthy of further consideration.

\begin{figure}
\centering
\includegraphics[width=0.85\textwidth]{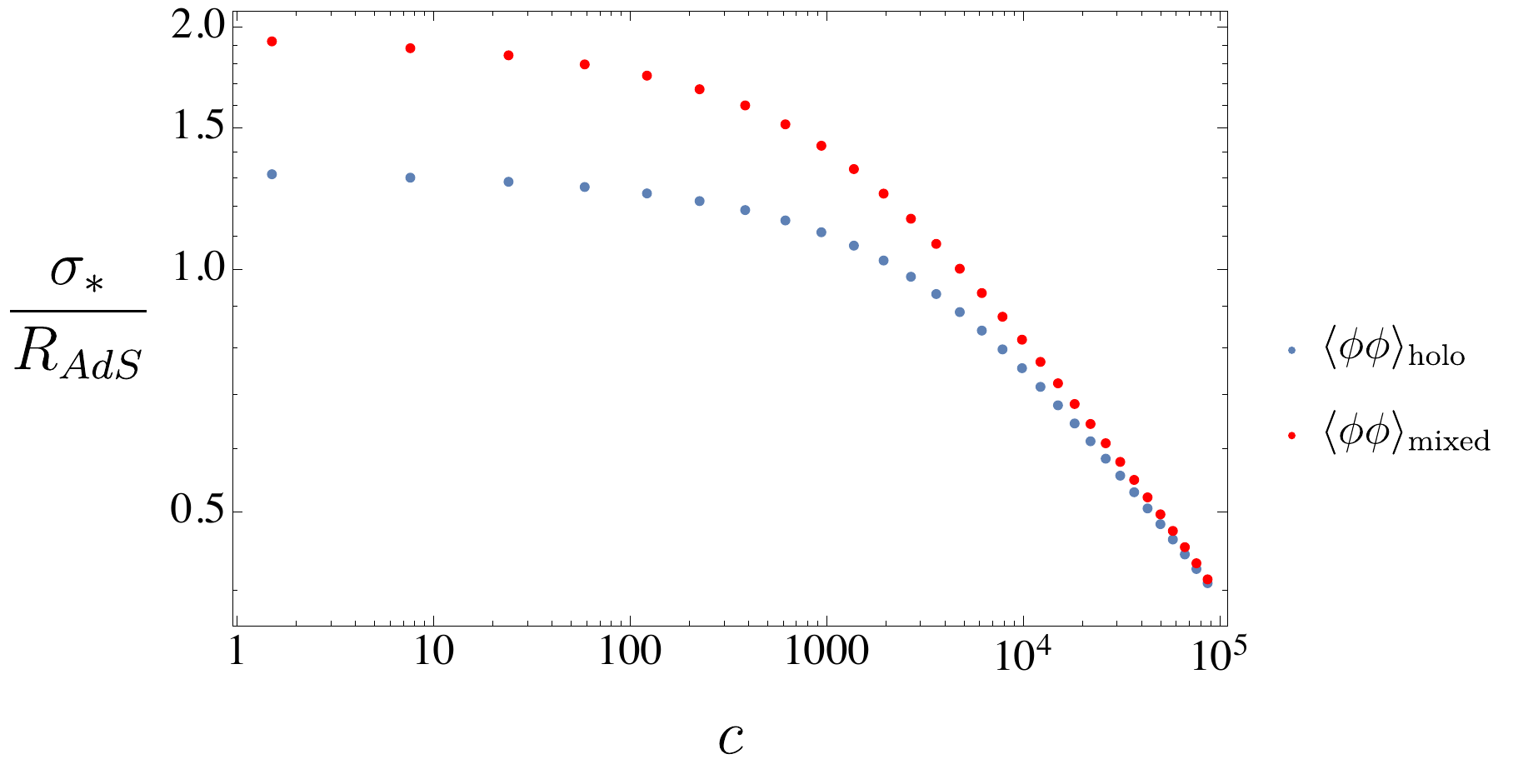}	
\caption{ This plot shows the same  data as figure \ref{fig:NumericFitsFinalDataZeroh} for the radius of convergence $\sigma_*$ as a function of $c$, except here we compare the radii of convergence of the holomorphic and the mixed contributions to the full correlator.  Due to numerical limitations we only include the coefficients $a_n$ and $b_n$ up to  $n=300$, which means that these data are not as reliable as those of figure \ref{fig:NumericFitsFinalDataZeroh}.  Nevertheless, this plot provides clear evidence that the parametric scaling of $\sigma_*$ with $c$ is the same for the full and holomorphic propagators when we restrict to the $z$-$\bar z$ plane.   }
\label{fig:HoloVsMixedFit}
\end{figure}

We can also study the radius of convergence in the case of general $h$.  At large enough $c$, we expect to enter the semiclassical regime where the radius of convergence should depend only on the ratio $h/c$, and indeed this is what we see in figure \ref{fig:SemiclassicalNumericalPlot}.  In fact, we find the radius of convergence in this limit exactly matches the results of section \ref{sec:LargehLimit}, for the critical value for  $\sigma$ where an imaginary piece develops.  One might have expected a distinctive feature at the BTZ black hole threshold, $h = \frac{c}{24}$, i.e. the value where the corresponding primary state develops a horizon in AdS, but we do not see any such feature, and it is not until $h/c \sim 1$ that the curve starts to flatten out towards its asymptotic large $h$ value.  

Interestingly, at small $c$, regardless of $h/c$, the radius of convergence also approaches the large $h$ value $R_{\rm AdS} \log( 2 + \sqrt{3})$. So it appears that this is a fairly generic ``strong coupling'' result, valid either at small $c$ or at large $h/c$.  It would be very interesting to understand a more physical origin of this scale.

Finally, in figure \ref{fig:HoloVsMixedFit} we compare the radius of convergence of the holomorphic propagator to that of the mixed terms (recall that the full propagator $K = 2 K_\hol + K_{\mathrm{mixed}} - K_{\mathrm{global}}$) in the $z$-$\bar z$ plane, following up on the preliminary analysis in section \ref{subsec:FullVSHolo}.  We only used the coefficients $a_n$ and $b_n$ up to  $n=300$, and so the precise $\sigma_*(c)$ from this plot is not as reliable as that of figure \ref{fig:NumericFitsFinalDataZeroh}.  However, we can see from the scaling of $\sigma_*$ with $c$ at large $c$ that the radius of convergence of the full correlator seems to scale in the same way as that of $K_\hol$.  At small $c$ the behavior also appears similar insofar as both correlators break down at the AdS scale, though the precise radius of convergence differs by an order one factor.  This result largely justifies our focus in this paper on the simpler $K_\hol$, but it would still be interesting to study the full propagator in more detail in future work.  We have not studied the full propagator away from the $z$-$\bar z$ plane in detail, so it would be very interesting to explore that regime.

\subsection{Interpreting the $c^{-1/4}$ Length Scale}
\label{sec:InterpretationBreakdown}

In the previous section, we found evidence that the $\< \phi \phi\>$ correlator develops a singularity or branch cut at a scale $\sim \CO(c^{-1/4})$ when $h$ is fixed as we take the large $c$ limit.\footnote{We showed results for vanishing $h$, but have found the same  behavior for fixed small $h$. }  This scale is truly quantum, invisible in the semiclassical limit, and in some sense represents an irreducible distance below which locality breaks down.  It is natural to ask if this length appears as a fundamental  scale in the AdS$_3$ gravity theory itself, and not just in the $\< \phi \phi\>$ correlator.  

From an effective theory viewpoint, such a fundamental scale would be extremely surprising,\footnote{For instance, an analogous result in de Sitter in $d=4$ would imply that quantum gravity effects become relevant at some geometric average length scale between the Planck length and the Hubble radius, long before they would be expected to be important.  However, we do not expect UV/IR mixing in higher dimensional theories, so we do not believe this phenomenon can occur.} since  it involves both the Planck length $\ell_{\rm pl}$ and the AdS  radius $R_{\rm AdS}$. Restoring dimensionful quantities,
\be
\sigma_* \sim c^{-1/4} \sim R_{\rm AdS}^{3/4} \ell_{\rm pl}^{1/4}.
\ee
One possibility is certainly that this length scale is not fundamental, but rather is an artifact of our definition of the proto-field $\phi$.  As we have mentioned, perhaps the correct lesson is that one should attempt to define an improved bulk field that has a good flat space limit.  

However, it may  be that this scale is truly indicative of the underlying physics of quantum gravity in AdS$_3$.  From this point of view, it is interesting to note that the scale $c^{-1/4}$ also arises as the smallest  string length $\ell_s$ in known stable, controlled string compactifications in perturbative string theory.   The basic reason that $c^{-1/4}$ appears in this context is straightforward to understand.  In compactifications of the form AdS$_3 \times S^3 \times M_4$ where $M_4$ is a 4d compact manifold, the radius of the $S^3$ is the same as the AdS$_3$ length scale $R_{\rm AdS}$, and the size of the $M_4$ cannot be taken smaller than $\ell_s$.  So, the 3d Planck length $\ell_{\rm pl}$ is related to the 10d Planck length $\ell_{10}$ by
\be
\ell_{\rm pl} R_{\rm AdS}^3 \ell_s^4 \lesssim \ell_{10}^8 \lesssim \ell_s^8 ,
\ee 
where we have used the fact that the $M_4$ volume is greater than $\ell_s^4$, and  the 10d Planck scale must be smaller than the string scale.  Therefore,
\be
\ell_s \gtrsim R_{\rm AdS}^{3/4} \ell_{\rm pl}^{1/4} \sim \sigma_* .
\ee
As far as we know, no stable AdS$_3$ string compactifications violate this inequality.  This may just be a ``lamp-post'' effect, i.e. stable string compactifications with smaller compact dimensions may exist but simply be much more difficult to find.  And the inequality relating $\ell_s$ and $\sigma_*$ may be coincidental.  On the other hand, a tantalizing explanation is that spacetime itself breaks down at the scale $\sigma_*$, creating an obstacle to the existence of weakly coupled strings with a smaller string length.\footnote{It would be interesting to study the interplay in compactified theories between the AdS$_3$  gravitational contributions we have included and the contributions from Kaluza Klein modes of the large compact directions.  Since the KK modes encode the fact that the theory really involves  higher-dimensional gravity, they may soften or remove the UV/IR mixing, similarly to what occurs in computations of the free energy \cite{Giombi:2013fka}.}

\section{Discussion}  

If a theory's dynamics are fully non-local, then the underlying spacetime picture loses its meaning, becoming a mere book-keeping device --  and the values of fields at different spacetime points become arbitrary independent variables.  What makes spacetime more than just a label is some notion of locality, which can be diagnosed using correlations. In a physical spacetime, nearby observables should be highly correlated, whereas correlators of distant fields should be small.  In a theory of quantum gravity, however, spacetime may play a role intermediate between these two extremes, with observables becoming more highly correlated as they approach each other, up to a point, beyond which local spacetime is revealed as a mere approximation.

While the two-point function of fields in a complete theory of quantum gravity is beyond the scope of presently available techniques, in this paper we have settled for something simpler that contains much of the same physics.  We have computed the two-point function $\< \phi(X_1) \phi(X_2)\>$ of an exact `proto-field' $\phi$ that is reconstructed in the bulk of AdS$_3$ in terms of a boundary primary operator $\CO$ and all of its Virasoro descendants.  Equivalently, in perturbation theory $\phi$ correlators are fully dressed by all graviton loops, but without any quantum corrections from matter fields.  The proto-field is a quantity defined in the spirit of the conformal bootstrap, in that it leverages the non-perturbative power of the conformal symmetry in the CFT$_2$ by resumming all contributions in an irreducible representation of the Virasoro algebra.  

In this paper we have developed techniques to compute the correlator $\< \phi(X_1) \phi(X_2)\>$ and characterized some of its most striking features.  We have analyzed the distance scale where it develops singularities and imaginary pieces, and we have interpreted these phenomena as an indication of the breakdown of bulk locality, as summarized in section \ref{sec:SummaryPhysics}.  For light bulk fields at large $c$, quantum effects produce the most imporant non-perturbative pathologies.
But in the semiclassical limit of fixed $h/c$ (or $G_N m_\phi$) and large central charge, branch cuts and imaginary pieces were already visible.  We have found that these semiclassical pathologies are not misleading, as they appear to persist in the exact quantum propagator.   

It should be possible to derive our semiclassical results, including the imaginary parts, from a bulk gravity calculation.  Such a derivation may shed light on the nature of any physical instabilities associated with these imaginary pieces.  Moreover, while the full generalization of our approach to higher dimensions is probably impossible (since in higher dimensions graviton interactions are not fixed by symmetry), semiclassical gravity computations are likely to be tractable. It would also be interesting to connect our results with other work \cite{Almheiri:2014lwa, Donnelly:2015hta, Jafferis:2017tiu, Ghosh:2017pel} on the breakdown of locality in quantum gravity.

Our results tentatively suggest a more general lesson -- when we attempt to define an exact bulk observable, we may induce small  violations of unitarity, even if the underlying CFT is healthy.  To test this idea we will need to better understand more general bulk correlators, their dependence on  CFT data, and their implications for physical bulk measurements.

We have studied the $\phi$ propagator at spacelike separations, as is most natural when we take a Euclidean CFT$_2$ as our starting point.  The Lorentzian correlators of any number of local CFT operators can be precisely determined via the analytic continuation of Euclidean correlators \cite{Luscher:1974ez, Hartman:2015lfa}.  It is much less clear whether Lorentzian $\phi$ correlators can be determined in the same way, because $\phi$ arises from an infinite sum of local operators in the CFT and carries an emergent bulk coordinate label.  This question may be connected with the gauge-dependence of $\phi$, since any analytic continuation in a bulk coordinate will clearly depend on our choice of the coordinate system!  At a pragmatic level, the most obvious next step would be to analytically continue $(z, \bar z)$  to Lorentzian signature, as these coordinates have a natural correspondence with the locations of operators in the boundary CFT$_2$.  This simply leads to the continuation from $\sigma > 0$ to $\sigma < 0$.
There are several formulas that hint at a simple analytic relation between the correlator at $\sigma$ and $-\sigma$. For example, the semiclassical potential $T$ (\ref{eq:Tphiphi}) is  $g'(\sigma)$ times an anti-symmetric function of $\sigma$.\footnote{Interestingly, the infinite $h$ result for $\< \phi \phi\>$ is formally invariant under $\sigma \rightarrow -\sigma$, though the path from positive to negative $\sigma$ passes around a branch cut that can break the symmetry and introduce dependence on the operator ordering.}  In any case, it will be very interesting to understand how the non-perturbative non-localities that we have discovered manifest in Lorentzian signature.

We have focused on the propagator of $\phi$ because it is the most tractable non-trivial $\phi$ correlator.  But another quantity that would be extremely interesting to compute is the heavy-light correlator
\be
\label{eq:NextStep}
\< \CO_H \CO_H \phi_L \CO_L\>  ,
\ee
where $\CO_H$ is a heavy operator and $\phi_L$ is the bulk proto-field made from $\CO_L$.  This correlator computes the bulk-to-boundary propagator for $\phi_L$ in the background of a heavy state such as a BTZ black hole, and therefore could be used to probe what happens when $\phi_L$ approaches a black hole horizon.  Many of the methods used in this paper to study $\< \phi \phi\>$ should be applicable to this heavy-light correlator as well, though the calculations will be more complicated because of the extra operator insertions.  Roughly speaking, each invariant contribution to the heavy-light bulk-boundary correlator will have the complexity of a 5-pt conformal block, since  $\phi$ involves a sum over an infinite set of Virasoro descendants. 

By computing equation (\ref{eq:NextStep}), we may begin to study the properties of black hole microstates without relying on bulk perturbation theory.  Major technical and conceptual challenges remain, but it appears that a direct investigation of the horizon may be possible.  As a first step, it will be interesting to explore features that emerge at the Euclidean horizon (the tip of the `cigar') as a consequence of the failure of the KMS condition \cite{Balasubramanian:2007qv, Fitzpatrick:2016ive, Chen:2017yze, Bena:2017upb, Tyukov:2017uig} in black hole microstate backgrounds.

 A distinct  line of inquiry will be to search for an improved observable that is free of the UV/IR mixing we observe in $\< \phi \phi\>$.  Along these lines, it would be rewarding to obtain a definition of bulk fields in other gauges.  Another possibility is that rather than evaluating the propagator in the vacuum, one ought to introduce a sum over boundary graviton configurations, along the lines of the way soft photons resolve IR divergences in 4d QED.   That is, it may be that the non-IR-safety of $\phi$ is similar to the physics of Sudakov factors, and when one attempts to produce and detect $\phi$ particles, one unavoidably produces some gravitons in the process.  It would be interesting to try to construct an IR-safe observable, directly related to local measurements in the bulk, and to see to what extent the behavior of $\< \phi \phi\>$ is modified.  

The UV/IR mixing behavior of the propagator might have a gauge invariant footprint in CFT observables.  For example, at the level of diagrammatics one would expect $\< \phi \phi \>$ to contribute as an intermediate propagator in CFT correlators computed using Witten diagrams.  Thus it would be interesting to examine the one-loop gravitational corrections \cite{Fitzpatrick:2015dlt, Chen:2016cms} to a 4-pt CFT$_2$ correlator including a scalar exchange.  At a deeper level, there is a simple relationship between the Zamolodchikov recursion relations that compute the exact bulk propagator and the relations that compute Virasoro blocks for 4-pt correlators.  Along with the idea of geodesic Witten diagrams \cite{Hijano:2015zsa} for conformal blocks, this may provide a direct avenue for further exploration.

\section*{Acknowledgments} 
 
We would like to thank Nikhil Anand, Nima Arkani-Hamed, Ibou Bah, Steve Carlip, Ethan Dyer, Simone Giombi, David Gross, Simeon Hellerman, Shamit Kachru, Ami Katz, Igor Klebanov, Markus Luty, Juan Maldacena, Shiraz Minwalla, Silviu Pufu, Suvrat Raju, Dan Roberts, Edgar Shaghoulian, Herman Verlinde, and Xi Yin  for discussions or correspondence.   We would also like to thank and the participants of the KITP Resurgence program and Simons Bootstrap Conference for discussions. JK would like to thank the KITP for hospitality while some of this work was completed. DL would like to thank Princeton University for hospitality while some of this work was completed.
ALF was supported in part by the US Department of Energy Office of Science under Award Number DE-SC-0010025.   JK and HC  have been supported in part by NSF grant PHY-1454083.  ALF, JK and DL were also supported in part by the Simons Collaboration Grant on the Non-Perturbative Bootstrap.

\appendix

\addtocontents{toc}{\protect\setcounter{tocdepth}{1}}
\section{Perturbative Computations of the Propagator}
\label{app:PerturbativePropagator}

In this section we will show that our first order result for the full propagator 
\begin{small}
\be
\< \phi \phi \> &=& \frac{\rho^h}{1-\rho}  \left(1 + \frac{12}{c } \left(\frac{\rho  \left(2 h^2 (\rho -1)^2+h (\rho  (3 \rho
   -11)+2) (\rho -1)+\rho ^2 ((\rho -5) \rho +10)\right)}{(\rho -1)^4} \right. \right.
\nn \\   && \left. \left. +
   2 h \rho ^2 \Phi (\rho ,1,2 h+1)+h \rho ^{1-2 h} B_{\rho }(2 h+1,-1)+2 (h-1) h
   \log (1-\rho )\right) + \CO \left( \frac{1}{c^2} \right) \right)
   \nn \\
\ee
\end{small}
follows directly from perturbation theory.  Note that this is the full propagator, which receives equal contributions from $T$ and $\bar T$, and so its $1/c$ correction is enhanced by a factor of $2$ compared to the purely holomorphic propagator. Primarily, we will be showing how this result matches an AdS$_3$ gravitational loop calculation (similar calculations in higher dimensions were recently studied in \cite{Giombi:2017hpr}, but as far as we know this calculation has not been carried out previously in AdS$_3$).  However, we will also demonstrate how the important $h$-independent $\frac{1}{c}$ terms arise directly from our definition of $\phi$ using a unitarity-based argument (an explicit sum over intermediate states).  We also provide a comparison with $U(1)$ Chern-Simons theory at short-distances, which does not display power-law UV/IR mixing.

\subsection{AdS$_3$ Gravity at One-Loop}
\label{app:GravitonLoop}

The only non-vanishing contribution to $\< \phi(X_1) \phi(X_2) \>$ from bulk perturbation theory at order $1/c$ comes from the diagram of Fig. \ref{fig:ScalarGravitonLoopDiagram}, as our regulator sets contact diagrams to zero.  In position space in AdS$_3$, this contributes 
\be
\< \phi(X_1) \phi(X_2) \> = \int d^3 X d^3 Y \, G(X_1,Y_1) G(Y_1,Y_2)  H(Y_1,Y_2) G(Y_2,X_2) V_1 V_2
\ee
where $G$ is a scalar propagator, $H$ is a graviton propagator, and $V_1$ and $V_2$ are vertex factors associated with vertices and index contractions at $Y_1$ and $Y_2$, which we will specify below.  

We can greatly simplify the computation by acting on  the correlator with $(\nabla^2 + m^2)$, the Klein-Gordon operator associated with both $X_1$ and $X_2$ \cite{Howtozintegrals}.  This collapses both of the external propagators to delta functions, giving
\be
\left( \nabla_1^2 + m^2 \right) \left( \nabla_2^2 + m^2 \right) \< \phi(X_1) \phi(X_2) \> = V_1 H(X_1,X_2) G(X_1,X_2)  V_2
\ee
so now there are no integrals to do.  The tree-level scalar propagator is simply
\be
G(X_1, X_2) = \frac{\rho^h}{1 - \rho}
\ee
as usual.  The graviton propagator in our gauge is identical to the stress tensor correlator $\< T(z_1) T(z_2) \>$, so it is simply
\be
H(X_1, X_2) = \frac{1}{2 c (z_1 - z_2)^4} \to \frac{1}{2c} \frac{\rho ^2}{\left(\sqrt{\rho }-1\right)^8}
\ee
where we have re-written $z_{12}$ in terms of $\rho = e^{-2 \sigma(X_1, X_2) }$ by fixing all of the parameters other than $z_{12}$.  Here we are only computing the holomorphic part, but of course the anti-holomorphic part makes an equal anti-holomorphic contribution.
The vertex factors arise entirely from differentiating $G(X_1, X_2)$ by $\partial_{\bar z_1}^2$ and $\partial_{\bar z_2}^2$, and this leads to
\be
V_1 G(X_1, X_2) V_2 &=& 
-\frac{16 h^4 \left(\sqrt{\rho }-1\right)^3 \rho ^h}{\left(\sqrt{\rho}+1\right)^5}+\frac{16 h^3 \left(\sqrt{\rho }-1\right)^2 (7 \rho +3) \rho^h}{\left(\sqrt{\rho }+1\right)^6}
\nn \\
&&
-\frac{4 h^2 \left(\sqrt{\rho }-1\right) (\rho (71 \rho +98)+11) \rho ^h}{\left(\sqrt{\rho }+1\right)^7}
 -\frac{120 (\rho +1)   (\rho  (\rho +5)+1) \rho ^{h+1}}{\left(\sqrt{\rho }-1\right) \left(\sqrt{\rho }+1\right)^9}
 \nn \\
&&
+\frac{4 h (\rho  (\rho  (77 \rho +239)+101)+3) \rho^h}{\left(\sqrt{\rho }+1\right)^8}
\ee
when written in terms of $\rho$.  Altogether, this means that we should expect 
\be
V_1 H_{12} G_{12} V_2 &=&  -\frac{16 h^4 \rho ^{h+2}}{(\rho -1)^5}+\frac{16 h^3 (7 \rho +3) \rho ^{h+2}}{(\rho-1)^6}-\frac{4 h^2 \left(71 \rho^2+98 \rho +11\right) \rho ^{h+2}}{(\rho-1)^7}
 \\
&& 
+\frac{4 h \left(77 \rho ^3+239 \rho ^2+101 \rho +3\right) \rho^{h+2}}{(\rho -1)^8}-\frac{120 \left(\rho ^3+6 \rho ^2+6 \rho +1\right) \rho^{h+3}}{(\rho -1)^9} \nn
\ee
This should be equal to $\left( \nabla_1^2 + m^2 \right) \left( \nabla_2^2 + m^2 \right) \< \phi \phi \>$.  This quantity can also be re-written in terms of $\rho$; writing $K(\rho)$ for the propagator, we find
\be
\label{eq:DoubleKG}
\left( \nabla_1^2 + m^2 \right) \left( \nabla_2^2 + m^2 \right) \< \phi \phi \>
&=&
16 (h-1)^2 h^2 K(\rho) + \frac{64 \left(-h^2+h+1\right) \rho ^2 K'(\rho )}{\rho -1}
\nn \\ &&
-\frac{32 \rho ^2 ((h-1) h (\rho -1)-7 \rho +1) K''(\rho )}{\rho -1}
\nn \\ &&
+\frac{64 \rho ^3 (2 \rho -1) K^{(3)}(\rho )}{\rho -1}+16 \rho ^4 K^{(4)}(\rho )
\ee
Apparently we are faced with the daunting task of solving a $4$th order ODE with a complicated source.  Fortunately, we already know part of the answer from semiclassical calculations (keeping the $h^2/c$ terms) and also from computations as $h \to 0$ in appendix \ref{app:PerturbativePropagatorfromDefinitionphi}.  After inputing these terms and then leaving the remaining terms in $K(\rho)$ as an unknown function, we were able to solve.  And given a proposed $K(\rho)$, it is very easy to verify that it is in fact valid by inputting it into the differential equation.

Using this method, we find that the full $1/c$ correction due to the holomorphic gravitons is
\be
&& \frac{6 \rho ^h}{1-\rho } \left(\frac{\rho  \left(2 h^2 (\rho -1)^2+h (\rho  (3 \rho
   -11)+2) (\rho -1)+\rho ^2 ((\rho -5) \rho +10)\right)}{(\rho -1)^4} \right.
\nn \\   && \left. +
   2 h \rho ^2 \Phi (\rho ,1,2 h+1)+h \rho ^{1-2 h} B_{\rho }(2 h+1,-1)+2 (h-1) h
   \log (1-\rho )\right)
\ee
where $\Phi$ is a Lurch and $B$ is the Beta function.  The anti-holomorphic gravitons make an equal contribution at order $1/c$, so we simply need to double this result.  Intriguingly, if we expand as $\rho = e^{-2\sigma}$ then the singular terms are
\be
\frac{9}{8 c \sigma ^5}-\frac{3 (5+2 (-1+h) h)}{8 c \sigma ^3}+\frac{12
   (-1+h) h \log (\sigma )}{c \sigma }
\ee
So we see that the AdS mass $2h(2h-2)$ appears prominently, and we only have odd powers of $1/\sigma$ appearing (we have dropped some terms that are simply $1/\sigma$, as these are no more singular than the free field theory result).  Restoring the AdS scale, we have
\be
\< \phi \phi \> \approx \frac{1}{\sigma} \left( \frac{3 G_N R^3 }{4 \sigma^4}-\frac{G_N R(10+m^2 R^2) }{8 \sigma^2}+ 2 G_N m^2 R \log \left(\frac{\sigma }{R}\right)  \right)
\ee
to leading order at short distances.  This makes it clear that the scale $\sigma \sim \sqrt[4]{G_N R^3}$ has made an explicit appearance.

\subsubsection*{Comparison with $U(1)$ Chern-Simons}

The one-loop AdS$_3$ gravity result displays a surprising UV/IR mixing.  To better understand this result, we will briefly compare it with a $U(1)$ Chern-Simons theory.

The double application of the Klein-Gordon equation in (\ref{eq:DoubleKG}) applies to loop computations of the AdS$_3$ propagator in other theories.  If we re-write this equation in terms of $\sigma$, and only keep the terms that dominate at short distances, we find 
\be
\left( \nabla_1^2 + m^2 \right) \left( \nabla_2^2 + m^2 \right) f(\sigma) \approx f^{(4)}(\sigma )+\frac{4 f^{(3)}(\sigma )}{\sigma }
\ee
where $f(\sigma)$ is the propagator at short distances.  In a $U(1)$ Chern-Simons theory, the propagator and vertices will be closely related to those that we found for gravity.  We expect $\< A_z(X_1) A_z(X_2) \> \propto \frac{1}{z_{12}^2}$ and the vertices can be obtained from $\partial_{\bar z_1} \partial_{\bar z_2}$ applied to the scalar field propagator.  In the short-distance limit, this leads to the differential equation
\be
F_{CS}^{(4)}(\sigma )+\frac{4 F_{CS}^{(3)}(\sigma )}{\sigma } \propto \frac{1}{\sigma^5} + \cdots
\ee
with the solution
\be
\< \phi \phi \>_{CS} \propto -\frac{\log ( \sigma )}{6 \sigma} + \frac{\kappa}{\sigma}  + \cdots
\ee
to leading order at short distances,  where $\kappa$ is a free parameter (which would be fixed in the full solution by boundary conditions) and the ellipsis denotes less singular terms.  Thus we see that unlike AdS$_3$ gravity, in perturbation theory the bulk $U(1)$ Chern-Simons theory does not exhibit power-law UV/IR mixing at short-distances.

\subsection{Unitarity-Based Calculation from the Definition of $\phi$}
\label{app:PerturbativePropagatorfromDefinitionphi}

In this section we will use the large $c$ expansion of the $\CL_{-N}$ that define the level $N$ contribution to $\phi$ in order to directly compute the $1/c$ correction to the propagator as $h \to 0$.  One can interpret this as a unitarity-based version of the calculation of the previous section, as we are decomposing each $\phi$ in $\< \phi \phi \>$ into a sum over the `double-trace' states in the $T(z) \CO(0)$ OPE.
We previously computed \cite{Anand:2017dav} the first $1/c$ corrections to $\CL_{-N}$, which are the coefficients $\eta_{N, k}$ of
\be
\label{eq:1OvercSquareInMathematicalL}
\CL_{-N}=L_{-1}^{N}+\frac{1}{c}\sum_{k=2}^{N}\eta_{N,k}L_{-k}L_{-1}^{N-k} + \cdots
\ee
and found that (see appendix D.5.3 of \cite{Anand:2017dav})
\be
\eta_{N, k} = -\frac{12(h(k+1)+N-k)}{k(k^2-1)} \frac{N!}{(N-k)!}
\ee
We can use this result to directly compute the $1/c$ terms in $\< \phi \phi \>$.  In the rest of this section, we will only keep the effects that survive in the limit $h \to 0$, which means that we can drop the term above proportional to $h$.
 
We will also need the matrix element 
\be
\CM_{k, p}^{N, M} & = & \< L_{-k} L_{-1}^{N-k} \CO(z) L_{-p} L_{-1}^{M-p} \CO(w) \>  
\nn \\
& \approx &
\frac{h c}{6}  \frac{ (-1)^{M+N} (k+p-1)! (M+N-k-p-1)! }{(k-2)!(p-2)! (z-w)^{2h + N+M } } 
\ee
where we have only kept the leading term at small $h$.  We need to multiply by $\eta_{N,k}$ factors and sum over $k$ and $p$, giving (setting $w=0$)
\be
\sum_{p, k=2}^{M,N} \eta_{N,k} \eta_{M, p} \CM_{k, p}^{N, M} 
= \frac{6h(N-2) (N-1) (M-2)(M-1) (M+N-3)!
}{cz^{N+M}} \ee
To see how to use this result, let us recall the computation to leading order and compare it to the $1/c$ correction we wish to calculate. The global correlator can be computed from the sums
\be
\< \phi\phi \>_\text{global} = \sum_{N, M} \frac{(-1)^{N+M} y_1^{2M} y_2^{2N} }{N! M! (2h)_N (2h)_M} \< L_{-1}^M \bar L_{-1}^M \CO  L_{-1}^N \bar L_{-1}^N \CO \>
\ee 
where we have
\be
\< L_{-1}^M \bar L_{-1}^M \CO  L_{-1}^N \bar L_{-1}^N \CO \>  &=&  \frac{(2h)_{M+N} (2h)_{M+N}}{(z\bar z)^{M+N}}
\nn \\
& \approx & \frac{ 4h^2 [(M+N-1)!]^2}{(z \bar z)^{M+N}}
\ee
One can easily verify directly that these formula agree with $\< \phi \phi \>_\text{global} = \frac{\rho^h}{1-\rho}$ when $h \to 0$.

To obtain the $1/c$ correction, we must make the replacement
\be
\< L_{-1}^M \bar L_{-1}^M \CO  L_{-1}^N \bar L_{-1}^N \CO \> \to \<  \bar L_{-1}^M \CO   \bar L_{-1}^N \CO \> \sum_{p, k=2}^{M,N} \eta_{N,k} \eta_{M, p} \CM_{k, p}^{N, M} 
\ee
Similarly, we find exact agreement between 
\be
\< \phi\phi \>_{\frac{1}{c}} = \sum_{N, M} \frac{(-1)^{N+M} y_1^{2M} y_2^{2N} }{N! M! (2h)_N (2h)_M}  \frac{12h^2   (N-2)_2 (M-2)_2 (M+N-3)!  (M+N-1)!
}{c(z \bar z)^{M+N}}\nn
\ee
in the limit $h \to 0$ and our result
\be
6 \rho^3 \frac{(\rho^2 - 5 \rho + 10)}{c (1-\rho)^5}
\ee
for the holomorphic part of the correction to $\< \phi \phi \>$.

To perform the relevant sums, it is useful to write $s = M+N$ and first sum over $M$ with fixed $s$.  Setting $y_i =1$ WLOG and working with the variable $z \bar z$, this leads to
\be
\< \phi\phi \>_{\frac{1}{c}} = \sum_{s=4}^\infty \frac{3 (-1)^s 4^{s-3} (s-5) (s-4) (s-3) (s-2) (s-1)  \Gamma
   \left(s-\frac{5}{2}\right)}{\sqrt{\pi } c \Gamma (s+1) (z \bar z)^{s}}
\ee
The sum over $s$ can now be performed exactly, and in the small $z \bar z$ limit it gives
\be
\< \phi\phi \>_{\frac{1}{c}}  \approx \frac{9}{8 c (z \bar z)^{\frac{5}{2}} } + \cdots
\ee
as expected. However, note that the sums defining $\< \phi \phi \>$ provide a long-distance (or near-boundary) expansion, whereas the interesting physics occurs at short distances in the bulk.  Thus connecting the two regimes requires an analytic continuation, meaning that we need to perform the full sum to observe the short-distance singularity.  Each term in the sums over $M, N$ is more singular than the total.

\section{Details of the Computation of $\CF_{n,\bar n}$}
\label{app:ComputingKdF}

In this section, we provide the details for computing $\left\langle \phi_{i,j}^{n,\overline{n}}\phi_{i,j}^{n,\overline{n}}\right\rangle $.
In Section \ref{sec:HolomorphicDefinitions}, we defined $\phi_{i,j}^{n,\overline{n}}$ to be (WLOG, assuming $n\ge \bar n$)
\begin{small}
\begin{equation}
\phi_{i,j}^{n,\overline{n}}\left(y,z,\overline{z}\right)\equiv y^{2h+2n}\sum_{m=0}^{\infty}\left(-1\right)^{n+m}y^{2m}\left|L_{-1}^{n+m}\mathcal{O}\right|^{2}\frac{L_{-1}^{m}\mathcal{L}_{-n}^{\text{quasi},i}}{\left|L_{-1}^{m}\mathcal{L}_{-n}^{\text{quasi},i}\mathcal{O}\right|^{2}}\frac{\overline{L}_{-1}^{m+n-\overline{n}}\mathcal{\overline{L}}_{-\overline{n}}^{\text{quasi},j}}{\left|\overline{L}_{-1}^{m+n-\overline{n}}\overline{\mathcal{L}}_{-\overline{n}}^{\text{quasi},j}\mathcal{O}\right|^{2}}\mathcal{O}\left(z,\overline{z}\right),
\end{equation}
\end{small}
and $\mathcal{F}_{n,\overline{n}}\left(h\right)$ to be 
\begin{equation}
\mathcal{F}_{n,\overline{n}}\left(h\right)\equiv\left\langle \phi_{i,j}^{n,\overline{n}}\left(y_{1},z_{1},\overline{z}_{1}\right)\phi_{i,j}^{n,\overline{n}}\left(y_{2},z_{2},\overline{z}_{2}\right)\right\rangle \left|\mathcal{L}_{-n}^{\text{quasi},i}\overline{\mathcal{L}}_{-\overline{n}}^{\text{quasi},j}\mathcal{O}\right|^{2}
\end{equation}
Since eventually we'll show that $\mathcal{F}_{n,\overline{n}}\left(h\right)$
only depends on the level of the quasi-primary $\left(n,\overline{n}\right)$,
we'll suppress the indexes $\left(i,j\right)$. Defining
\begin{equation}
\mathcal{O}^{n,\overline{n}}=\mathcal{L}_{-n}^{\text{quasi}}\overline{\mathcal{L}}_{-\overline{n}}^{\text{quasi}}\mathcal{O}
\end{equation}
and using the following identities, 
\begin{align}
\left|\mathcal{O}^{n,\overline{n}}\right|^{2} & =\left|\mathcal{L}_{-n}^{\text{quasi}}\mathcal{O}\right|^{2}\left|\mathcal{L}_{-\overline{n}}^{\text{quasi}}\mathcal{O}\right|^{2}\nn\\
\left|L_{-1}^{n+m}\mathcal{O}\right|^{2} & =\left(2h\right)_{n+m}\left(n+m\right)!\\
\left|L_{-1}^{m}\mathcal{L}_{-n}^{\text{quasi}}\mathcal{O}\right|^{2} & =\left(2h+2n\right)_{m}m!\left|\mathcal{L}_{-n}^{\text{quasi}}\mathcal{O}\right|^{2}\nn
\end{align}
we find
\begin{align}
\phi^{n,\overline{n}}\left(y,z,\overline{z}\right)= & \frac{\left(-1\right)^{n}y^{2h+2n}}{\left|\mathcal{O}^{n,\overline{n}}\right|^{2}}\sum_{m=0}^{\infty}\left(-1\right)^{m}y^{2m}\frac{\left(2h\right)_{n+m}\left(n+m\right)!}{\left(2h+2n\right)_{m}m!\left(2h+2\overline{n}\right)_{n-\overline{n}+m}\left(n-\overline{n}+m\right)!}.\nonumber \\
 & \times L_{-1}^{m}\overline{L}_{-1}^{m}\left(\overline{L}_{-1}^{n-\overline{n}}\mathcal{O}^{n,\overline{n}}\left(z,\overline{z}\right)\right)
\end{align}

For simplicity, we'll define
\begin{align}
h_{n} & \equiv h+n,\nonumber \\
h_{\overline{n}} & \equiv h+\overline{n},\\
l & \equiv n-\overline{n}.\nonumber 
\end{align}
Then $\left\langle \phi^{n,\overline{n}}(y_1,z_1, \bar z_1)\phi^{n,\overline{n}}(y_2,z_2,\bar z_2\right\rangle $
is given by 
\begin{small}
\begin{align}\label{eq:PhinnbPhinnbDetail}
\left\langle \phi^{n,\overline{n}}\phi^{n,\overline{n}}\right\rangle = & \frac{\left(y_{1}y_{2}\right)^{2h_{n}}}{\left|\mathcal{O}^{n,\overline{n}}\right|^{4}}\sum_{m,m'=0}^{\infty}\frac{\left(-1\right)^{m+m'}y_{1}^{2m}y_{2}^{2m'}\left(2h\right)_{n+m}\left(n+m\right)!}{\left(2h_{n}\right)_{m}\left(2h_{\overline{n}}\right)_{l+m}m!\left(l+m\right)!}\frac{\left(2h\right)_{n+m'}\left(n+m'\right)!}{\left(2h_{n}\right)_{m'}\left(2h_{\overline{n}}\right)_{l+m'}m'!\left(l+m'\right)!}\nonumber \\
 & \times\left\langle L_{-1}^{m}\overline{L}_{-1}^{m}\left[\overline{L}_{-1}^{l}\mathcal{O}^{n,\overline{n}}\right]L_{-1}^{m'}\overline{L}_{-1}^{m'}\left[\overline{L}_{-1}^{l}\mathcal{O}^{n,\overline{n}}\right]\right\rangle .
\end{align}	
\end{small}
The second line of above equation is given by
\begin{align}
 & \left\langle L_{-1}^{m}\overline{L}_{-1}^{m}\left[\overline{L}_{-1}^{l}\mathcal{O}^{n,\overline{n}}\right]L_{-1}^{m'}\overline{L}_{-1}^{m'}\left[\overline{L}_{-1}^{l}\mathcal{O}^{n,\overline{n}}\right]\right\rangle \nonumber \\
= & \partial_{z_{1}}^{m}\partial_{\overline{z}_{1}}^{m+l}\partial_{z_{2}}^{m'}\partial_{\overline{z}_{2}}^{m'+l}\frac{\left(-1\right)^{n+\overline{n}}\left|\mathcal{O}^{n,\overline{n}}\right|^{2}}{\left(z_{1}-z_{2}\right)^{2h_{n}}\left(\overline{z}_{1}-\overline{z}_{2}\right)^{2h_{\overline{n}}}}\label{eq:TwoPointBlockCalculation}\\
= & \left|\mathcal{O}^{n,\overline{n}}\right|^{2}\frac{\left(2h_{n}\right)_{m+m'}\left(2h_{\overline{n}}\right)_{2l+m+m'}}{z_{12}^{2h_{n}+m+m'}\overline{z}_{12}^{2h_{\overline{n}}+2l+m+m'}}\nonumber ,
\end{align}
where in the second line, we've used the fact that the two-point function of the quasi-primaries is given by
\begin{equation}
\left\langle \mathcal{O}^{n,\overline{n}}\left(z_{1}, \overline{z}_1\right)\mathcal{O}^{n,\overline{n}}\left(z_{2},\overline{z}_2\right)\right\rangle =\frac{\left(-1\right)^{n+\overline{n}}\left|\mathcal{O}^{n,\overline{n}}\right|^{2}}{\left(z_{1}-z_{2}\right)^{2h_{n}}\left(\overline{z}_{1}-\overline{z}_{2}\right)^{2h_{\overline{n}}}}.
\end{equation}
and the $\left(-1\right)^{n+\overline{n}}$ is canceled by the derivatives
acting on $z_{1}$ and $\overline{z}_{1}$ in the third line of equation
(\ref{eq:TwoPointBlockCalculation}). For later convenience, the factor
$\left(2h_{\overline{n}}\right)_{m+m'+2l}$ in the last line of equation
(\ref{eq:TwoPointBlockCalculation}) can be written as 
\begin{equation}
\left(2h_{\overline{n}}\right)_{2l+m+m'}=\left(2h_{\overline{n}}\right)_{2l}\left(2h_{n}\right)_{m+m'}.\end{equation}

Now let's simplify the first line of equation (\ref{eq:PhinnbPhinnbDetail}):
\begin{align}
 & \frac{\left(2h\right)_{n+m}\left(n+m\right)!}{\left(2h_{n}\right)_{m}\left(2h_{\overline{n}}\right)_{l+m}m!\left(l+m\right)!}\frac{\left(2h\right)_{n+m'}\left(n+m'\right)!}{\left(2h_{n}\right)_{m'}\left(2h_{\overline{n}}\right)_{l+m'}m'!\left(l+m'\right)!}\\
= & \left[\frac{n!\left(2h\right)_{n}}{l!\left(2h_{\overline{n}}\right)_{l}}\right]^{2}\frac{\left(2h+n\right)_{m}\left(2h+n\right)_{m'}\left(n+1\right)_{m}\left(n+1\right)_{m'}}{\left(2h_{n}\right)_{m}\left(2h_{\overline{n}}+l\right)_{m}\left(2h_{n}\right)_{m'}\left(2h_{\overline{n}}+l\right)_{m'}\left(l+1\right)_{m}\left(l+1\right)_{m'}}\frac{1}{m!m'!}.\nonumber 
\end{align}

Putting everything together, we have 
\begin{align}
\mathcal{F}_{n,\overline{n}}\left(h\right)\equiv & \left\langle \phi^{n,\overline{n}}\phi^{n,\overline{n}}\right\rangle \left|\mathcal{O}^{n,\overline{n}}\right|^{2}\\
= & \left(\frac{y_{1}y_{2}}{z_{12}\overline{z}_{12}}\right)^{2h_{n}}\left(2h_{\overline{n}}\right)_{2l}\left[\frac{n!\left(2h\right)_{n}}{l!\left(2h_{\overline{n}}\right)_{l}}\right]^{2}\sum_{m,m'=0}^{\infty}\frac{\left(-\frac{y_{1}^{2}}{z_{12}\overline{z}_{12}}\right)^{m}\left(-\frac{y_{2}^{2}}{z_{12}\overline{z}_{12}}\right)^{m'}}{m'!m!}\nonumber \\
 & \times\frac{\left(2h_{n}\right)_{m+m'}\left(2h_{n}\right)_{m+m'}\left(2h+n\right)_{m}\left(2h+n\right)_{m'}\left(n+1\right)_{m'}\left(n+1\right)_{m}}{\left(2h_{n}\right)_{m}\left(2h_{n}\right)_{m'}\left(2h_{\overline{n}}+l\right)_{m}\left(2h_{\overline{n}}+l\right)_{m'}\left(l+1\right)_{m}\left(l+1\right)_{m'}}\nonumber \\
= & \left(Y_{1}Y_{2}\right)^{h_{n}}\left(2h_{\overline{n}}\right)_{2l}\left[\frac{n!\left(2h\right)_{n}}{l!\left(2h_{\overline{n}}\right)_{l}}\right]^{2}\nonumber \\
 & \times F_{0,3}^{2,2}\left(\begin{array}{cc}
2h_{n},2h_{n}: & 2h+n,2h+n;n+1,n+1;\\
-: & 2h_{n},2h_{n};2h_{n}-l,2h_{n}-l;l+1,l+1;
\end{array},-Y_{1},-Y_{2}\right)\nonumber 
\end{align}where $F_{0,3}^{2,2}$ is a Kampe de Feriet series and we've defined
\begin{equation}
Y_{1}\equiv\frac{y_{1}^{2}}{z_{12}\overline{z}_{12}},\qquad Y_{2}\equiv\frac{y_{2}^{2}}{z_{12}\overline{z}_{12}}.
\end{equation}

Now we will discuss several properties of the function $\mathcal{F}_{n,\overline{n}}$. First, it is not just a function of the geodesic separation between the two points. In addition, it depends on a parameter encoding the angle between the two points and the $z$-$\bar{z}$ plane. Most of the results presented in section \ref{subsec:FullVSHolo} are computed when the two points lie on the same constant $y$-plane. We note that if we take the other limit, where the separation is purely in the $y$ direction, the small $\frac{-y_{i}^{2}}{z_{12}\bar{z}_{12}}$ expansion presented above is not useful. To explore the behavior of $\mathcal{F}_{n,\overline{n}}$ in this configuration, we need to re-sum the series. We are not aware of existing results that fully solve this problem. However, we can partially re-sum the series using a Borel style procedure, yielding the following integral representation:
\be
\hspace{-1cm}\mathcal{F}_{n,\overline{n}}=2R_{h,n,\bar{n}}\eta^{h+n}\int_{0}^{\infty}daK_{0}\left(2\sqrt{\frac{a}{Y_1}}\right)a^{2\left(h+n\right)-1}W\left(h,n,\bar{n};a\right)W\left(h,n,\bar{n};\eta a\right)
\ee
where $K_{0}$ is the modified Bessel function and we've defined $\eta=\frac{Y_2}{Y_1}$, 
\be
W\left(h,n,\bar{n};a\right) =\,_{2}F_{3}\left(n+1,2h+n;2h+2n,n-\bar{n}+1,2h+n+\bar{n};-a\right),
\ee
and
\be
R_{h,n,\bar{n}}=(2h_{\bar n})_{2(n-\bar n)}\left(\frac{n!(2h)_n}{\Gamma\left(2h_n\right)\left(n-\bar{n}\right)!(2h_{\bar n})_{n-\bar{n}}}\right)^{2}.
\ee
This integral is typically convergent at large values of $Y_1$, making it useful for computing $\mathcal{F}_{n,\overline{n}}$ when the two points are only separated on the $y$-direction.

In the case the $\overline{n}=0$ (or $n=0$) we have $l=n$, and
the expression for $\mathcal{F}_{n,0}$ is simplified to be 
\begin{align}
\mathcal{F}_{n,0}\left(h\right) & =\left(2h\right)_{2n}\left(Y_{1}Y_{2}\right)^{2h_{n}}\sum_{m,m'=0}^{\infty}\frac{\left(2h_{n}\right)_{m+m'}\left(2h_{n}\right)_{m+m'}}{\left(2h_{n}\right)_{m}\left(2h_{n}\right)_{m'}}\frac{\left(-1\right)^{m+m'}}{m!m'!}Y_{1}^{m}Y_{2}^{m'}\nonumber \\
 & =\left(2h\right)_{2n}\left(Y_{1}Y_{2}\right)^{2h_{n}}F_{4}\left(2h_{n},2h_{n},2h_{n},2h_{n},-Y_{1},-Y_{2}\right)\\
 & =\left(2h\right)_{2n}\frac{\rho^{h+n}}{1-\rho}\nonumber 
\end{align}
with $\rho=\frac{\xi^{2}}{\left(1+\sqrt{1-\xi^{2}}\right)^{2}},\xi=\frac{2\sqrt{Y_{1}Y_{2}}}{1+Y_{1}+Y_{2}}$ and $F_{4}$ is the Appell hypergeometric function.

\section{Correlators of Stress Tensors with $\phi \phi$} \label{app:OPEblockCalculation}

In \cite{Anand:2017dav},  we used the OPE blocks of $\phi\mathcal{O}$ to compute
correlation functions of the form $\left\langle \phi\mathcal{O}T\cdots T\overline{T}\cdots\overline{T}\right\rangle $.
Similarly, we can derive the OPE block for two bulk operators $\phi\phi$,
and use it to compute the correlation functions of the form $\left\langle \phi\phi T\cdots T\overline{T}\cdots\overline{T}\right\rangle$ with the regulator proposed in Appendix B of \cite{Anand:2017dav}.
Notice that this method will only give the first several terms of
the large $c$ limit of $\left\langle \phi\phi T\cdots T\overline{T}\cdots\overline{T}\right\rangle $, up to order $\CO(c^0)$,
in contrast to the cases in \cite{Anand:2017dav}, where the correlation functions
$\left\langle \phi\mathcal{O}T\cdots T\overline{T}\cdots\overline{T}\right\rangle $ computed in that paper are exact. This is because this bulk-bulk OPE block does include the gravitational dressing of the $\phi$ operators.

In the vacuum AdS$_3$ metric
\begin{equation}
ds^{2}=\frac{du^{2}+dwd\overline{w}}{u^{2}},
\end{equation}
the bulk-bulk propagator is given by 
\begin{equation}\label{eq:Vacuum2pt}
\left\langle \phi\left(u_{0},w_{0},\overline{w}_{0}\right)\phi\left(u_{1},w_{1},\overline{w}_{1}\right)\right\rangle =\frac{e^{-2h\Sigma}}{1-e^{-2\Sigma}}
\end{equation}
where the geodesic length $\Sigma$ between the two bulk operators
is given by 
\begin{equation}
\Sigma=\log\frac{1+\sqrt{1-\Xi^{2}}}{\Xi},\qquad\Xi=\frac{2u_{0}u_{1}}{u_{0}^{2}+u_{1}^{2}+\left(w_{0}-w_{1}\right)\left(\overline{w}_{0}-\overline{w}_{1}\right)}
\end{equation}
Now, we can view the coordinates $\left(u,w,\overline{w}\right)$
as the result of an operator valued diffeomorphism from a general
vacuum metric of the form
\begin{equation}\label{eq:GeneralVacuumMetric}
ds^{2}=\frac{dy^{2}+dzd\overline{z}}{y^{2}}-\frac{6T\left(z\right)}{c}dz^{2}-\frac{6\overline{T}\left(z\right)}{c}d\overline{z}^{2}+y^{2}\frac{36T\left(z\right)\overline{T}\left(z\right)}{c^{2}}dzd\overline{z}.
\end{equation}
The diffeomorphism \cite{Roberts:2012aq} is given by 
\begin{align}
w & \rightarrow f\left(z\right)-\frac{2y^{2}\left(f'\left(z\right)\right)^{2}\overline{f}''\left(\overline{z}\right)}{4f'\left(z\right)\overline{f}'\left(\overline{z}\right)+y^{2}f''\left(z\right)\overline{f}''\left(\overline{z}\right)},\nonumber \\
\overline{w} & \rightarrow\overline{f}\left(z\right)-\frac{2y^{2}\left(\overline{f}'\left(\overline{z}\right)\right)^{2}f''\left(z\right)}{4f'\left(z\right)\overline{f}'\left(\overline{z}\right)+y^{2}f''\left(z\right)\overline{f}''\left(\overline{z}\right)},\\
u & \rightarrow y\frac{4\left(f'\left(z\right)\overline{f}'\left(\overline{z}\right)\right)^{\frac{3}{2}}}{4f'\left(z\right)\overline{f}'\left(\overline{z}\right)+y^{2}f''\left(z\right)\overline{f}''\left(\overline{z}\right)}.\nonumber 
\end{align}
And $T\left(z\right)$ (and similary for $\overline{T}\left(\overline{z}\right)$)
satisfies 
\begin{equation}
\frac{12T\left(z\right)}{c}=\frac{f'''\left(z\right)f'\left(z\right)-\frac{3}{2}\left(f''\left(z\right)\right)^{2}}{\left(f'\left(z\right)\right)^{2}},
\end{equation}
which can be solve order by order in $\frac{1}{c}$ and the first
two terms are 
\begin{equation}
f\left(z\right)=z+\frac{f_{1}\left(z\right)}{c}+\mathcal{O}\left(\frac{1}{c^{2}}\right)
\end{equation}
with 
\begin{equation}
f_{1}\left(z\right)=-6\int_{0}^{z}dz'\left(z-z'\right)^{2}T\left(z'\right).
\end{equation}

Suppose that the positions of the two operators in the general vacuum background are at $\left(y,0,0\right)$ and $\left(y,z,\overline{z}\right)$
\footnote{Here, we consider the case that the two bulk operators are at the
same bulk depth $y$ for simplicity.}, that is, $u_{0}=u\left(y,0,0\right)$ and $u_{1}=u\left(y,z,\overline{z}\right)$,
and similarly for $w_{0},\overline{w}_{0},w_{1}$ and $\overline{w}_{1}$, then as in the bulk-boundary case, we can expand  the geodesic separation in terms of large $c$
as follows
\begin{align}\label{eq:GeodesicLargecExpandion}
\log\frac{\Xi}{1+\sqrt{1-\Xi^{2}}}= & \log\frac{\xi}{1+\sqrt{1-\xi^{2}}}+K_{T}^{b}+K_{\overline{T}}^{b}+\mathcal{O}\left(\frac{1}{c^{2}}\right)
\end{align}
where $\xi=\frac{2y^{2}}{2y^{2}+z\overline{z}}$  and 
\begin{equation}
K_{T}^{b}=\frac{z\bar{z}f_{1}'(z)-2\bar{z}f_{1}(z)+y^{2}zf_{1}''(z)}{2c\sqrt{z\overline{z}\left(z\bar{z}+4y^{2}\right)}}.
\end{equation}
Here, we use superscribe $b$ in $K^{b}$ to denote that these are
the OPE blocks for two bulk operators, in contract to the case in \cite{Anand:2017dav}, where one of the operators is on the boundary. So plugging in the expression of $f_{1}$, we get
\begin{equation}
K_{T}^{b}=\frac{1}{c}\int_{0}^{z}dz'\frac{6\left(\bar{z}\left(z-z'\right)z'+y^{2}z\right)}{\sqrt{z\overline{z}\left(z\bar{z}+4y^{2}\right)}}T\left(z'\right)
\end{equation}
When sending $y$ to $0$, $K_{T}^{b}$ reduces to the OPE block
of the two operators on the boundary \cite{Anand:2017dav}.

Now, expanding the RHS of equation (\ref{eq:Vacuum2pt}) in terms of large $c$ using equation (\ref{eq:GeodesicLargecExpandion}), we get the OPE block of two bulk operators
\begin{align*}
\phi\left(y,0,0\right)\phi\left(y,z,\overline{z}\right)\sim & \frac{\rho^{h}}{1-\rho}\left[1+2\left(h+\frac{\rho}{1-\rho}\right)\left(K_{T}^{b}+K_{\overline{T}}^{b}\right)+\mathcal{O}\left(\frac{1}{c^{2}}\right)\right]
\end{align*}
with $\rho=\frac{\xi^2}{(1+\sqrt{1-\xi^2})^2}$. 
So using $\left\langle \phi\left(y,0,0\right)\phi\left(y,z,\overline{z}\right)\right\rangle _{\text{global}}=\frac{\rho^{h}}{1-\rho}$,
we find
\begin{align}\label{eq:PhiPhiTFromOPEBlock}
\frac{\left\langle \phi\left(y,0,0\right)\phi\left(y,z,\overline{z}\right)T\left(z_{1}\right)\right\rangle }{\left\langle \phi\left(y,0,0\right)\phi\left(y,z,\overline{z}\right)\right\rangle _{\text{global}}}= & 2\left(h+\frac{\rho}{1-\rho}\right)\left\langle K_{T}^{b}T\left(z_{1}\right)\right\rangle \\
= & 2\left(h+\frac{\rho}{1-\rho}\right)12\int_{0}^{z}dz'\frac{\bar{z}\left(z-z'\right)z'+zy^{2}}{2c\sqrt{z\bar{z}}\sqrt{z\bar{z}+4y^{2}}}\left\langle T\left(z'\right)T\left(z_{1}\right)\right\rangle \nonumber \\
= & 2\left(h+\frac{\rho}{1-\rho}\right)\frac{z^{2}\left[\left(6y^{2}+z\bar{z}\right)z_{1}\left(z_{1}-z\right)+2y^{2}z^{2}\right]}{2z_{1}^{3}\left(z_{1}-z\right)^{3}\sqrt{z\overline{z}\left(z\bar{z}+4y^{2}\right)}}\nonumber 
\end{align}

In Section 3.1, we shown that (equation (\ref{eq:TzForAppendix}))
\begin{equation}
\frac{\left\langle \phi\left(1,0,0\right)\phi\left(1,1,1\right)T\left(z_{1}\right)\right\rangle }{\left\langle \phi\left(1,0,0\right)\phi\left(1,1,1\right)\right\rangle}=\frac{\xi\left(\xi+\left(2\xi+1\right)\left(z_{1}-1\right)z_{1}\right)g'\left(\xi\right)}{2\left(z_{1}-1\right)^{3}z_{1}^{3}}.\label{eq:PhiPhiTFromSection3}
\end{equation}
This is actually equivalent to equation (\ref{eq:PhiPhiTFromOPEBlock})
if we replace $g(\xi)$ with its leading large $c$ limit, that
is
\begin{equation}
\lim_{c\rightarrow\infty}g\left(\xi\right)=\log\left\langle \phi\phi\right\rangle _{\text{global}}=\log\left(\frac{\rho^{h}}{1-\rho}\right).
\end{equation}
To see this, notice that $\frac{d}{d\xi}\log\left(\frac{\rho^{h}}{1-\rho}\right)=2\left(h+\frac{\rho}{1-\rho}\right)\frac{1}{\xi\sqrt{1-\xi}}$,
so that equation (\ref{eq:PhiPhiTFromOPEBlock}) can be writen as
\begin{small}
\begin{align}
\frac{\left\langle \phi\left(y,0,0\right)\phi\left(y,z,\overline{z}\right)T\left(z_{1}\right)\right\rangle }{\left\langle \phi\left(y,0,0\right)\phi\left(y,z,\overline{z}\right)\right\rangle _{\text{global}}} & =\left[\frac{d}{d\xi}\log\left(\frac{\rho^{h}}{1-\rho}\right)\right]\xi\sqrt{1-\xi^{2}}\frac{z^{2}\left(\left(6y^{2}+z\bar{z}\right)z_{1}\left(z_{1}-z\right)+2y^{2}z^{2}\right)}{2z_{1}^{3}\left(z_{1}-z\right)^{3}\sqrt{z\overline{z}\left(z\bar{z}+4y^{2}\right)}}\nn\\
 & =\xi\left[\frac{d}{d\xi}\log\left(\frac{\rho^{h}}{1-\rho}\right)\right]\frac{z^{2}\left(\xi z^{2}+\left(2\xi+1\right)\left(z_{1}-z\right)z_{1}\right)}{2\left(z_{1}-z\right)^{3}z_{1}^{3}}.
\end{align}
\end{small}
Setting $z=1$, we get exactly equation (\ref{eq:PhiPhiTFromSection3}). 

One can continue this procedure to compute correlators with more $T(z)$ (and $\bar T(z)$) insertions. But the result will not capture the $\CO(\frac{1}{c})$ terms of the exact correlator $\left\langle \phi\phi T\cdots T\overline{T}\cdots\overline{T}\right\rangle$ because it does not include the gravitational dressing of $\phi$.

\section{Algorithms for Implementing the Recursion Relations}
\label{app:AlgorithmForRecursion}
\subsection{$c$-recursion Algorithm}
The $c$-recursion relation is 
\begin{equation}
F\left(h,c\right)=1+\sum_{m\ge1,n\ge2}-\frac{\partial c_{m,n}\left(h\right)}{\partial h}\frac{A_{m,n}^{c_{m,n}}\left(2h\right)_{2mn}}{c-c_{m,n}\left(h\right)}\rho^{mn}F\left(h+mn,c_{m,n}\left(h\right)\right)\label{eq:cRecursionInAppendix}
\end{equation}
We know that the above recursion will give $F\left(h,c\right)$ as
the following expansion
\begin{equation}
F\left(h,c\right)=\sum_{N=0}^{\infty}C_{N}\left(2h\right)_{N}\rho^{N}.
\end{equation}
The factor $\left(2h\right)_{2mn}$ in the resiude will eventually
give $\left(2h\right)_{N}$, so for now let's consider how the coefficients
$C_{N}$ are contructed from the above recursion. Let's denote the residue
without the factor $\left(2h\right)_{2mn}$ as $R_{m,n}\left(h\right)=-\frac{\partial c_{m,n}\left(h\right)}{\partial h}A_{m,n}^{c_{m,n}}$.

The recursion (\ref{eq:cRecursionInAppendix}) is actually saying that every time we can write $N$
as a sum of products of intergers, i.e. 
\begin{equation}\label{eq:NDecomposition}
N=m_{1}n_{1}+m_{2}n_{2}\cdots+m_{i}n_{i},
\end{equation}
 then we get a contribution to $C_{N}$ from the recursion. In the above decomposition,
each term represents one iteration of the recursion. Denote the contribution
to $C_{N}$ from the decompsition whose last term is $m_{i}n_{i}$
as $C_{N,m_{i},n_{i}}$, then we can write $C_{N}$ as the following
sum
\begin{equation}\label{eq:CNSum}
C_{N}=\sum_{2\le m_{i}n_{i}\le N}C_{N,m_{i},n_{i}}.
\end{equation}
Then $C_{N,m_{i},n_{i}}$ will satisfy the following equation
\begin{small}
\begin{align}\label{eq:CNmini}
C_{N,m_{i},n_{i}}= & \frac{R_{m_{i},n_{i}}\left(h\right)}{c-c_{m_{i},n_{i}}\left(h\right)}\delta_{N,m_{i}n_{i}}\\
+ & \sum_{2\le m_{j}n_{j}\le N-m_{i}n_{i}}C_{N-m_{i}n_{i},m_{j},n_{j}}\frac{R_{m_{i},n_{i}}\left(h+N-m_{i}n_{i}\right)}{c_{m_{j},n_{j}}\left(h+N-m_{i}n_{i}-m_{j}n_{j}\right)-c_{m_{i},n_{i}}\left(h+N-m_{i}n_{i}\right)}.\nn
\end{align}
\end{small}
The first term can be thought of as the boundary condition, which
is just the case that there is only one term in the decomposition (\ref{eq:NDecomposition}).
The second term\footnote{The terms in the parentheses are the arguments of the functions $R_{m_i,n_i}$ and $c_{m,n}$, not to be confused as a factor times $R_{m_i,n_i}$ and $c_{m,n}$. }
sums over all the contributions from the cases where
there are more than one term in (\ref{eq:NDecomposition}) and supposes that the second last
term is $m_{j}n_{j}$: $N=m_{1}n_{1}\cdots+m_{j}n_{j}+m_{i}n_{i}$.

To actually implement the above algorithm to compute the coefficients
$C_{n}$ with $n\le N$ , we can first comput all the boundary terms
$C_{m_{i}n_{i},m_{i},n_{i}}=\frac{R_{m_{i},n_{i}}\left(h\right)}{c-c_{m_{i},n_{i}}\left(h\right)}$.
Then we increase $n$ from $n=m_{i}n_{i}+2$ to $N$. For each $n$,
we compute all the $C_{n,m_{i},n_{i}}$ via equation (\ref{eq:CNmini}).
We are able to do this because all the information (i.e. $C_{n-m_{i}n_{i},m_{j},n_{j}}$)
needed to compute $C_{n,m_{i},n_{i}}$ has already been computed.
The complexity for this algorithm will be roughly $N^{4}(\log N)^2$. 
\subsection{$h$-recursion Algorithm}
The algorithm for implementing $h$-recursion will be faster than
the $c$-recursion. The reason is that in the $h$-recursion
\begin{equation}
H\left(h,c\right)=1+\sum_{m,n}^{\infty}\frac{q^{mn}\left(2h_{m,n}\right)_{2mn}A_{m,n}^{c}}{h-h_{m,n}\left(c\right)}H\left(h_{m,n}+mn,c\right),
\end{equation}
each time each time we only change $h\rightarrow h_{m,n}+mn$, whereas
in the $c$-recursion, we change both $h\rightarrow h+mn$ and $c\rightarrow c_{m,n}\left(h\right)$.
Denoting the coefficients of $q^{N}$ in $H$ as $H_{N}$, i.e. $H=1+\sum_{N=2}^{\infty}H_{N}q^{N}$,
then we can write the solution of $H_{N}$ as in equations (\ref{eq:CNSum})
and (\ref{eq:CNmini}). But here, we'll think of the problem in another way. In equation (\ref{eq:CNSum})
and (\ref{eq:CNmini}), we were working  backward from the last step to arrive at $N$ from $N-m_{i}n_{i}$. But since in the $h$-recursion, $H\left(h_{m,n}+mn,c\right)$ only depends on $m,n$ and $c$,\footnote{In fact,  $F\left(h+mn,c_{m,n}\left(h\right)\right)$  depends on the value of $h+mn$, so it actually depends on the ``history'' of the recursion. For example, in the decomposition (\ref{eq:NDecomposition}), the first term will involve $F(h+m_1n_1,c_{m_1,n_1}(h))$, but the second term will involve $F\left(h+m_1n_1+m_2n_2,c_{m_2,n_2}(h+m_1n_1)\right)$. } it's 
actually easier to consider the problem here forward from the first
step, that is, we can write $H_{N}$ as the following sum (define
$\tilde{R}_{m,n}\equiv\left(2h_{m,n}\right)_{2mn}A_{m,n}^{c}$)
\begin{equation}
H_{N}=\sum_{2\le mn\le N}\frac{\tilde{R}_{m,n}}{h-h_{m,n}\left(c\right)}H_{m,n}^{\left(N-mn\right)}
\end{equation}
where $H_{m,n}^{\left(N-mn\right)}$ is the coefficient of $q^{N-mn}$
in $H\left(h_{m,n}+mn,c\right)$. Then it's easy to see that $H_{m,n}^{\left(N-mn\right)}$
is given by
\begin{equation}
H_{m,n}^{\left(N-mn\right)}=\sum_{2\le m_{i}n_{i}\le N-mn}\frac{\tilde{R}_{m_{i},n_{i}}}{h_{m,n}+mn-h_{m_{i},n_{i}}}H_{m_{i},n_{i}}^{\left(N-mn-m_{i}n_{i}\right)}
\end{equation}
The complexity for the $h$-recursion will be roughly  $N^{3}\left(\log N\right)^{2}$. We've described
the algorithm for implementing the $h$-recursion for Virasoro blocks in detail in \cite{Chen:2017yze}, and the $h$-recursion for $\left\langle \phi\phi\right\rangle _{\text{holo}}$ is almost the same (except that the residues are different, which doesn't affect the algorithm), so we refer the reader to Appendix A of that paper.

\newpage

\bibliographystyle{utphys}
\bibliography{VirasoroBib}

\end{document}